\documentclass[journal]{IEEEtran}
\ifCLASSINFOpdf
\else
\fi
\usepackage{url}
\usepackage{amsmath,amssymb,amsbsy}
\usepackage{bbm}
\usepackage{bm}
\usepackage{wrapfig}
\usepackage{graphicx}
\usepackage{epstopdf}
\usepackage{caption}
\usepackage{subcaption}
\usepackage{makecell}
\usepackage{algorithm}
\usepackage{algorithmicx}
\usepackage{algpseudocode}

\usepackage{multirow,booktabs}
\usepackage{threeparttable}
\usepackage{array}
\usepackage[numbers,sort&compress]{natbib}

\newcommand{\SODS}{\mathcal{SODS}}
\newcommand{\NB}{\operatorname{NB}}
\newcommand{\GLM}{\operatorname{GLM}}
\newcommand{\GLMs}{\operatorname{GLMs}}
\newcommand{\LVMs}{\operatorname{LVMs}}

\begin{document}
\graphicspath{ {./figures/} }
\title{An Efficient and Flexible Spike Train Model via Empirical Bayes}
%
%
%

\author{Qi~She\textsuperscript{*},~\IEEEmembership{Member,~IEEE,}
        Xiaoli~Wu\textsuperscript{*},
        Beth~Jelfs,~\IEEEmembership{Member,~IEEE,}
        Adam~S.~Charles,~\IEEEmembership{Member,~IEEE,} 
        and\\~Rosa~H. M.~Chan,~\IEEEmembership{Senior Member,~IEEE }
\thanks{Qi She was previously with the Department of Electrical Engineering, City University of Hong Kong, Hong Kong, and is currently with Bytedance AI Lab, Beijing, China,  e-mail: (sheqi1991@gmail.com)}
\thanks{Xiaoli Wu and Rosa H. M. Chan are with the State Key Laboratory of Terahertz and Millimeter Waves, Department of Electrical Engineering, City University of Hong Kong, Hong Kong, e-mail: (spikeliuliu@gmail.com, rosachan@cityu.edu.hk).}
\thanks{Beth~Jelfs is with the School of Engineering, RMIT University, Australia, e-mail: (beth.jelfs@rmit.edu.au).}
\thanks{Adam~S.~Charles is with the Biomedical Engineering department, Mathematical Institute for Data Science, Center for Imaging Science, and Kavli Neuroscience Discovery Institute at The Johns Hopkins University, Maryland, USA, e-mail: (adamsc@jhu.edu).}
\thanks{* These authors contributed equally to
the work.}}

%
%

\markboth{Journal of \LaTeX\ Class Files,~Vol.~14, No.~8, August~2015}%
{Shell \MakeLowercase{\textit{et al.}}: Bare Demo of IEEEtran.cls for IEEE Journals}
%



\maketitle

\begin{abstract}
Accurate statistical models of neural spike responses can characterize the information carried by neural populations. But the limited samples of spike counts during recording usually result in model overfitting. Besides, current models assume spike counts to be Poisson-distributed, which ignores the fact that many neurons demonstrate over-dispersed spiking behaviour. Although the Negative Binomial Generalized Linear Model (NB-GLM) provides a powerful tool for modeling over-dispersed spike counts, the maximum likelihood-based standard NB-GLM leads to highly variable and inaccurate parameter estimates. Thus, we propose a hierarchical parametric empirical Bayes method to estimate the neural spike responses among neuronal population. Our method integrates both Generalized Linear Models (GLMs) and empirical Bayes theory, which aims to (1) improve the accuracy and reliability of parameter estimation, compared to the maximum likelihood-based method for NB-GLM and Poisson-GLM; (2) effectively capture the over-dispersion nature of spike counts from both simulated data and experimental data; and (3) provide insight into both neural interactions and spiking behaviours of the neuronal populations. We apply our approach to study both simulated data and experimental neural data. The estimation of simulation data indicates that the new framework can accurately predict mean spike counts simulated from different models and recover the connectivity weights among neural populations. The estimation based on retinal neurons demonstrate the proposed method outperforms both NB-GLM and Poisson-GLM in terms of the predictive log-likelihood of held-out data. Codes are available in \url{https://doi.org/10.5281/zenodo.4704423}
\end{abstract}


\ifCLASSOPTIONpeerreview
\begin{center} \bfseries EDICS Category: 3-BBND \end{center}
\fi
%
\IEEEpeerreviewmaketitle

\section{Introduction}
%
%
%
\label{sec:Intro}
\IEEEPARstart{U}{nderstanding} the statistical dependencies between neural time series (\emph{e.g.}, spike counts, membrane potential, local field potential, EEG and fMRI) is vital to deducing how populations of neurons process information ~\cite{park2013kernel,chen2011statistical,okatan2005analyzing,song2015identification}. With the recent increase in accessibility of datasets containing spiking activities from large-scale neural populations, it is now possible to test the effectiveness of different methods for extracting functional dependencies at the neuronal level. \textit{Here, we consider the problem of recovering the connectivity weights between neurons in a network by merely observing their simultaneous spiking activity (\emph{e.g.}, spike counts)}.

Two of the most commonly used models for simultaneously recorded spiking activity are Generalized Linear Models ($\GLMs$)~\cite{pillow2008spatio,seghouane2014sparse,she2015reconstruction,she2016evaluating,she2016effective} and Latent Variable Models ($\LVMs$)~\cite{lakshmanan2015extracting,cichocki2015tensor,palmerston2018weighted,she2018reduced,she2020neural}. In the supervised setting, $\GLMs$ have used stimuli and spiking histories as covariates driving the spiking of a neural population~\cite{pillow2003maximum}. $\GLMs$ are also closely related to the well-known \emph{Hawkes process} model~\cite{hawkes1971spectra}, which has been used extensively for network inference~\cite{blundell2012modelling,moore2016hawkes,chen2017multivariate,mark2018network,linderman2014discovering}. $\GLMs$ essentially introduce a nonlinearity to the Hawkes process that ensures positive rates and allows for super- or sub-linear influences between nodes. Spike counts can then be generated using a count-valued distribution via selecting a certain bin size. In the unsupervised setting, $\LVMs$ focus on extracting a low-dimensional, smooth, and time-evolving latent structure that can capture the variability of the recorded data, both temporally and spatially. However, in both these settings, the spike counts in each time bin are often assumed to be conditionally Poisson, given the shared signal \cite{macke2011empirical}. 

While the Poisson assumption gives algorithmic conveniences, it implies the conditional mean and variance of spike counts are equal. This ignores the fact that in some cases the variance of spike counts could be much larger than its mean~\cite{churchland2010stimulus,goris2014partitioning}, that is, the data is over-dispersed. Various models have been proposed for representing the non-Poisson spike counts~\cite{gao2015high,stevenson2016flexible}. The Negative Binomial ($\NB$) model has been proposed as a solution to handling over-dispersed spike counts~\cite{lawless1987negative,chen1998mean,gao2016linear}. \textit{Here we intend to extract functional dependencies among neurons and give insights over neural interactions. Thus, $\NB$-$\GLM$ is a natural extension to achieve this goal while simultaneously capturing the over-dispersion of each neuron}.

Despite the ease of implementation of maximum likelihood estimation for the $\NB$-$\GLM$, when the recorded length of spike-train data is short, and a large number of neurons are recorded simultaneously, the accuracy of the estimated coefficients using $\GLMs$ with $\NB$ responses is low~\cite{brown2004multiple,okatan2005analyzing,chen2011statistical}. Unfortunately, in typical experimental settings, we cannot obtain long sequences of high-quality neural data due to (i) the short lifetime of some neurons, (ii) the limited viable time of recording materials and (iii) the micro-movement of recording electrodes during the activity of the animal~\cite{spira2013multi}. Hence, dataset size is often small due to either the short experiment length or the need for real-time inference~\cite{mauk2004neural,bertschinger2004real,she2016evaluating}. In this case, the maximum likelihood estimator of the parameters in the $\NB$ distribution leads to a large mean square error (MSE) under a standard $\GLM$. To alleviate this problem, one can employ regularizing priors in the form of a hierarchical model, as a trade-off between bias and variance. \textit{The key challenges of hierarchical modeling are how to design flexible prior structures and efficiently solve the non-trivial inference problem, which are the main focuses of our work}. 

In this paper, we propose a hierarchical empirical Bayes estimator for the probability parameters of $\NB$-$\GLM$, which helps to model Short Over-Dispersed Spike-Trains (we call ``$\SODS$''). Finally, it can capture accurate spiking behaviour of neurons and meanwhile recover connectivity among neurons under the $\GLM$ framework. Our hierarchical framework places a prior distribution on the parameters of the $\NB$ distribution, which can be estimated using empirical Bayes. The hyperparameters of the prior distribution are estimated using maximum marginal likelihood methods. The estimated value can then be used to obtain the mean spike counts. In summary, our main contributions are four-fold: 
\begin{enumerate} 
	\item  \textit{Provide a hierarchical extension of the $\NB$-$\GLM$ for modeling the statistical dependences among neural responses including a flexible link function};  
	\item  \textit{Develop an efficient empirical Bayes method for inference of the hierarchical $\NB$-$\GLM$ parameters}; 
	\item  \textit{Present more accurate prediction performance on retinal ganglion cells compared with state-of-the-art methods};
	\item  \textit{Give insightful findings on both neural interactions and spiking behaviours of real retina cells}. 
\end{enumerate}

This paper is organized as follows. In Section~\ref{ssec:Review}, we review the properties of the Negative Binomial Distribution and the differences between full and empirical Bayes approaches. In Section~\ref{sec:Model}, we introduce the ``$\SODS$" and the roles of the parameters. Section~\ref{sec:EmpBayes} discusses parameter estimations in $\SODS$, via numerical optimization of the maximum marginal likelihood. Section~\ref{sec:simulated} introduced different data simulation methods we used for estimators evaluation. Results for both simulated and experimental data are presented in Section~\ref{sec:Results}. Discussion of our contributions and findings are concluded in Section~\ref{sec:Discus}.

\section{Review}
\label{ssec:Review}

\subsection{Negative Binomial Distribution}
\label{ssec:Prior}
The Negative Binomial distribution can be seen as an extension of the Poisson distribution. The mean of Poisson distribution $\lambda$ represents the mean spike counts, which can be heterogeneous within different time intervals. By assuming the rate parameter $\lambda$ is generated from the Gamma distribution, we have:
\begin{eqnarray} 
Y\mid\lambda & \sim & \mathrm{Poisson}(\lambda), \label{eq:Poisson}\\
\lambda\mid \bm{r},\theta & \sim & \mathrm{Gamma}\left(\bm{r},\frac{\theta}{1-\theta}\right).
\label{eq:gamma}
\end{eqnarray}
where $\bm{r}$ is the shape parameter and $\theta$ is the probability parameter. Then the discrete random variable $Y$ follows the Negative Binomial distribution 
$\NB(\bm{\bm{r}}, \theta)$,
\begin{equation} \label{eq:nb}
P(Y=y\mid \bm{r},\theta) =\dbinom{\bm{r}+y-1}{y}\theta^{\bm{r}}(1-\theta)^{y}.
\end{equation}
Therefore, we can calculate $\mathbb{E}[Y]=\frac{(1-\theta)\bm{r}}{\theta}$, and $\mathrm{Var}[Y]=\frac{(1-\theta)\bm{r}}{\theta^{2}}$, with $\mathrm{Var}[Y] > \mathbb{E}[Y]$ since $0<\theta<1$. 

\begin{figure}[tbp]
\centering
\includegraphics[width=\columnwidth]{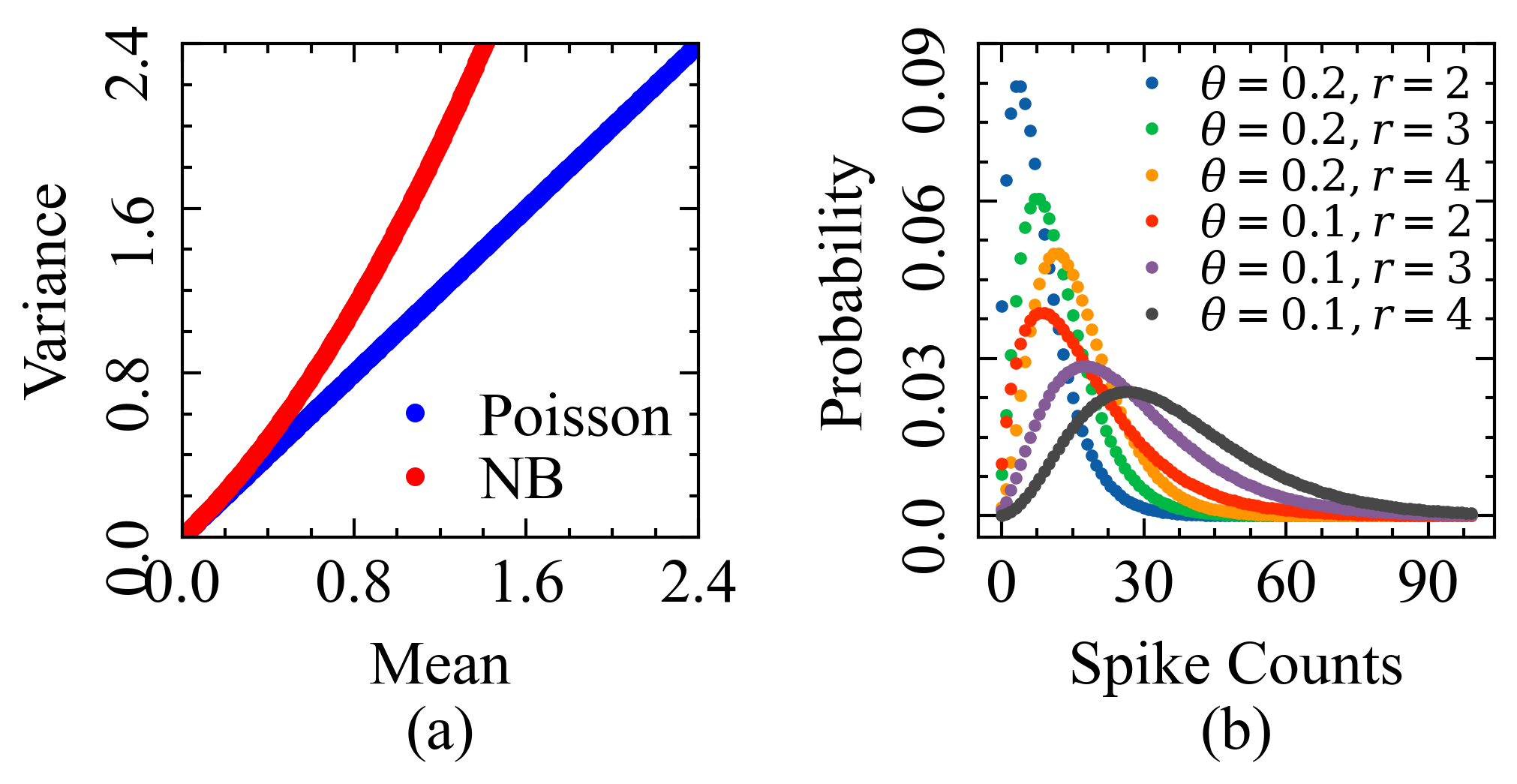}
\caption{(a) The relationship between variance and mean of the Poisson and Negative Binomial distributions. Negative Binomial shows super-Poisson variability (variance larger than mean). (b) The probability mass function of $\NB$ distribution with different parameters ($\theta$ = $\{0.1, 0.2\}$, $\bm{r}$ = $\{2, 3, 4\}$). Larger $\bm{r}$ and smaller $\bm{\theta}$ lead to a higher probability to generate large count values.}
\label{fig:review_nbglm}
\end{figure}

Fig.~\ref{fig:review_nbglm}a shows the relationship between variance and mean of Negative Binomial and Poisson distributions. The variance of the $\NB$ distribution is larger than the mean, which shows super-Poisson variability~\cite{goris2014partitioning,Charles165670}. Fig.~\ref{fig:review_nbglm}b shows the probability mass function of $\NB$ distribution with different combinations of parameters $\bm{r}$ and $\theta$.
\subsection{Empirical Bayes Inference}
\label{ssec:bayes}
Neuronal connectivity is modeled as an input-output system, which links the Negative Binomial output and spiking activities of input neurons via a hierarchical model. In the hierarchical setting, we can use either fully Bayesian inference or empirical Bayes to estimate the model parameters. Fully Bayesian inference assumes a specific hyperprior over the hyperparameters, which needs to be integrated out. As we often cannot obtain the closed form of this marginalization, fully Bayesian inference requires a sampling strategy to approximate this distribution. Correspondingly, this comes at a high computational cost, especially for high-dimensional data~\cite{scott2012fully}. 

On the other hand, empirical Bayes inference sets the parameters in the highest level of the hierarchical model with their most likely value. Setting the hyperparameters by maximizing the marginal likelihood function incurs a much lower computational cost. Hence, by combining empirical Bayes with the Negative Binomial $\GLM$ we can produce an estimator for the parameters of the Negative Binomial distribution which should efficiently handle both over-dispersion and smaller datasets. \textit{The key is to establish a network model in this framework and still capture super-Poisson spiking behaviour}.

\section{Proposed Method For ``$\SODS$"}
\label{sec:Model}
\subsection{Hierarchical Negative Binomial Model}
\label{sec:model}
\begin{figure}[tbp]
	\centering
	\includegraphics[width=0.45\textwidth]{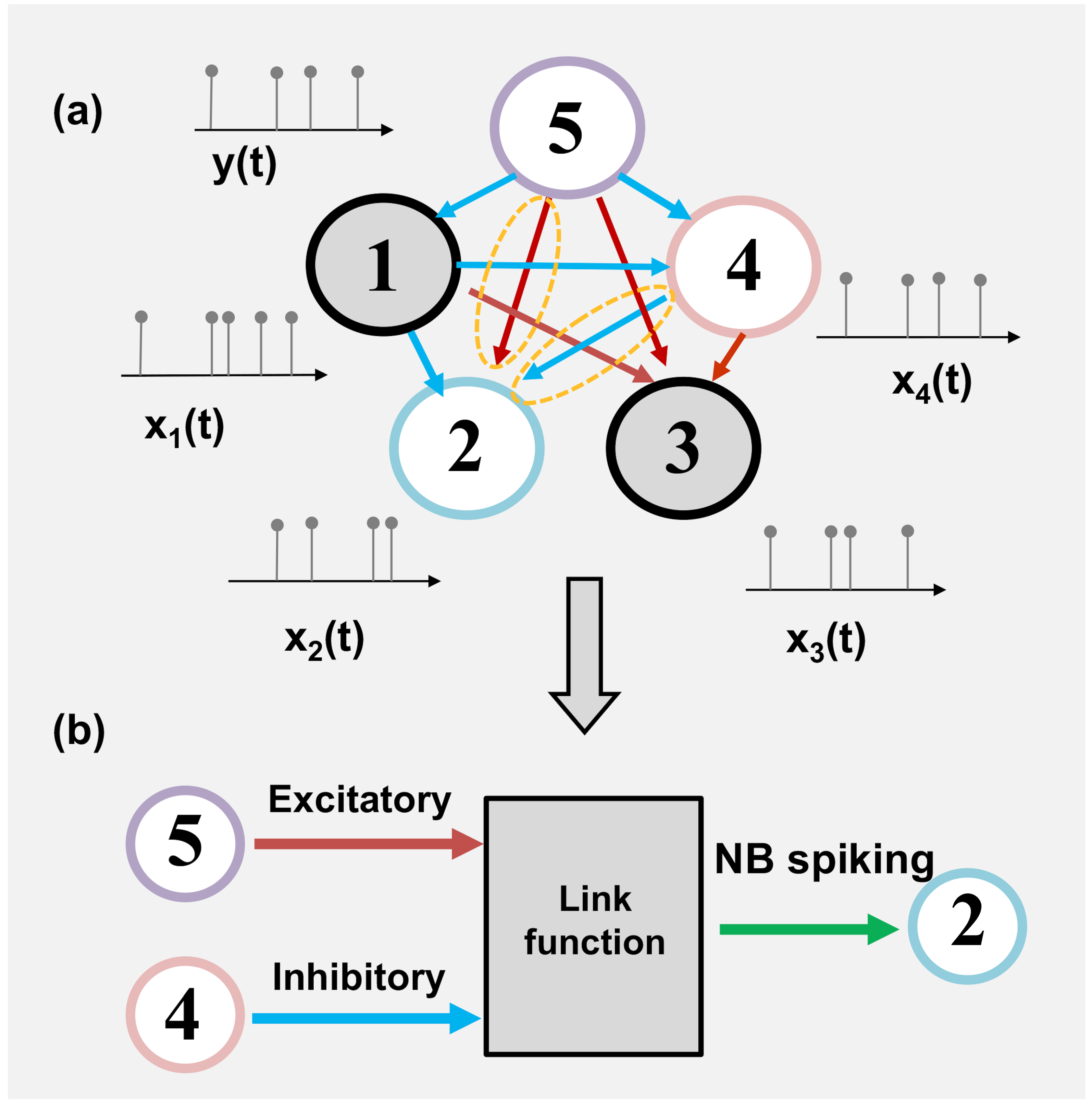}
	\caption{(a) A simple network model considered in our work, with excitatory and inhibitory effects and which generates $\NB$ spiking behaviour. (b) An illustration that neuron \#4 and \#5 have effects on \#2 through a flexible link function and then a NB distribution. The observed data are multiple spike-train data recorded simultaneously, which are presented as $x_{1:4}(t)$ and $y(t)$. Each grey line in $x$ and $y$ signals indicates one spike obtained.}
	\label{fig:model1}
\end{figure}

Fig.~\ref{fig:model1}a illustrates an example of a simple network considered in this work. We represent functional dependencies in this graph with the connection strengths (weights) between neurons. Note that we can use input neurons' spiking activities (\emph{e.g.}, neurons \#1: $x_{1}(t)$, \#2: $x_{2}(t)$, \#3: $x_{3}(t)$, \#4: $x_{4}(t)$) as regressors to predict an output neuron's spike counts (\emph{e.g.}, neuron \#5: $y(t)$). Fig.~\ref{fig:model1}b presents how neurons \#5 and \#4 have excitatory and inhibitory effects on neuron \#2 via a flexible link function and a $\NB$ distribution model, respectively. \textit{The accurate modeling of both link functions and the $\NB$ model can help to retrieve intrinsic coupling strengths effectively}. 

Let $Y_{ij}$ be the spike counts recorded from the $j$th experimental trial at time $i$. We assume that $\{Y_{i}\}_{j=1}^K$ is generated from the Negative Binomial distribution (with shape parameter $\bm{r}$ and probability parameter $\theta_{i}$). Furthermore, $\{Y_{i}\}_{j=1}^K$ are conditionally independent given the shared $\theta_{i}$ across different trials:
\begin{equation} \label{eq:y}
Y_{ij}\mid \bm{r},\theta_{i} \sim \NB(\bm{r}, \theta_{i}). 
\end{equation}
We use the beta distribution, the conjugate prior of the Negative Binomial distribution, as the prior for $\theta_{i}$:
\begin{equation} \label{eq:theta}
 \theta_{i} \sim \mathrm{Beta}(\alpha_{i}, \beta_{i}),
\end{equation} 
\emph{i.e.},
\[ p(\theta_{i}) = 
\dfrac{(\theta_{i})^{\alpha_{i}-1}(1-\theta_{i})^{\beta_{i}-1}}{\mathrm{B}(\alpha_{i},\beta_{i})},
\]
where $\alpha_{i}, \beta_{i}$ are the hyperparameters, and 
\begin{equation} \label{eq:beta}
\mathrm{B}(\alpha_{i},\beta_{i})=\int_{0}^{1}x^{\alpha_{i}-1}(1-x)^{\beta_{i}-1}\mathrm{d}x=\dfrac{\Gamma(\alpha_{i})\Gamma(\beta_{i})}{\Gamma(\alpha_{i}+\beta_{i})},
\end{equation} 
is the beta function, and $\Gamma(t)$ is the Gamma function.

\begin{figure}[t]
\centering
\begin{subfigure}{0.7\columnwidth}
\centering
\includegraphics[width=\columnwidth]{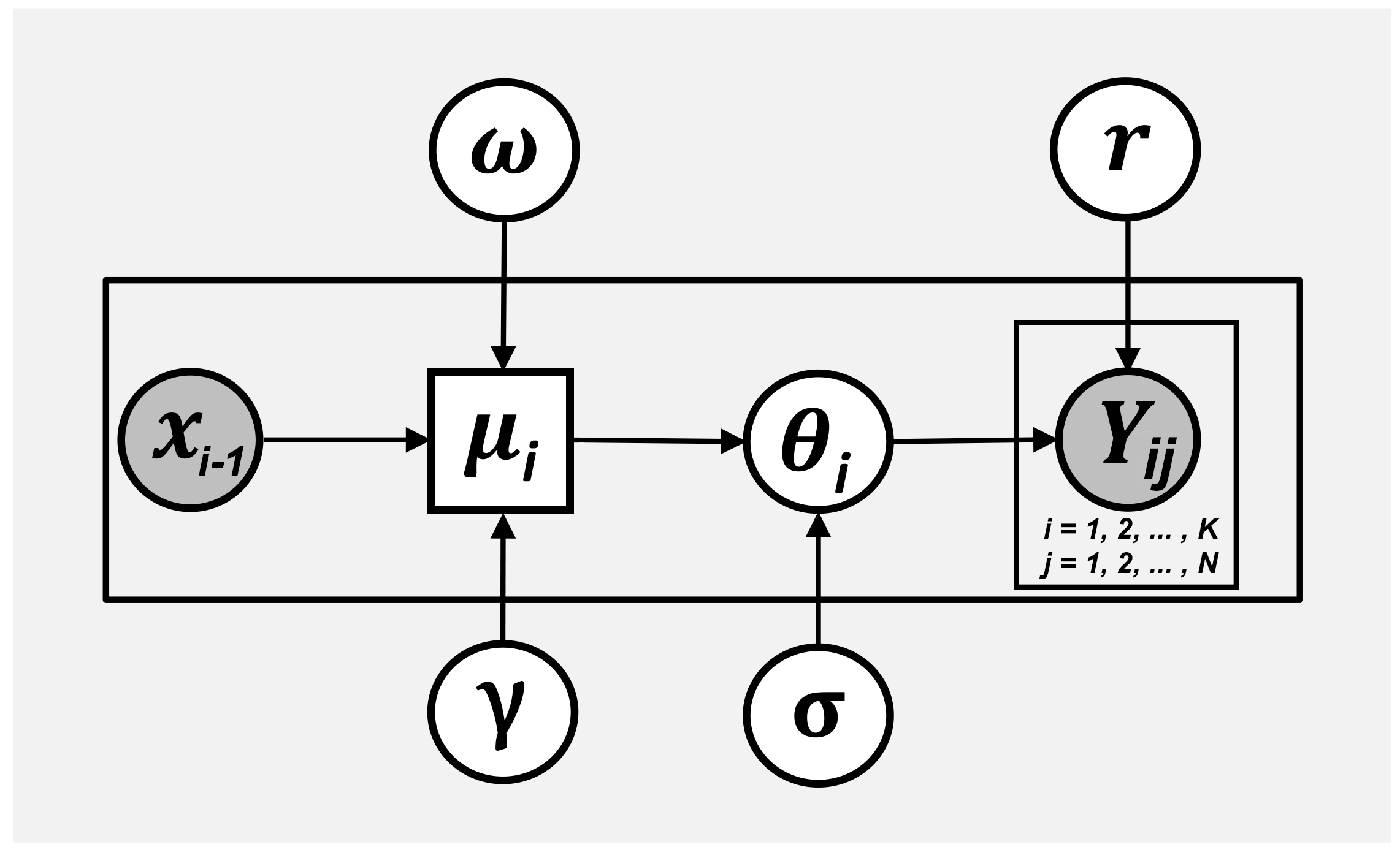}
\caption{}
\label{fig:h_nbglm}
\end{subfigure}
\begin{subfigure}{0.7\columnwidth}
\centering
\includegraphics[width=\columnwidth]{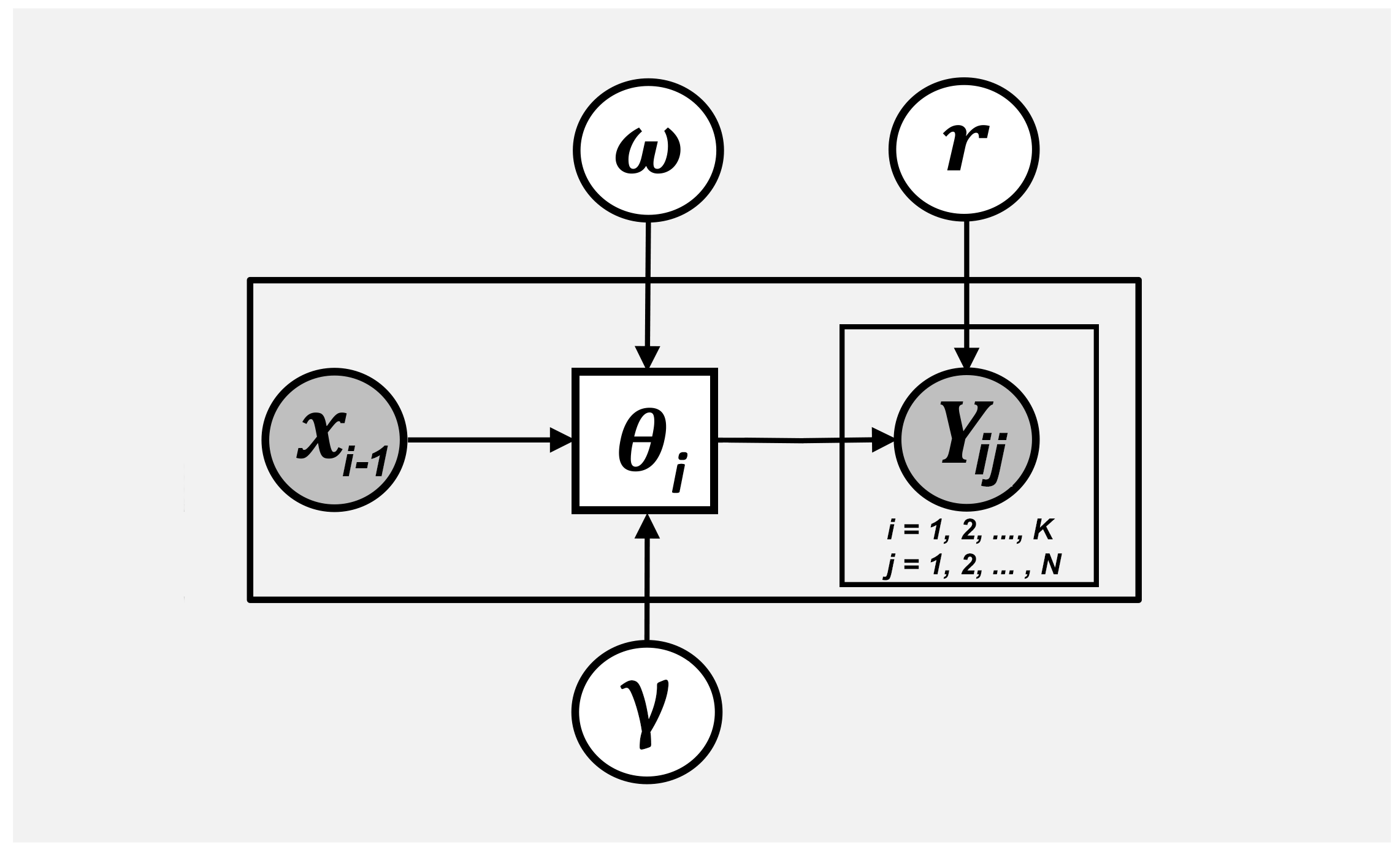}
\caption{}
\label{fig:standard_nbglm}
\end{subfigure}
\caption{Graphical representation of the proposed model and the NB-$\GLM$ model. (a) For the proposed model, the prior mean $\mu_{i}$ is formed from the $\GLM$ of the input regressors $\bm{x}_{i-1}$, the weight vector $\bm{\omega}$, and link function parameterized by $\bm{\gamma}$. $\mu_{i}$ is the mean of the beta prior of the $\NB$ probability parameter $\theta_{i}$. $\bm{\sigma}$ is the precision of the prior beta distribution. Finally, $\theta_{i}$, together with the shape parameter for the $\NB$ distribution $\bm{r}$, generate the observed spike counts $Y_{ij}$. (b) For the NB-$\GLM$ model, $\theta_{i}$ is derived from the $\GLM$ of the input regressors $\bm{x}_{i-1}$, the weight vector $\bm{\omega}$, and link function parameterized by $\bm{\gamma}$. Then $\theta_{i}$, together with the shape parameter for the $\NB$ distribution $\bm{r}$, generate the observed spike counts $Y_{ij}$. Shaded nodes $\bm{x}_{i-1}$ and $Y_{ij}$ denote observed random variables; $\mu_{i}$ (in the proposed model) and $\theta_{i}$ (in both models) are latent random variables. $\bm{r}$, $\bm{\omega}$, $\bm{\sigma}$, and $\bm{\gamma}$ are hyperparameters for the proposed model, while $\bm{r}$, $\bm{\omega}$, and $\bm{\gamma}$ are hyperparameters for the standard $\NB$-$\GLM$ model. The bigger rectangular box is ``plate notation", which denotes replication; the smaller rectangular box is ``inner plate'', which denotes the $Y_{ij}$ from different trials which share the same $\theta_{i}$.}
\label{fig:model2}
\end{figure}

We introduce the hyperparameter $\bm{\sigma} \equiv \alpha_{i}+\beta_{i}$, which can be interpreted as a precision parameter that reflects the degree of prior belief in the $\GLM$, and is fixed across different time bins. The prior mean is $ \mu_{i} \equiv \mathbb{E}(\theta_{i}|\alpha_{i},\beta_{i})=\frac{\alpha_{i}}{\bm{\sigma}}$, and $\alpha_{i} = \bm{\sigma}\mu_{i}$, $\beta{i} = \bm{\sigma}(1-\mu_{i})$. 
We can thus determine the beta distribution by learning $\mu_{i}$ and $\bm{\sigma}$. In particular, we learn $\mu_{i}$ by using a $\GLM$ with the mean spike counts of the input neurons at the previous time step ($\bm{x}_{i-1}$) (see the graphical model in Fig.~\ref{fig:model2}). A vector of functional weights, $\bm{\omega}$, capture the directed effects of input neurons on the output neuron. $\mu_{i}$ is modeled as:
\begin{equation}
g(\mu_{i}) = \bm{x}_{i-1}^\top\bm{\omega}.
\end{equation}
Here $\bm{\omega}$ is the vector of coupling weights, which captures how the input neurons affect the spiking behaviour of the output neurons. Positive or negative weights represent the excitatory or inhibitory effects on the output neurons. As biological neural networks usually have sparse topology, most weights are zero or closed to zero\cite{pessoa2014understanding}.

Usually, the link function $g(\cdot)$ is predefined using a specific form such as $log$, $logit$, $probit$, $identity$, and $log-log$~\cite{nelder1972generalized}.
However, we do not want to constrain the link function to be a fixed form. Hence, we propose a family of link functions governed by a hyperparameter, $\bm{\gamma}$, such that,
\begin{equation}
\label{eq:link}
g\left(\mu_{i}, \bm{\gamma} \right) = \log \left( \dfrac{\left(\mu_{i}\right)^{-\bm{\gamma}}-1}{\bm{\gamma}}\right).
\end{equation}
We design this link family with three considerations: (1) it can represent many widely used link functions. For instance, the $logit$ function when $\bm{\gamma}=1$, the complementary $log-log$ link function when $\bm{\gamma} \approx 0$;
(2) It should constrain the prior mean, modeled as the mean value of the probability parameter, to $\mu_{i} > 0$ and (3) it can be inverted to provide gradients for the hyperparameters $\bm{\gamma}$ and $\bm{\omega}$ (discussed in Section~\ref{ssec:Optimization}) easily. Note that the hyperparameter $\bm{\gamma}$, is a flexible parameter which determines the specific form of the link function, $g(\cdot)$, therefore ensuring the flexibility of the nonlinear transformation from the regressors to the output.
Denoting the inverse link function by $g^{-1}\left(\bm{x}_{i-1}^\top\bm{\omega}, \bm{\gamma} \right)$, the prior mean becomes 
\begin{equation}
\mu_{i}=g^{-1}\left(\bm{x}_{i-1}^\top\bm{\omega}, \bm{\gamma} \right) = \left(\bm{\gamma} e^{\bm{x}_{i-1}^\top\bm{\omega}}+1\right)^{-\frac{1}{\bm{\gamma}}}.
\label{eq:inv}
\end{equation}

In the sequel, we let $\bm{\zeta} \equiv \{\bm{r},\bm{\omega},\bm{\sigma},\bm{\gamma}\}$ denote the full set of model parameters. Table~\ref{tab:explanation} provides a complete summary of all the variables used in the ``$\SODS$" estimator and Fig.~\ref{fig:model2}a shows the graphical model of the proposed hierarchical structure. The observation data are $Y_{i}$ and $\bm{x}_{i-1}$; $\mu_{i}$ and $\theta_{i}$ are latent variables; $\bm{\zeta} \equiv \{\bm{r},\bm{\omega},\bm{\sigma},\bm{\gamma}\}$ are global parameters, which are consistent across all time steps.
\begin{table}[htbp]
	\centering
	\begin{threeparttable}
		\caption{\label{tab:explanation}Summary of variable definitions.}
		\begin{tabular}{c|l}
			\toprule 
			Variable						&	Definition	\\
			\hline 
			$y_{ij}$						&	Spike counts of $j$-th trial at $i$-th time bin	\\
			$\bm{x}_{i-1}$				&	Vector of regressors at ($i$-$1$)-th time step\\   
			$\lambda_{i}$				&	Mean of Poisson distribution (firing rate of neurons)	\\           
			$\theta_{i}$				&	Probability parameter of Negative Binomial distribution\\  
			$\bm{r}$							&	\# failures in Negative Binomial distribution \\	       
			$\alpha_{i}, \beta_{i}$	&	Parameters of beta distribution	\\  
			$\bm{\sigma}$						&	Precision of prior distribution ($\bm{\sigma} \equiv \alpha_{i}+ \beta_{i}$)	\\ 
			$\bm{\omega}$				&	Vector of weights	\\
			$g(\cdot)$					&	Family of link functions	\\
			$\bm{\gamma}$						&	Parameter determining specific form of link function	\\
			$\mu_{i}$					&	Mean of prior distribution	\\
			$n_{i}$						&	Number of trials at $i$-th time bin	\\    
			$\bar{y}_{i}$				&	Mean spike counts across all trials	\\
			$\pi_{i}$					&	Weight of the observation component in our estimator	\\           
			$K$							&	Data length (the total number of bins)	\\         
			$A_{i}, B_{ij}, C_{ij}$ &	Components of the gradients	\\  
			$p$							&	Element number of $\bm{\omega}$	\\ 
			$N$							&	Total number of neurons	\\
			$\bm{\zeta}$				&	$(\bm{r},\bm{\omega},\bm{\sigma},\bm{\gamma})$	\\
			$\lambda_{e}$               &   Constant that multiplies the elastic-net penalty terms\\
			$\alpha_{e}$                &  L1 ratio in elastic-net regularization \\
			\bottomrule
		\end{tabular}
	\end{threeparttable}
\end{table}

\subsection{Empirical Bayes Estimator: $\SODS$}
\label{ssec:SODS}
First, we study the posterior distribution of $\theta_i$. As the Beta distribution is the conjugate prior of the Negative Binomial likelihood function, the posterior distribution of $\theta_{i}$ given $Y_{ij} = y_{ij}$ follows the beta distribution \cite{lawless1987negative}:
\begin{align}
	\theta_{i}\mid y_{ij}, \bm{x}_{i-1} &\sim \mathrm{Beta}\Big(\bm{\sigma}\mu_{i}+n_{i}\bm{r},\bm{\sigma}\left(1-\mu_{i}\right)+n_{i}\overline{y}_{i}\Big),
	\label{eq:betaNegative Binomial}
\end{align}
where $n_{i}$ is the number of trials in the $i$th time bin, and $\overline{y}_{i}$ is the
mean count across all training trials at bin $i$. Substituting (\ref{eq:inv}) into \eqref{eq:betaNegative Binomial}, we get
\begin{eqnarray}
\theta_{i}\mid y_{ij}, \bm{x}_{i-1},\bm{\zeta}  \sim \mathrm{Beta}\Big(\bm{\sigma} g^{-1}(\bm{x}_{i-1}^\top\bm{\omega},\bm{\gamma}) +n_{i}\bm{r},\nonumber\\ \bm{\sigma}\big[1-g^{-1}(\bm{x}_{i-1}^\top\bm{\omega},\bm{\gamma})\big]+n_{i}\overline{y}_{i}\Big).
\label{eq:posterior}
\end{eqnarray}
We take the mean of this posterior distribution as the estimator for $\theta_{i}$, we call this estimator derived from our model as ``$\SODS$" estimator, and denoted as $\theta^{\mathrm{SODS}}$:
\begin{equation}
\theta^{\mathrm{SODS}}_{i} = \mathbb{E}(\theta_{i}\mid y_{ij},\bm{\zeta})=\dfrac{n_{i}\bm{r}+\bm{\sigma} g^{-1}(\bm{x}_{i-1}^\top\bm{\omega},\bm{\gamma})}{n_{i}\bm{r}+n_{i}\overline{y}_{i}+\bm{\sigma}},
	\label{eq:SODS}
\end{equation}
which can be rewritten as
\begin{equation}
\theta^{\mathrm{SODS}}_{i}=\pi_{i}\left(\dfrac{\bm{r}}{\bm{r}+\overline{y}_{i}}\right)+(1-\pi_{i})g^{-1}(\bm{x}_{i-1}^\top\bm{\omega},\bm{\gamma}),
\label{eq:convexsimp}
\end{equation}
where $\pi_{i}=\frac{\bm{r}+\overline{y}_{i}}{\bm{r}+\overline{y}_{i}+\bm{\sigma}/n_{i}} \in (0,1)$.
Hence, $\theta^{\mathrm{SODS}}_{i}$ is a convex combination of the data-driven estimate of $\theta_{i}$
and the prior mean of the $\GLM$. We can consider $\pi_{i}$ as the parameter that trades off between bias and variance. $\bm{\sigma}$ can be viewed as a precision parameter. When $\bm{\sigma} \to 0$, thus $\pi_{i} \to 1$, it results in $\theta_{i}^{\mathrm{SODS}}$ only reflecting the observed data. When $\bm{\sigma} \to \infty$, thus $\pi_{i} \to 0$, the estimator reduces to be standard Negative Binomial $\GLM$, which links the probability parameter with the input regressors via a link function
\begin{equation}
\mathbb{E}(\theta_{i}\mid y_{ij},\bm{\zeta})=  g^{-1}(\bm{x}_{i-1}^\top\bm{\omega},\bm{\gamma}).
\label{eq:NBglm}
\end{equation}

With the estimator $\theta^{\mathrm{SODS}}_{i}$,
the mean spike counts can then
be obtained from Eq.~\eqref{eq:nb}:
\begin{eqnarray}
\mathbb{E}[Y_{i}\vert\theta^{\mathrm{SODS}}_{i}] &= &
\bm{r}\Big(\dfrac{1}{\theta^{\mathrm{SODS}}_{i}}-1\Big) \nonumber \\
&=& \bm{r}\dfrac{n_{i}\overline{y}_{i}+\bm{\sigma}-\bm{\sigma}g^{-1}\left(\bm{x}_{i-1}^\top\bm{\omega},\bm{\gamma}\right)}{n_{i}\bm{r}+\bm{\sigma} g^{-1}\left(\bm{x}_{i-1}^\top\bm{\omega},\bm{\gamma}\right)}.
	\label{eq:spikecount}
\end{eqnarray}


\subsection{Maximum Marginal Likelihood}
\label{sec:EmpBayes}
$\theta^{\mathrm{SODS}}$ depends on $\bm{\zeta} \equiv \{\bm{r},\bm{\omega},\bm{\sigma},\bm{\gamma}\}$. To estimate $\bm{\zeta}$, we use the empirical Bayes approach. We first derive the marginal likelihood function, where the marginal distribution is the spike counts conditioned only on the parameters. Then we minimize the objective function, which combines the negative marginal log-likelihood function and the elastic-net regularization. Finally, we discuss how to use prior knowledge to set the initial value for a more stable and accurate optimization result.

Since using the maximum marginal likelihood approach does not include any assumptions on the parameters, we have the benefit of relatively low computational cost for estimating high-dimensional parameters. To derive the marginal likelihood, we need to integrate out the probability parameter $\theta_{i}$, as $p(y_{ij}) = \int p(\theta_{i}) p(y_{ij} \mid \theta_{i}) d\theta_{i}$. Reformulating the Negative Binomial likelihood as 
\begin{align}
p(y_{ij}\mid \theta_{i}) &= \dfrac{\Gamma(\bm{r}+y_{ij})}{\Gamma(y_{ij}+1)\Gamma(\bm{r})} \theta_{i}^{\bm{r}} (1-\theta_{i})^{y_{ij}}\nonumber\\ 
& =  \dfrac{\Gamma(\bm{r}+y_{ij})}{\Gamma(y_{ij})\Gamma(\bm{r})} \dfrac{\Gamma(y_{ij})}{\Gamma(y_{ij}+1)} \theta_{i}^{\bm{r}} (1-\theta_{i})^{y_{ij}}\nonumber\\
& = \dfrac{\theta_{i}^{\bm{r}} (1-\theta_{i})^{y_{ij}}}{\mathrm{B}(\bm{r}, y_{ij})y_{ij}},
\label{eq:likelihood}
\end{align}
then, the marginal likelihood is 
\begin{align}
p(y_{ij})\!&=\!\int_{0}^{1} p(\theta_{i}) \dfrac{\theta_{i}^{\bm{r}} (1-\theta_{i})^{y_{ij}}}{\mathrm{B}(\bm{r}, y_{ij})y_{ij}} d\theta_{i}\nonumber\\
&=\!\dfrac{1}{\mathrm{B}(\bm{r}, y_{ij})\mathrm{B}(\alpha_{i}, \beta_{i})y_{ij}}\!\int_{0}^{1} \!\theta_{i}^{\bm{r}+\alpha_{i}-1} (1\!-\!\theta_{i})^{y_{ij}+\beta_{i}-1} d\theta_{i}\nonumber\\
&=\!\dfrac{\mathrm{B}(\bm{r}+\alpha_{i},y_{ij}+\beta_{i})}{\mathrm{B}(\bm{r}, y_{ij})\mathrm{B}(\alpha_{i}, \beta_{i})y_{ij}}.
\label{eq:Mlikelihood}
\end{align}
Substituting $\alpha_{i}$ and $\beta_{i}$ into Eq.~\eqref{eq:Mlikelihood} with
$\alpha_{i} = \bm{\sigma} g^{-1}(\bm{x}_{i-1}^\top\bm{\omega},\bm{\gamma})$, $\beta_{i} = \bm{\sigma} -  \bm{\sigma} g^{-1}(\bm{x}_{i-1}^\top\bm{\omega},\bm{\gamma}) $, 
the marginal density of the spike counts conditioned on $\bm{\zeta}$ is
{\footnotesize
\begin{equation}
	p_{i}(y_{ij}\vert \bm{\zeta},\bm{x}_{i})\!=\! \dfrac{\mathrm{B}\Big(\bm{r}\!+\!\bm{\sigma} g^{-1}(\bm{x}_{i-1}^\top\bm{\omega},\bm{\gamma}),y_{ij}\!+\!\bm{\sigma}\! -\! \bm{\sigma} g^{-1}(\bm{x}_{i-1}^\top\bm{\omega},\bm{\gamma})\Big)}{\mathrm{B}(\bm{r},y_{ij}) \mathrm{B}\Big(\bm{\sigma} g^{-\!1}(\bm{x}_{i-1}^\top\bm{\omega},\bm{\gamma}),\bm{\sigma}\!-\!\bm{\sigma} g^{-\!1}(\bm{x}_{i-1}^\top\bm{\omega},\bm{\gamma})\Big)y_{ij}}\nonumber
\end{equation}}
and conditioning on $y_{ij}$, the log marginal likelihood $\ell(\bm{\zeta}) = \sum_{i=1}^{K}\sum_{j=1}^{n_{i}} \log p_{i}(y_{ij})$ of the conditional posterior is
\begin{align}
	\ell(\bm{\zeta}) & \propto \sum_{i=1}^{K}\sum_{j=1}^{n_{i}} \bigg[\log\Gamma\Big(\bm{r}+\bm{\sigma} g^{-1}(\bm{x}_{i-1}^\top\bm{\omega},\bm{\gamma})\Big) + \log\Gamma(\bm{\sigma}) \nonumber \\
	&+ \log\Gamma(\bm{r}+y_{ij})-\log\Gamma(\bm{r}+y_{ij}+\bm{\sigma})-\log\Gamma(\bm{r}) \nonumber \\
	&+ \log\Gamma\Big(y_{ij}+\bm{\sigma}-\bm{\sigma} g^{-1}(\bm{x}_{i-1}^\top\bm{\omega},\bm{\gamma})\Big)\nonumber\\
	&-\log\Gamma\Big(\bm{\sigma}-\bm{\sigma} g^{-1}(\bm{x}_{i-1}^\top\bm{\omega},\bm{\gamma})\Big)\nonumber\\&-\log\Gamma\Big(\bm{\sigma} g^{-1}(\bm{x}_{i-1}^\top\bm{\omega},\bm{\gamma})\Big)\bigg].
	\label{eq:MLL}
\end{align}
To obtain the objective function, we combine the elastic-net regularization \cite{friedman2010regularization} with the negative log marginal likelihood -$\ell(\bm{\zeta})$, as
\begin{align}
H(\bm{\zeta}) = -\ell(\bm{\zeta}) +\lambda_{e}\left(\alpha_{e}\|\bm{\omega}\|_{1}+\frac{1-\alpha_{e}}{2}\|\bm{\omega}\|^{2}\right)
	\label{eq:MLL_Elastic}
\end{align}
where $\alpha_{e}$ and $\lambda_{e}$ are parameters for the elastic-net regularization. During optimization, we set $\alpha_{e} = 0.5$, and the optimal $\lambda_{e} \in [0.1, 1, 10]$ is chosen through 5-fold cross validation.
\subsection{Prior Knowledge of the Parameters for Optimization}
\label{ssec:LatentVar}
By minimizing the objective $H(\bm{\zeta})$ in Eq.~\eqref{eq:MLL_Elastic}, we obtain the $\bm{r},\bm{\omega},\bm{\sigma}$ and $\bm{\gamma}$ for the ``$\SODS$" estimator. These parameters play different roles in explaining the neural spiking behaviour. During the optimization with our simulation data, we randomly select initial value of these parameters from different distribution, as shown in Table~\ref{tab:initial_value}. But for optimization using real neuronal recordings, tuning of the initial value should be considered carefully to improve the estimation. In this section we discuss each of them in turn, and present rules to tune the initial values used in the optimization procedure.
\begin{itemize}
	\item $\bm{r}$ is the Negative Binomial response's shape parameter. Physically, it contributes to the underlying firing rates of neurons together with $\theta$. As shown in Fig.~\ref{fig:review_nbglm}b, larger values of $\bm{r}$ give larger spike counts. In real situations, the actual firing rates of the underlying neural population may not be very high, e.g., in hippocampal areas. In such cases, to get reasonable mean spike counts, we should ensure that the initial value of $\bm{r}$ is small, as this helps the spike count observations match the low firing rates. Accordingly, if we believe a brain area has a high firing rate, e.g., in motor cortex, we can initialize $\bm{r}$ to a higher value. During optimization, $\bm{r}$ should be constraint as positive value.
	\item $\bm{\omega}$ is a vector of coupling weights, which help to capture the directed effects of input neurons on the output neuron. \textit{This is the core element to introduce connectivity into our hierarchical model}. This vector can also include other factors such as the spiking history of the output neuron or external stimuli \emph{e.g.}, if we have prior knowledge, for instance the pixels of an image shown to excite the retinal neurons. These weights can be positive or negative, which can be explained as neurons having either an excitatory or inhibitory effect on the output neuron. In section~\ref{sec:simulated}, using simulated data, we test the ability of proposed estimator to capture these excitatory and inhibitory effects of neuronal connectivity. The initial values of elements in $\bm{\omega}$ are randomly chosen from the uniform distribution on the interval $\left(-1,1\right)$ and $\bm{\omega}$ is limited on the interval $\left(-1,1\right)$ during optimization. 
	
	\item $\bm{\sigma}$ is the precision of the beta distribution. It controls the balance between our limited data sample and prior knowledge. From Eq.~\eqref{eq:convexsimp}, we can see that our proposed estimator is the weighted combination of the observed data $\theta_{i}^{obs}=\frac{\bm{r}}{\bm{r}+\overline{y}_{i}}$ and the standard $\GLM$ estimation $\theta_{i}^{\GLM}=g^{-1}\left(\bm{x}_{i-1}^\top\bm{\omega},\bm{\gamma}\right)$. The weights of each component are $\pi_{i}=\frac{n_{i}\bm{r}+n_{i}\overline{y}_{i}}{n_{i}\bm{r}+n_{i}\overline{y}_{i}+\bm{\sigma}}$ and $1-\pi_{i}=\frac{\bm{\sigma}}{n_{i}\bm{r}+n_{i}\overline{y}_{i}+\bm{\sigma}}$. Thus, if $\bm{\sigma}$ is large, the proposed method is close to the $\GLM$ of the prior mean; when it is small, the estimator is approaching the observed data. The initial value of $\bm{\sigma}$ should be determined by the number of trials $n_{i}$, such that, the more trials we have, the smaller $\bm{\sigma}$ should be, which means we can have more confidence in the observed data. During optimization, $\bm{\sigma}$ should be constraint as positive value.
	
	\item $\bm{\gamma}$ conveys the nonlinear effects of input neurons on the output neurons, which selects the best fit link function for the dataset. Regularly, $\GLMs$ choose the link function by specifying a parametric link. Our work, however, determines the unknown parameter automatically. Learning from the dataset itself allows our approach to select a suitable link function automatically. The initial value of $\bm{\gamma}$ is determined so as to result in relatively low firing rate, which has empirically been shown to give good performance for spike count prediction. As the inverse link function Eq.~\eqref{eq:inv} needs to map into the range [0,1], the $\bm{\gamma}$ should be kept positive during optimization. The initial value of $\bm{\gamma}$ should not be too large, as the corresponding inverse link function Eq.~\eqref{eq:inv} would converge to 1 with large $\bm{\gamma}$ value.
 
\end{itemize}
\subsection{Optimization of Parameters}
\label{ssec:Optimization}
In hierarchical modeling, closed-form estimators are often elusive. Thus, we use numerical optimization. Here, we apply the Quasi-Newton method, Limited-memory $BFGS$ \cite{shanno1985broyden,liu1989limited} with global optimization method Basin-Hopping. 
We derive the gradients w.r.t. $\bm{r},\bm{\omega},\bm{\sigma},\bm{\gamma}$ as
\begin{eqnarray}
\dfrac{\partial H(\bm{\zeta})}{\partial \bm{r}} &= &
-\sum_{i=1}^{K}\sum_{j=1}^{n_{i}}\{\Psi\left[\bm{r}+\bm{\sigma}
g^{-1}(\bm{x}_{i-1}^\top\bm{\omega},\bm{\gamma})\right] \nonumber \\
&& +\Psi(\bm{r}+y_{ij})-\Psi(\bm{r}+y_{ij}+\bm{\sigma})-\Psi(\bm{r})\}\nonumber\\
\dfrac{\partial H(\bm{\zeta})}{\partial \omega_{p}} &= & -\bm{\sigma} \sum_{i=1}^{K}\sum_{j=1}^{n_{i}}\dfrac{\partial g^{-1}(\bm{x}_{i-1}^\top\bm{\omega},\bm{\gamma})}{\partial \omega_{p}} \left(A_{i}-B_{ij}\right) \nonumber \\
&& + \lambda_{e}\Big[(1-\alpha_{e}) \omega_{p} + \alpha_{e}\operatorname{sgn}(\omega_{p})\Big] \nonumber\\
\dfrac{\partial H(\bm{\zeta})}{\partial \bm{\sigma}} &= &
-\sum_{i=1}^{K}\sum_{j=1}^{n_{i}}\left\{A_{i}g^{-1}(\bm{x}_{i-1}^\top\bm{\omega},\bm{\gamma})\right. \nonumber \\
&& \left. +B_{ij}\Big[1-g^{-1}(\bm{x}_{i-1}^\top\bm{\omega},\bm{\gamma})\Big]+C_{ij}\right\}\nonumber\\
	\dfrac{\partial H(\bm{\zeta})}{\partial \bm{\gamma}} &= & -\sum_{i=1}^{K}\sum_{j=1}^{n_{i}}\dfrac{\partial g^{-1}(\bm{x}_{i-1}^\top\bm{\omega},\bm{\gamma})}{\partial \bm{\gamma}} \left(A_{i}-B_{ij}\right)
	\label{eq:gradients}
\end{eqnarray}
where
\begin{eqnarray}
A_{i} &= & \Psi\left(\bm{r}+\bm{\sigma} g^{-1}(\bm{x}_{i-1}^\top\bm{\omega},\bm{\gamma})\right)-
\Psi\left(\bm{\sigma} g^{-1}(\bm{x}_{i-1}^\top\bm{\omega},\bm{\gamma})\right)\nonumber\\
B_{ij} &=& \Psi\left(y_{ij}+\bm{\sigma}-\bm{\sigma}
g^{-1}(\bm{x}_{i-1}^\top\bm{\omega},\bm{\gamma})\right)\nonumber\\&-& \Psi\left(\bm{\sigma}-\bm{\sigma}
g^{-1}(\bm{x}_{i-1}^\top\bm{\omega},\bm{\gamma})\right)\nonumber\\
C_{ij} &=& \Psi(\bm{\sigma})- \Psi(\bm{r}+y_{ij}+\bm{\sigma})
\label{eq:ABC}
\end{eqnarray}
$\Psi(x)=\frac{\partial \log\Gamma(x)}{\partial x}$ is the Digamma function,
$\omega_{p}$ is the individual element in the vector $\bm{\omega}$ with $p=(1,2,\dots,N)$ and $N$ is the number of neurons. Moreover,
in Eq.~\eqref{eq:gradients}, the ease of implementation of gradient calculations gives 
\begin{equation}
\dfrac{\partial g^{-1}\left(\bm{x}_{i-1}^\top\bm{\omega},\bm{\bm{\gamma}}\right)}{\partial
\omega_{p}} = -x_{p}\mathrm{e}^{\bm{x}_{i-1}^\top\bm{\omega}}\left(\bm{\gamma}\mathrm{e}^{\bm{x}_{i-1}^\top\bm{\omega}}+1\right)^{-\frac{1}{\bm{\bm{\gamma}}}-1}
\end{equation}
\hspace*{-3mm}\begin{align}
\dfrac{\partial g^{-1}\left(\bm{x}_{i-1}^\top\bm{\omega},\bm{\gamma}\right)}{\partial \bm{\gamma}} &= \left(\bm{\gamma}\mathrm{e}^{\bm{x}_{i-1}^\top\bm{\omega}}+1\right)^{-\frac{1}{\bm{\gamma}}} \nonumber\\
& \left[\dfrac{\log\left(\bm{\gamma}\mathrm{e}^{\bm{x}_{i-1}^\top\bm{\omega}}+1\right)}{\bm{\gamma}^{2}} - \dfrac{\mathrm{e}^{\bm{x}_{i-1}^\top\bm{\omega}}}{\bm{\gamma}(\bm{\gamma}\mathrm{e}^{\bm{x}_{i-1}^\top\bm{\omega}}+1)}\right]\nonumber\\
\end{align}

We check the convexity of $H(\bm{\zeta})$ by simulation, and found it did not satisfy the Jensen's inequality \cite{jensen1906fonctions}. It is difficult to find the global minimum for a non-convex problem, especially for multivariate models, as multiple distinct local minima could exist. Basin hopping is a global optimization framework designed for multivariable multimodal optimization problems \cite{olson2012basin,wales1997global}. During optimizations, we use Basin hopping with the local minimization optimizer, the Limited-memory Broyden–Fletcher–Goldfarb–Shanno with Bound constraints (L-BFGS-B), to find the global minimum of the objective function $H(\zeta)$. 
During each iteration, the values in $\zeta$ are randomly perturbed and used as the initial value for the L-BFGS-B local optimizer. The new local minimum is accepted as a global minimum if it is smaller than the old local minimum, and the $\zeta$ will be updated with the new values. The iteration of Basin-hopping will stop when the same $H(\zeta)$ is obtained for consecutive 20 iterations. The maximum iteration for Basin-hopping is 50. For L-BFGS-B, we only store the most recent m=10 gradients for the approximation of the Hessian matrix and the maximum iteration for the local minimizer is 5000. Iteration of L-BFGS-B will stop when the reduction ratio of the objective value is less than $ftol$ or when the maximum component of the projected gradient is less than $gtol$ ($ftol =2.22e-09, gtol=1e-05$). We implement L-BFGS-B and Basin-Hopping using the Scipy Optimization library \cite{scipytool}. The steps for empirical Bayes inference for ``$\SODS$" estimator are summarized in Algorithm~\ref{alg:steps}. 

\begin{table}[tbp]
    \centering
    \caption{The parameters of ``$\SODS$" estimator and the probability density function used for parameters initialization.}
    \begin{tabular}{c|c}
         Parameters & Distribution \\ \hline
         $\bm{r}$ & $\it{Uniform}(1,100)$ \\
         $\bm{\sigma}$ & $\it{Uniform}(1,100)$ \\
         $\bm{\gamma}$ &  $\it{Uniform}(1,100)$ \\
         $\bm{\omega}$ & $\it{Uniform}(-1,1)$ \\\hline
    \end{tabular}
    \label{tab:initial_value}
\end{table}

\begin{algorithm}[tbp]
	\renewcommand{\algorithmicrequire}{\textbf{Input:}}
	\renewcommand{\algorithmicensure}{\textbf{Output:}}
	\caption{The Hierarchical Parametric Empirical Bayes Framework for Short Over-Dispersed Spike-Trains}
	\label{alg:steps}
	\begin{algorithmic}[1]
		\Require $\bm{x}_{i-1}=[x_{i-1,1},x_{i-1,2},\dots,x_{i-1,N}]^\top$ and $y_{ij}$ $\left(i = 1,...,K\text{ and }j=1,...,N_s\right)$.
		\Ensure  $\bm{r},\bm{\omega},\bm{\sigma},\bm{\gamma}$.
		\State Initialize $\bm{r}_{0},\bm{\omega}_{0},\bm{\sigma}_{0},\bm{\gamma}_{0}$ based on Section~\ref{ssec:LatentVar}
		\State Based on Eq.~\eqref{eq:MLL_Elastic} and L-BFGS-B, calculate the local minimum $H_{min}$ and the corresponding $\bm{r},\bm{\omega},\bm{\sigma},\bm{\gamma}$,
		\For{$iter = 0;\ iter < 50;\ iter = iter + 1$}
    		\State Perturb the parameters $\bm{r},\bm{\omega},\bm{\sigma},\bm{\gamma}$, 
    		\State Calculate new local minimum $H_{new}$ and the corresponding $\{\bm{r},\bm{\omega},\bm{\sigma},\bm{\gamma}\}_{new}$.
    		\If{$H_{new} < H_{min}$}
    		    \State
    		    $H_{min} = H_{new}$;
    		    $\{\bm{r},\bm{\omega},\bm{\sigma},\bm{\gamma}\} = \{\bm{r},\bm{\omega},\bm{\sigma},\bm{\gamma}\}_{new}$.
    		\EndIf
    		\If{$H_{min}$ is the same for 20 iteration}
    		    \State \textbf{Break}
    		\EndIf
    	\EndFor
    	\State \textbf{return} $\bm{r},\bm{\omega},\bm{\sigma},\bm{\gamma}$
		\State Calculate the empirical Bayes estimation of probability parameter $\mathbb{E}(\theta_{i}\mid \bm{r},\bm{\omega},\bm{\sigma},\bm{\gamma})$ from Eq.~\eqref{eq:SODS} as
		\begin{equation*}
		\theta^{\mathrm{SODS}}_{i} =
		\mathbb{E}(\theta_{i}|\bm{r},\bm{\omega},\bm{\sigma},\bm{\gamma})=\dfrac{n_{i}\bm{r}+\bm{\sigma} g^{-1}\left(\bm{x}_{i-1}\top\bm{\omega},\bm{\gamma}\right)}{n_{i}\bm{r}+n_{i}\overline{y}_{i}+\bm{\sigma}}.
		\end{equation*}
		\State Obtain the mean spike counts based on Eq.~\eqref{eq:nb}: 
		\begin{equation*}
		\mathbb{E}[Y_{ij}\vert\theta_{i}] =\bm{r}\left( \dfrac{n_{i}\overline{y}_{i}+\bm{\sigma}-\bm{\sigma} g^{-1}\left(\bm{x}_{i-1}\top\bm{\omega},\bm{\gamma}\right)}{n_{i}\bm{r}+\bm{\sigma} g^{-1}\left(\bm{x}_{i-1}\top\bm{\omega},\bm{\gamma}\right)}\right).
		\end{equation*}
	\end{algorithmic}
\end{algorithm}
\begin{table*}[tbp] 
	\centering
	\begin{threeparttable}
		\caption{\label{tab:config}Sets of the parameters used for the ``$\SODS$"  simulations.}
		\begin{tabular}{c|c|c|c|c|c|c}
			\toprule
			$N_s$ & $K$ & $\bm{\omega}$ & $\bm{r}$ & $\bm{\sigma}$ & $\bm{\gamma}$ \\
			(\# of simulated trials)& (data length) & (weights) & (shape parameter of $\NB$) & (precision) & (link function) \\
			\hline
			\multirow{2}{*}{10, 50, 100}	& \multirow{2}{*}{100, 500, 1000, 2000} 	& \multirow{2}{*}{(-1,1)}	& \multirow{2}{*}{3, 5, 7}	& \multirow{2}{*}{50, 100, 1000, 1e9}	& \multirow{2}{*}{5, 7, 9}\\   
			& 					& 		& 				&			& 	&  \\
			\bottomrule
		\end{tabular}
	\end{threeparttable}
\end{table*}

\section{Methods}
\label{sec:simulated}

\subsection{Simulated Data from ``$\SODS$" models}
The simulated data is generated via the process outlined in Fig.~\ref{fig:model2}a, and the general setup of simulations is listed in Table~\ref{tab:config}. To simulate a sparse neural network, we generate the parameter \bm{$\omega$} as a sparse vector, which has a density of 0.2. The non-zero items in \bm{$\omega$} are generated from uniform distribution on the interval $(-1,1)$.
The nonzero values of \bm{$\omega$} are randomly selected from the uniform distribution on the interval $(-1, 1)$. We simulate the spiking data with several combinations of simulation trials $N_{s}$, data length per trial $K$, and the global parameters of $\bm{r},\bm{\sigma},\bm{\gamma},\bm{\omega}$. For each parameters set, we simulated ten pairs of training and testing data, which shared the same simulation parameters, except for the weights \bm{$\omega$}. Only training and testing data from the same pair had the same weights. Different regressors $\bm{x}$ were assigned to the training and testing data separately. The regressors $\bm{x}$ were 100-dimensional vector representing the activities from 100 independent input neurons. The value of each simulated input neurons was generated from a standard normal distribution.

\subsection{Simulated Data from NB-$\GLM$ and Poisson-like GLM}
We also generate simulated data from NB-$\GLM$ and Poisson-like $\GLMs$. Similar to the ``$\SODS$" model, we simulate spiking data with different data length, simulation trials, and various combinations in the parameters. For NB-$\GLM$ simulation, as outlined in Fig.~\ref{fig:model2}b, we use the similar flexible link function from the ``$\SODS$" model and estimate the probability parameter $\theta$ directly without the conjugate prior. The inverse link function for NB-$\GLM$ is:
\begin{equation}
\label{eq:nb-inv-link}
\theta_{i}=g^{-1}\left(\bm{x}_{i-1}^\top\bm{\omega}, \bm{\gamma} \right) = \left(\bm{\gamma} e^{\bm{x}_{i-1}^\top\bm{\omega}}+1\right)^{-\frac{1}{\bm{\gamma}}}.
\end{equation}

where the $\bm{\gamma}$ is the flexible parameters for the link function, the $\theta_{i}$ is the probability parameters of NB distribution at time $i$. The NB-$\GLM$ spiking data is generated based on this inverse link function Eq.~\eqref{eq:nb-inv-link} and Eq.~\eqref{eq:y}.

For Poisson-like $\GLM$, we utilize the smooth rectifier with Poisson-$\GLM$, as in \cite{park2014encoding}. Only $\bm{\omega}$ is needed in the Poisson-like $\GLM$, the inverse link function is:
\begin{equation}
\label{eq:poisson_link}
\mu_{i} = \log(1 + \mathrm{e}^{\bm{x}_{i-1}\top\bm{\omega}}).
\end{equation}
where $\mu_{i}$ is the rate parameter for output neuron at time $i$, $\bm{x}_{i-1}$ are the mean spike counts for the regressor neurons at time $i-1$, the $\bm{\omega}$ denotes the linear coefficients. The Poisson-like $\GLM$ spiking data is generated based on this inverse link function Eq~\eqref{eq:poisson_link}.

Ten pairs of training and testing data are simulated for NB-$\GLM$ and Poisson-like $\GLM$ separately. The parameters and regressor $\bm{x}$ are also initiated in similar ways to those simulated from ``$\SODS$" model.

\subsection{Estimation from NB-$\GLM$ and Poisson-like GLM}
We compared the performance of the ``$\SODS$" with the NB-$\GLM$ and Poisson-like $\GLM$ estimators. The estimating process from NB-$\GLM$ and Poisson-like $\GLM$ are similar with ``$\SODS$". We use the marginal log-likelihood approach combined with the elastic-net regularization on weights. The optimization procedures, the initial guess and bounds for parameters are similar with those in the ``$\SODS$" estimator.

\subsection{Simulated Data from Spiking Neural Network}
In addition to simulation from GLMs, we also simulate data from a simple spiking neural network (SNN) adapted from \cite{brette2007simulation}. The simple spiking neural network includes two layers which have 100 neurons per layer. Spikes from neurons in the first layer are generated from a Poisson process with the rate of 50Hz. The neurons in the second layer are built with the leaky integrate-and-fire (LIF) model with exponential conductance and a stochastic current. The following stochastic differential equations describe the neuronal model:
\begin{align}
\frac{d V}{d t} & = \big[g_{e} + g_{i} - (V - E_{l})\big]/\tau_{m} + \frac{1}{2}\xi(V_{t}-V_{r})\sqrt{1/\tau_{m}} \nonumber \\
\frac{d g_{e}}{d t} & =  -g_{e}/\tau_{e} \nonumber \\
\frac{d g_{i}}{d t} & =  -g_{i}/\tau_{i} \nonumber \\
g_{e} & \gets g_{e} + w_{e}, \textit{ upon spike arriving at excitatory synapse}\nonumber \\
g_{i} & \gets g_{i} + w_{i}, \textit{ upon spike arriving at inhibitory synapse} \nonumber
\end{align}
where $V$ is the membrane potential; $V_t=-50$mV is the action potential threshold; $E_{l}=-49$mV is the leak potential of the membrane; $V_{r}=-60$mV is the resting membrane potential; $g_{e}$, $g_{i}$ are the synaptic conductance for excitatory and inhibitory synapse, both are initiated as $0$mV; $\tau_{m}=20$ms is the time constant for the membrane potential, $\tau_{e}=5$ms, $\tau_{i}=10$ms are the time constants for the excitatory and inhibitory synaptic conductance; $w_{e}=1.62$mV, $w_{i}=-9$mV are the synaptic weight for excitatory and inhibitory synapse; $\xi$ denotes the stochastic current, which is a random variable generated from standard normal distribution. The membrane potentials of the target neurons are initiated randomly from the uniform distribution on the interval $\left(V_{r}, V_{t}\right)$.

There are 80 excitatory and 20 inhibitory neurons in the first layer of SNN. Each target neuron in the second layer receives excitatory and inhibitory inputs from neurons in the first layer; the connection probability for all synapse is 0.2. The simulation time lasted for 2500 seconds (biological time). We set the bin size as 100ms based on the target neurons' firing rates so that the mean spike counts would be within a reasonable range. Data from neurons in the first layer is used as regressors $\bm{x}$, and those from the second layer are used as targets $\bm{y}$. The SNN simulation was implemented by the Brian2 Python Package \cite{stimberg2019brian}.

\subsection{Model Performance}
We evaluate the ``$\SODS$" estimator on simulated data in two aspects:
\begin{itemize}
    \item Performance of the ``$\SODS$" estimator. In the simulation process, we have the underlying ground truth regarding the mean spike counts. We tested the performance of the ``$\SODS$" estimator by calculating the Mean Squared Error (MSE) and the R-squared value between the ground truth and the estimated mean spike counts, based on Eq.~\eqref{eq:nb}, $\mathbb{E}[Y_{i}\vert\theta_{i}]=\bm{r}(\frac{1}{\theta_{i}}-1)$. We compare the MSE and R-squared values of the ``$\SODS$" estimators with those estimated by the NB-$\GLM$ and Poisson-like $\GLM$ methods. A better estimator is the one that has lower MSE and higher R-squared value.
	\item Interaction estimation. For the data simulated with ``$\SODS$" model, we have the underlying ground truth regarding the weights. The goodness-of-fit between the true weights and the estimated weights is measured to evaluate how accurately our model can recover the weights of the interactions. 
	\end{itemize} 
\section{Results}
\label{sec:Results}
\subsection{Estimation Results of Simulated Data from the ``$\SODS$" model}
\begin{figure}[tbp]
    \centering
    \includegraphics[width=\columnwidth]{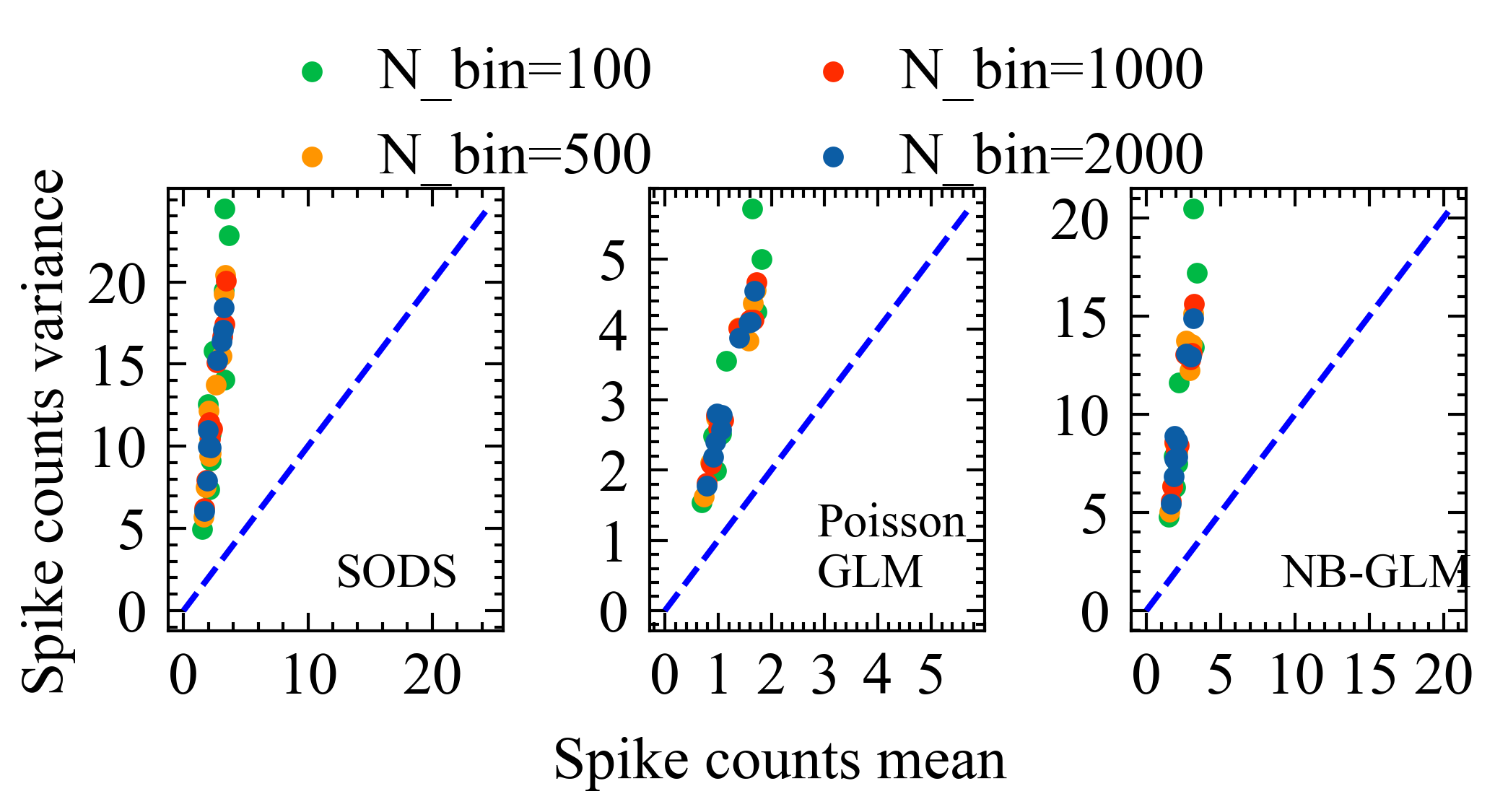}
    \caption{Spike count means and variances from simulated data of different data length. The simulated spiking data from all three $\GLMs$ have larger spike count variances than means. The blue dashed line indicates the identity line, where the variance equals to the mean. Simulated parameters used for ``$\SODS$" simulation: $\bm{r} = 5$, $N_{s}=50$, $\bm{\sigma} = 50$, $\bm{\gamma}=7$.}
    \label{fig:simulated_data_overdispersion}
\end{figure}

\begin{figure}[tbp]
\centering
\begin{subfigure}{0.8\columnwidth}
\centering
    \includegraphics[width=\columnwidth]{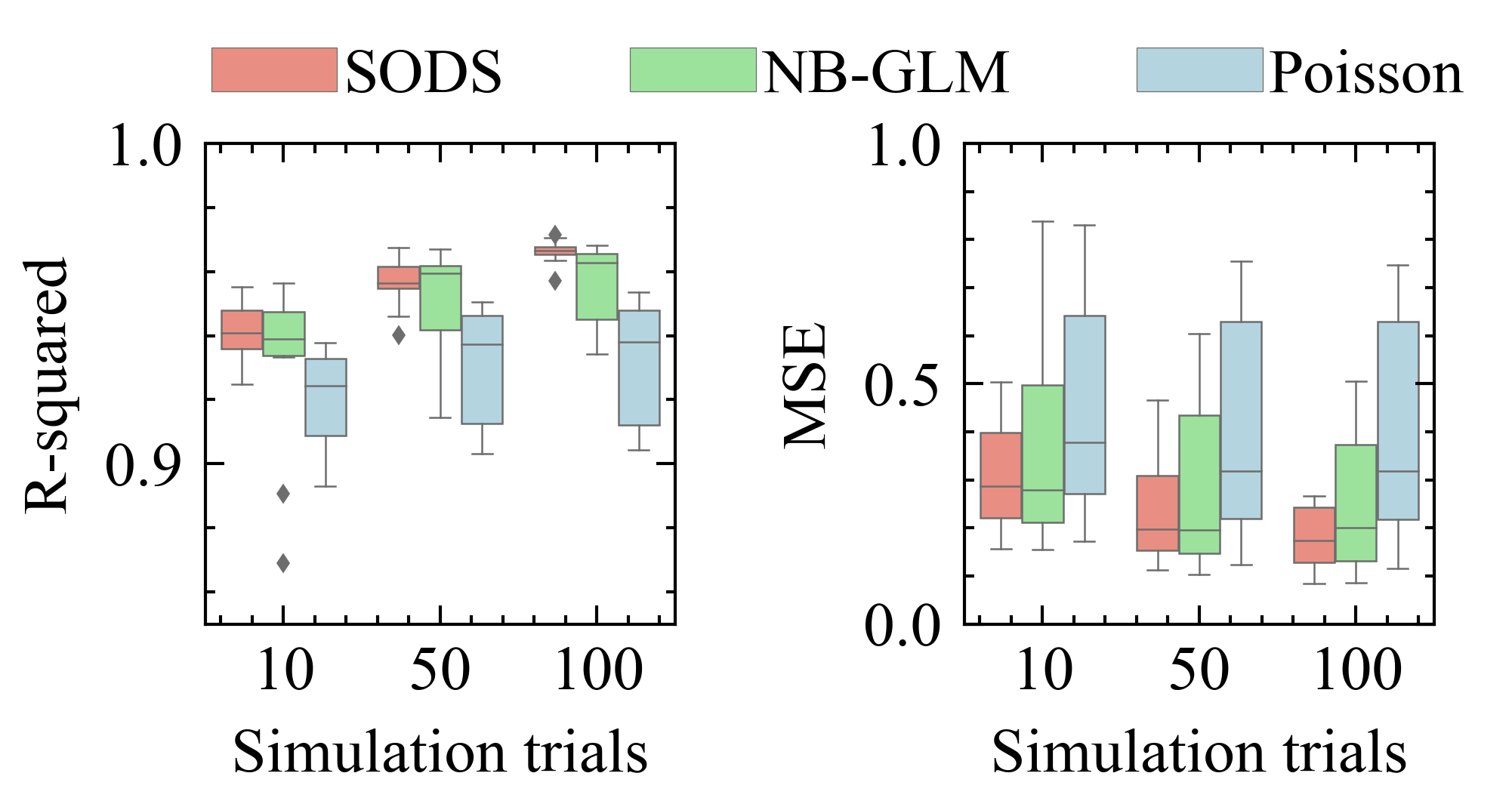}
    \vspace{-8mm}
    \caption{}
\end{subfigure}
\begin{subfigure}{0.8\columnwidth}
\centering
     \includegraphics[width=\columnwidth]{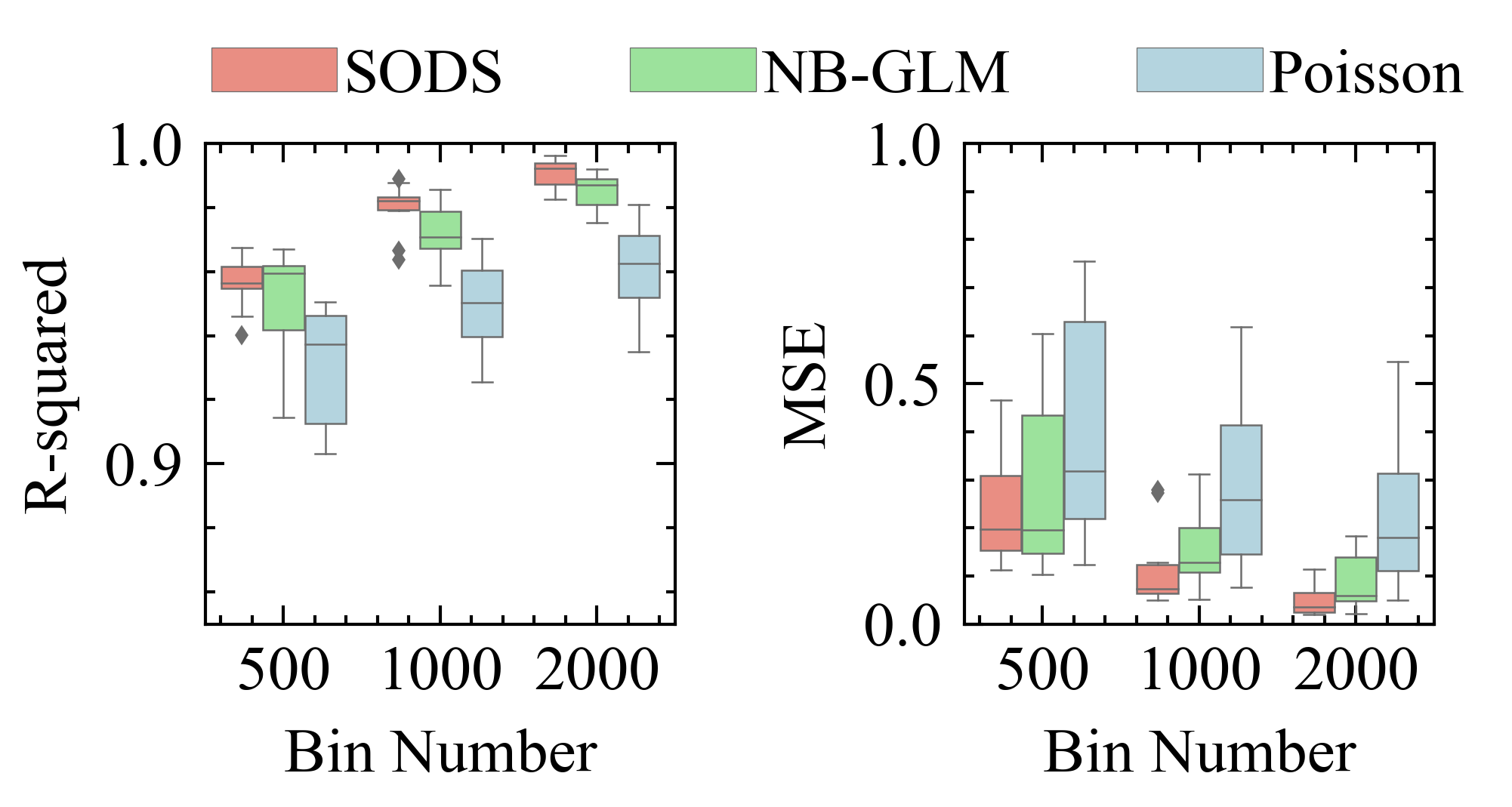}
     \vspace{-8mm}
     \caption{}
\end{subfigure}
\caption{Box-plots of the R-squared and MSE values for three $\GLM$ methods when estimating mean spike counts from ``$\SODS$" simulated data of different trial numbers $N_{s}$ (a) and different data length $K$ (b). Red:``$\SODS$" estimator; Green: NB-$\GLM$; Blue: Poisson-like $\GLM$. Other simulated parameters used for ``$\SODS$" simulation: $\bm{r} = 5, \bm{\sigma} = 50, \bm{\gamma}=7$. $K=500$ in (a), $N_{s}=50$ in (b).}
\label{fig:bin_sim_r2_mse}
\end{figure}

\begin{figure}[tbp]
    \centering
    \includegraphics[width=0.8\columnwidth]{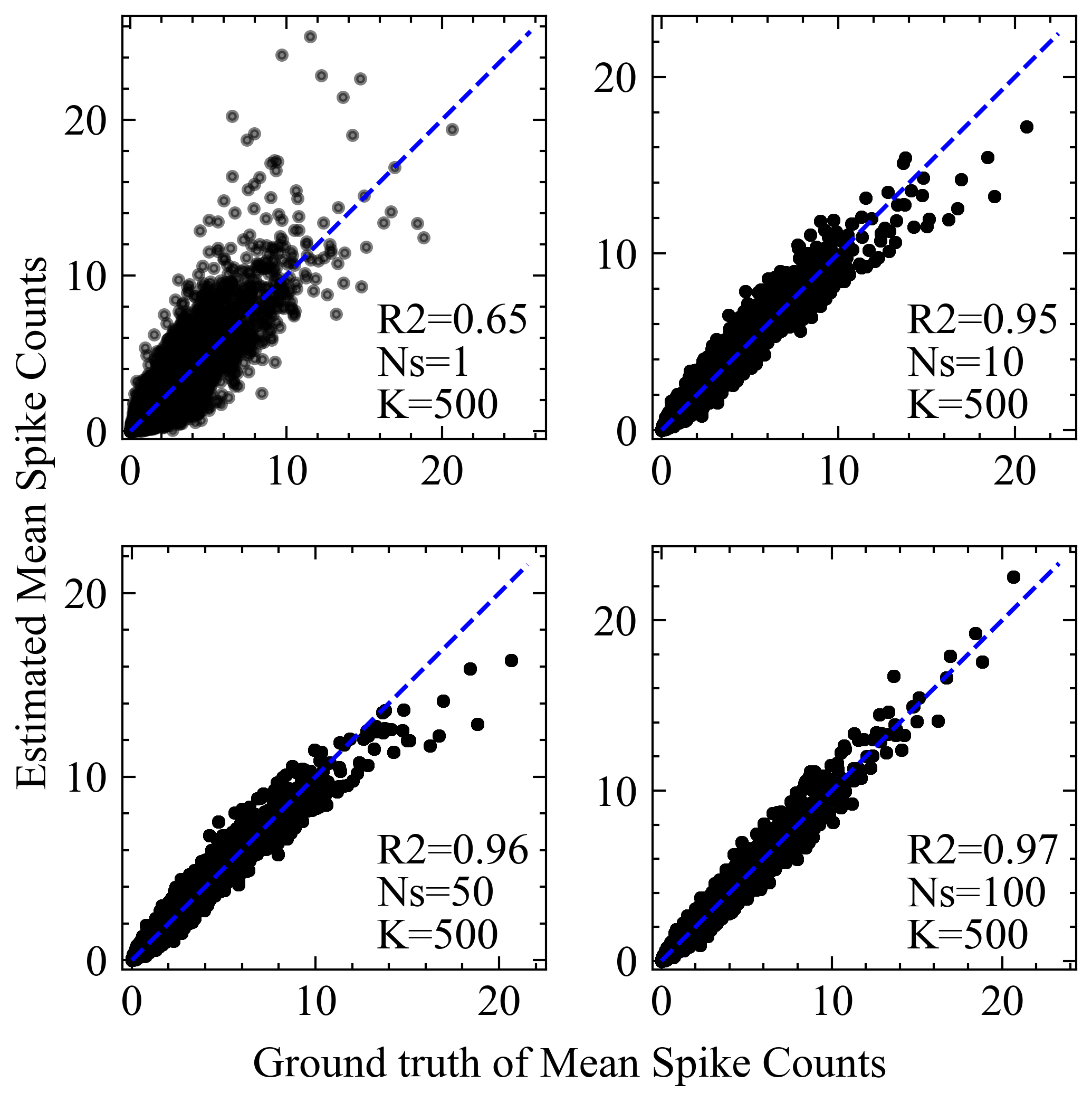}
    \caption{Comparison of the ground truth and estimated mean spike counts from ``$\SODS$" estimation of data simulated from ``$\SODS$" model. A larger number of simulation trials $N_{s}$ improves the estimation. Other simulated parameters used for ``$\SODS$" simulation: $\bm{r} = 5, \bm{\sigma} = 50, \bm{\gamma}=7$.}
    \label{fig:spike_counts}
\end{figure}

As shown in Fig.~\ref{fig:simulated_data_overdispersion}, the simulated data from all three $\GLMs$ had higher spike counts variance than spike counts mean. Fig.~\ref{fig:bin_sim_r2_mse} shows the performance of ``$\SODS$", NB-$\GLM$, and Poisson-like $\GLM$ estimators in simulated data of different simulation trials and data length. 
First, we evaluated the estimation of the mean spike counts. As expected, increasing simulation trials and data length reduced the MSE and improves R-squared values for all estimators. With flexible link function and capturing the over-dispersed spiking behaviour, the ``$\SODS$" and NB-$\GLM$ estimators showed better performance than Poisson-like $\GLM$. Moreover, ``$\SODS$" estimator outperforms the other two models with higher R-squared value and smaller MSE. Fig.~\ref{fig:bin_sim_r2_mse} shows that with only $10$ trials and $500$ bins, ``$\SODS$" still performed well with small MSE and high R-squared value. The scatter plots in Fig.~\ref{fig:spike_counts} provide a clear view of the comparison between ground truth and estimated mean spike counts under different simulation trials and data length.

\begin{figure}[tbp]
    \centering
    \includegraphics[width=0.8\columnwidth]{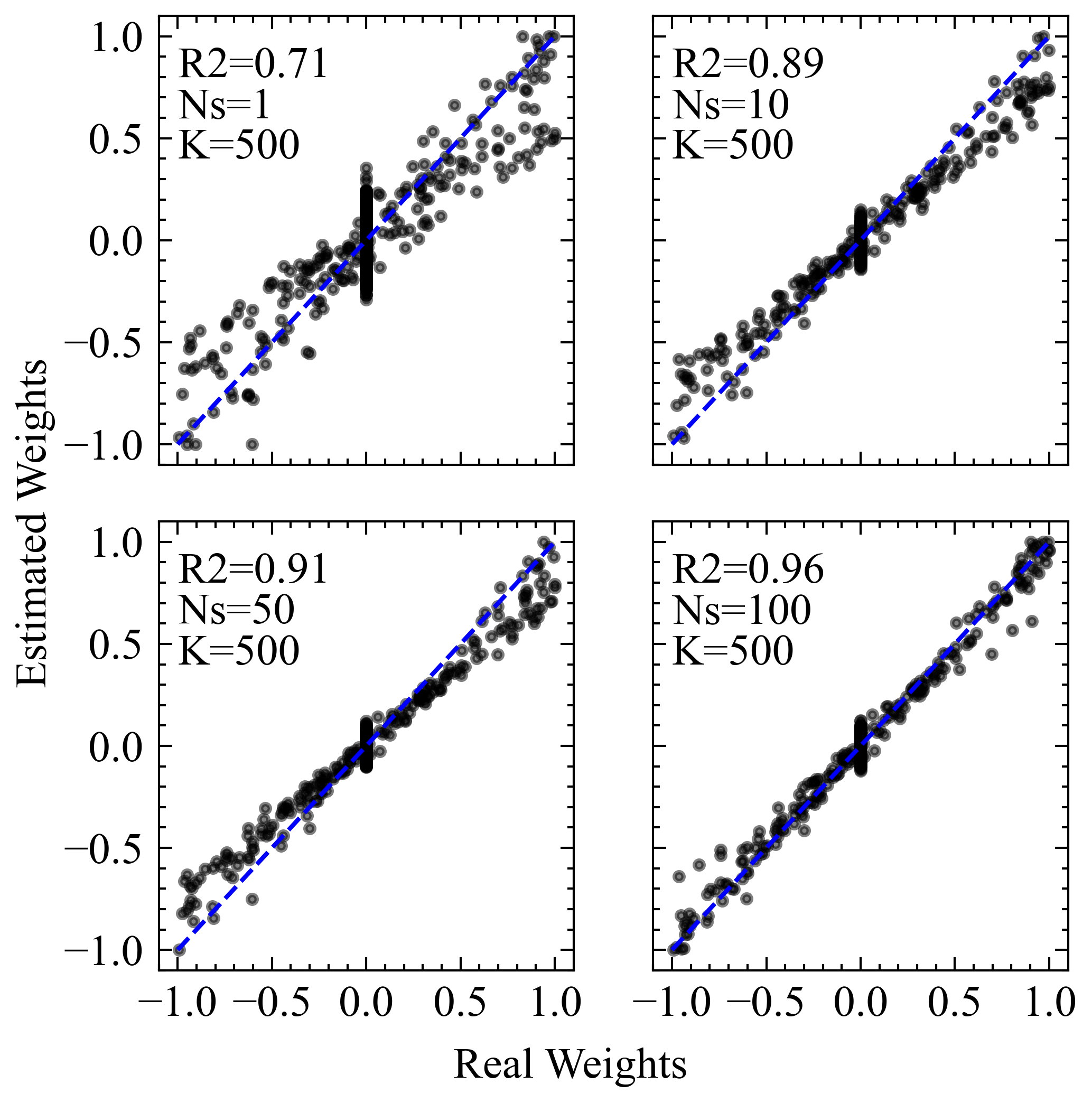}
    \caption{Comparison of the estimated weights from ``$\SODS$" estimator and the ground truth weights. Different number of trials $N_{s}$ were used for simulation from ``$\SODS$" model. Other simulated parameters used for ``$\SODS$" simulation: $\bm{r} = 5, \bm{\sigma} = 50, \bm{\gamma}=7, K = 500$.}
    \label{fig:weights}
\end{figure}
\begin{figure}[tbp]
\centering
\begin{subfigure}{0.8\columnwidth}
    \includegraphics[width=\columnwidth]{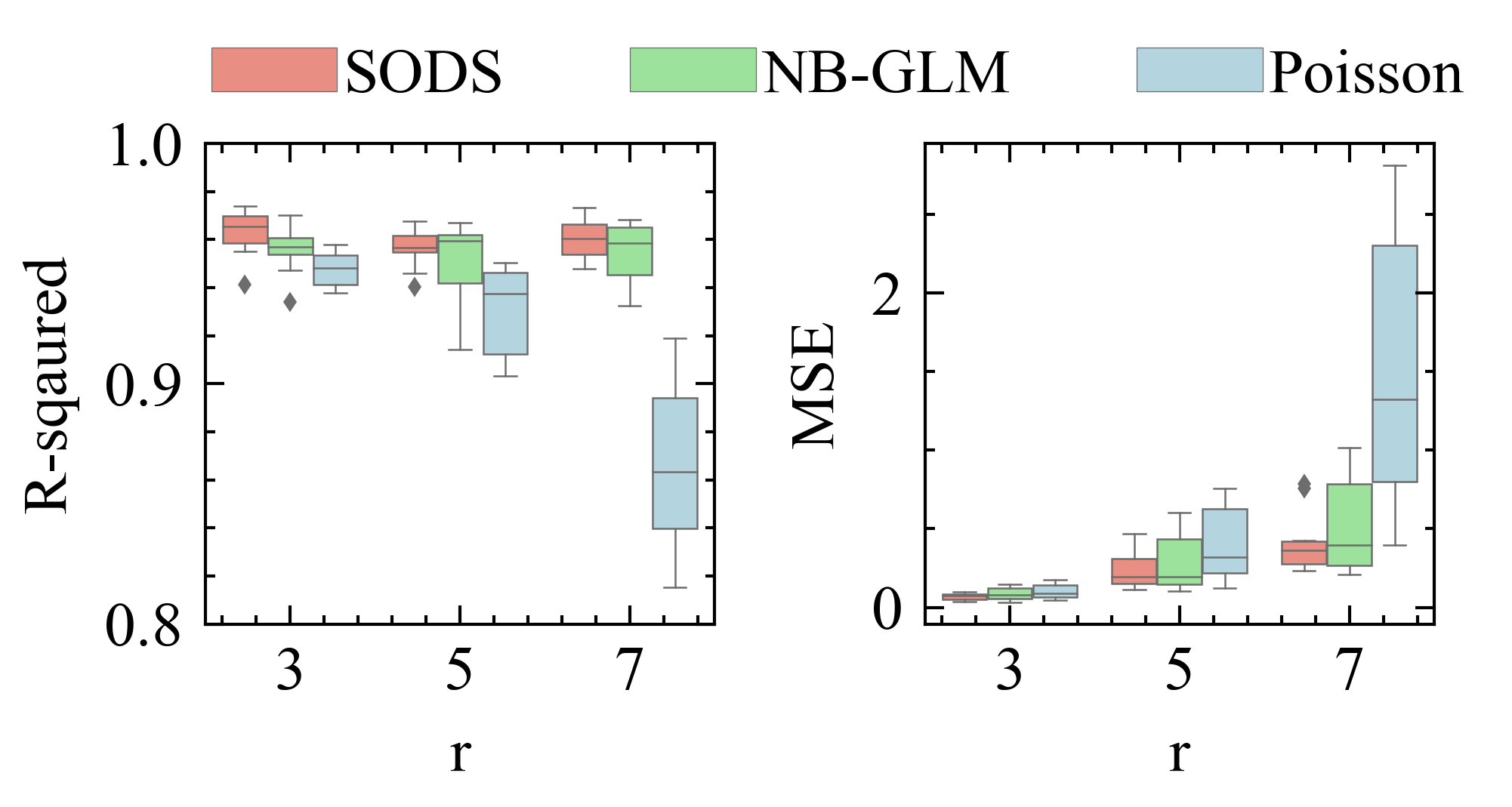}
    \vspace{-5mm}
    \caption{}
\end{subfigure}
\begin{subfigure}{0.8\columnwidth}
    \includegraphics[width=\columnwidth]{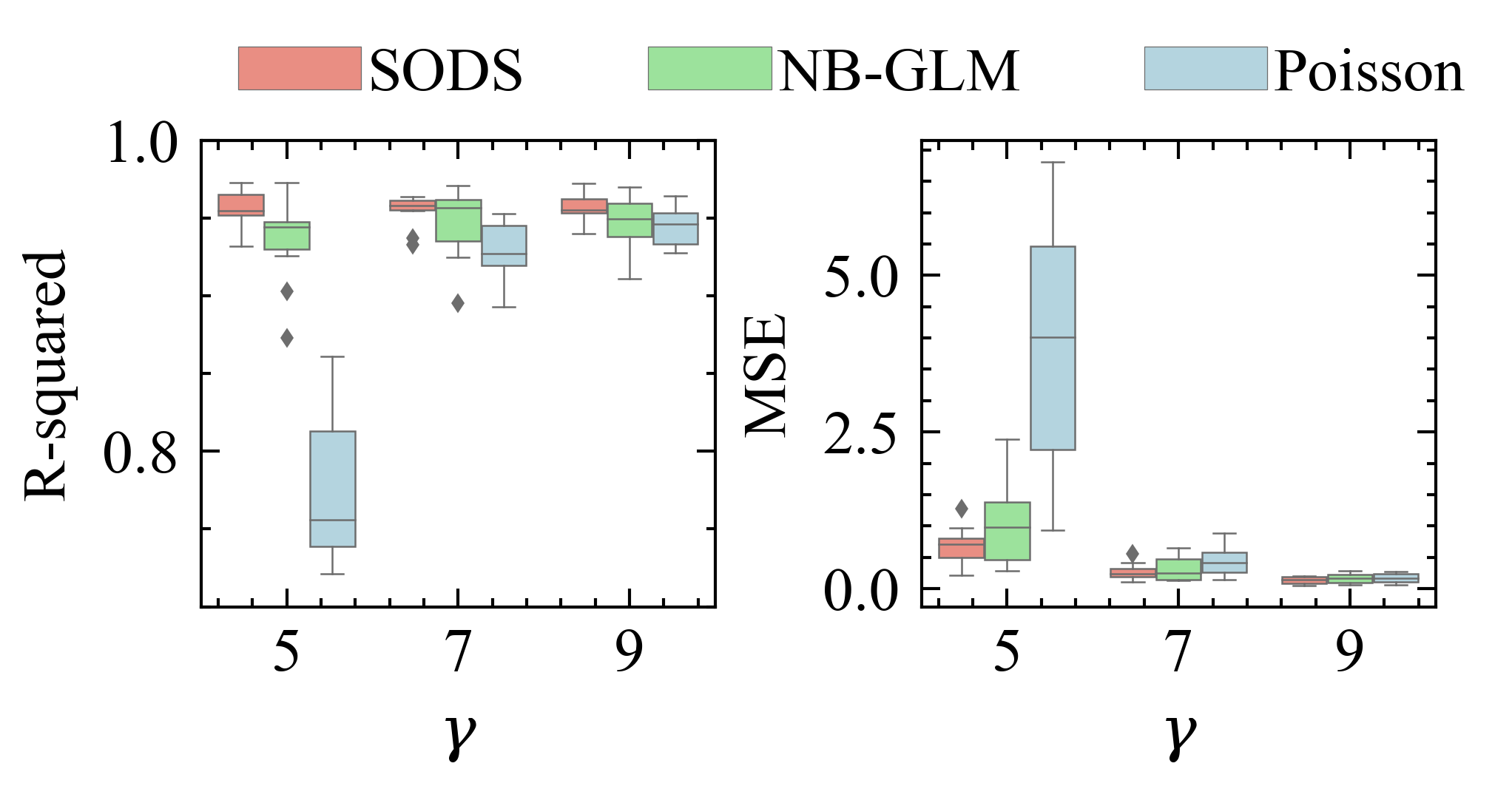}
    \vspace{-5mm}
    \caption{}
\end{subfigure}
\begin{subfigure}{0.8\columnwidth}
     \includegraphics[width=\columnwidth]{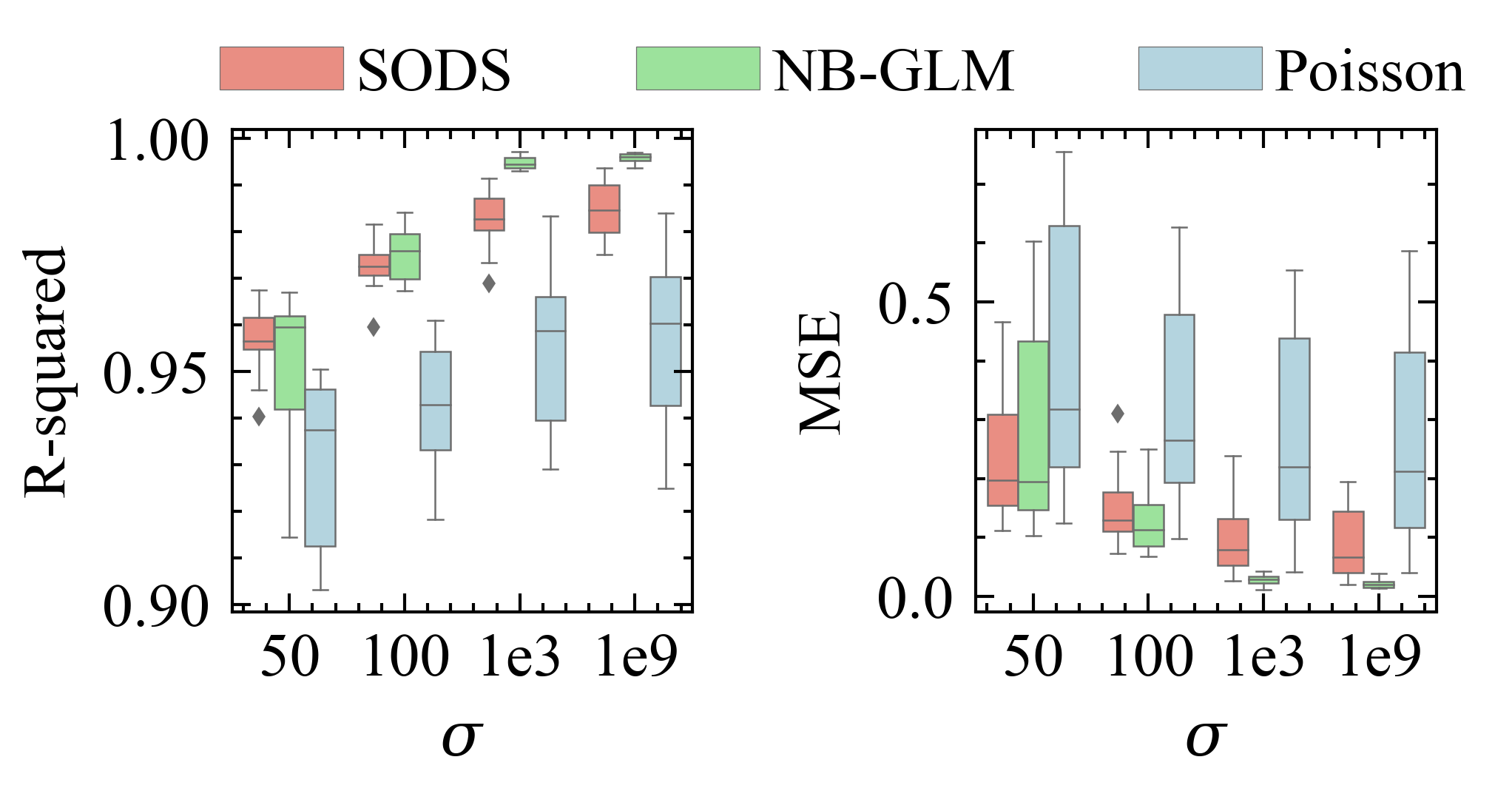}
     \vspace{-5mm}
     \caption{}
\end{subfigure}
\caption{Box-plots of the R-squared and MSE value for three $\GLM$ methods when estimating the mean spike counts from ``$\SODS$" simulated data of different parameters $\bm{r}$, $\bm{\sigma}$, and $\bm{\gamma}$. Red:``$\SODS$" estimator; Green: NB-GLM; Blue: Poisson-like GLM. Other simulated parameters used for ``$\SODS$" simulation: $N_{s} = 50, K = 500$.}
\label{fig:sod_s_r_gam_r2_mse}
\end{figure} 
Figures~\ref{fig:sod_s_r_gam_r2_mse}a and~\ref{fig:sod_s_r_gam_r2_mse}b show that ``$\SODS$" model consistently outperformed the other two models with different values of $\bm{r}$ and $\bm{\gamma}$. The shape parameter $\bm{r}$ affects the expected mean spike counts according to Eq.~\eqref{eq:nb}. ``$\SODS$" model better describes the spiking behaviour regardless of the true value of $\bm{r}$, which makes it a better estimator for neurons with different range of spike firing rate. Although the MSE increased with $\bm{r}$, the MSE increase in ``$\SODS$" was less than the other two models. 

When the simulation used a smaller $\bm{\sigma}$, all three estimators gained smaller R-squared values and larger MSE regarding the mean spike counts, as shown in Fig.~\ref{fig:sod_s_r_gam_r2_mse}(c). This is because $\bm{\sigma}$ represents the precision of the NB distribution. And since $\bm{\sigma}$ in ``$\SODS$" estimator was initiated within an interval of $(1,100)$, NB-$\GLM$ outperformed ``$\SODS$" when the simulated data was generated from $\bm{\sigma}$ far beyond that interval. However, the R-squared value of ``$\SODS$" estimator was still higher than 0.95, indicating the performance of ``$\SODS$" estimator was not greatly affected by the initial guess of $\bm{\sigma}$ during optimization.

Next, we evaluated the estimation of the weights $\bm{\omega}$ from the ``$\SODS$" simulated data. As we can see from Fig.~\ref{fig:weights}: (1) the ``$\SODS$" estimator performed better when more simulation trials were used; (2) estimation variation was large when the actual weights were close to 0; and (3) with as few as 50 simulation trials and 500 bins, the ``$\SODS$" estimator was sufficient to provide accurate weight estimations (R-squared value $>$ 0.9).
\subsection{Estimation Results of Simulated Data from Other $\GLM$ Models}
\begin{figure}[tbp]
\centering
\begin{subfigure}{0.8\columnwidth}
    \includegraphics[width=\columnwidth]{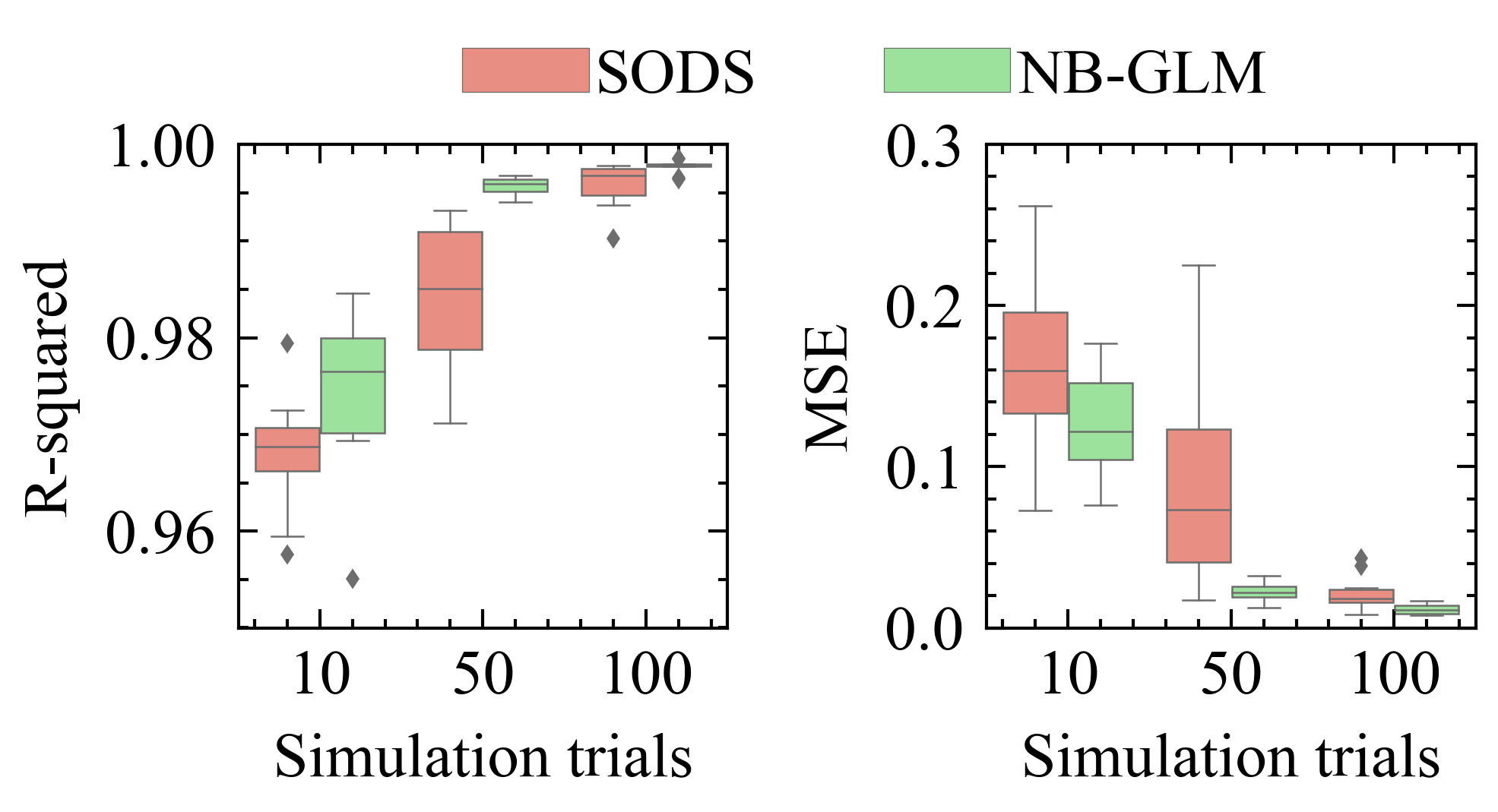}
    \vspace{-5mm}
    \caption{}
\end{subfigure}
\begin{subfigure}{0.8\columnwidth}
     \includegraphics[width=\columnwidth]{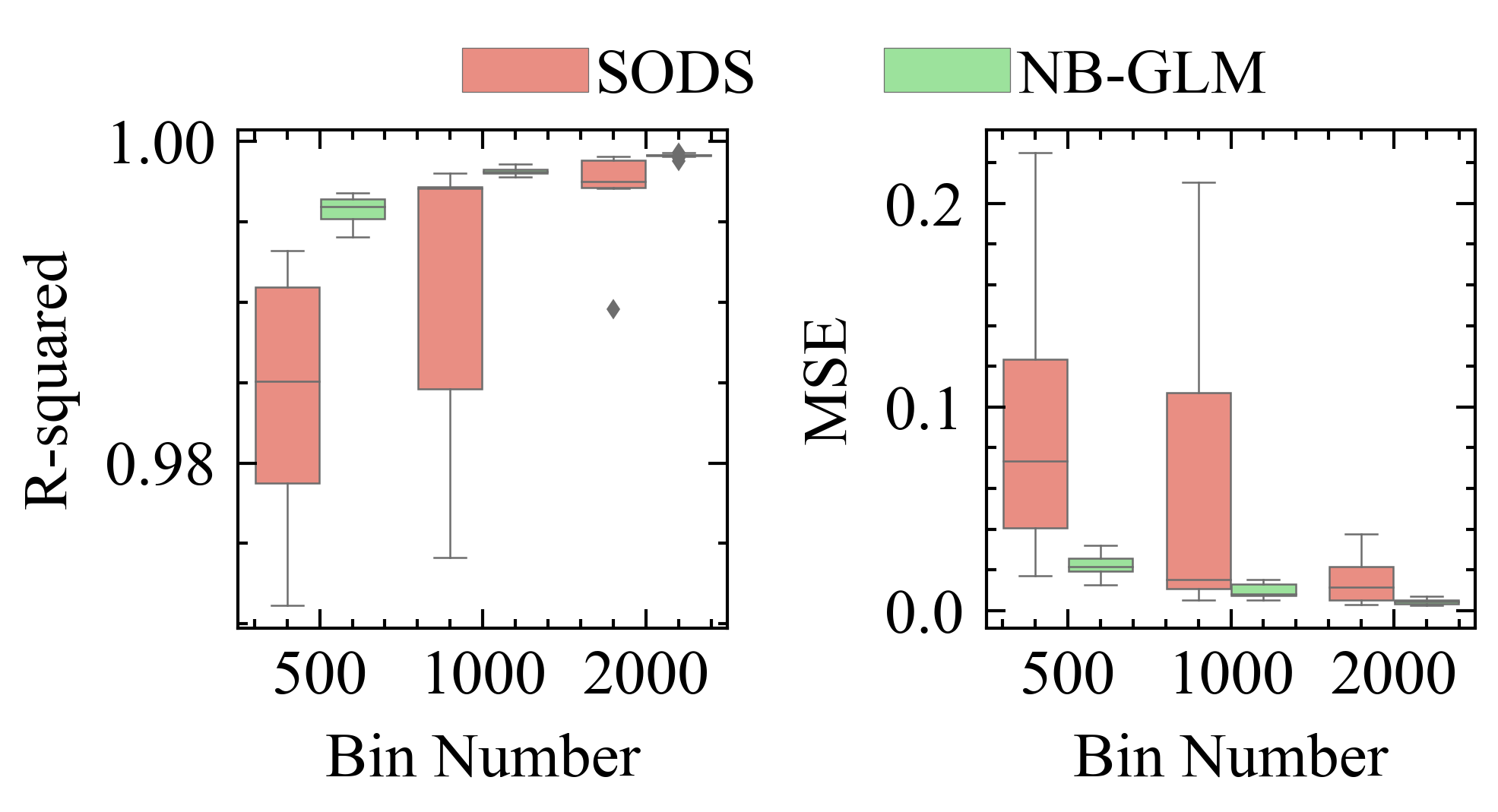}
     \vspace{-5mm}
     \caption{}
\end{subfigure}
\caption{Box-plots of the R-squared and MSE value for ``$\SODS$" and NB-$\GLM$ estimator, when estimating the mean spike counts from NB-$\GLM$ simulated data of different simulation trials (a) and different data length (b). Red:``$\SODS$" estimator; Green: NB-$\GLM$. Other simulated parameters: $\bm{r}=5, \bm{\gamma}=7$. $K=500$ in (a), $N_{s}=50$ in (b).}
\label{fig:nbglm_bin_sim_r2_mse}
\end{figure}

\begin{figure}[tbp]
\centering
\begin{subfigure}{0.8\columnwidth}
    \includegraphics[width=\columnwidth]{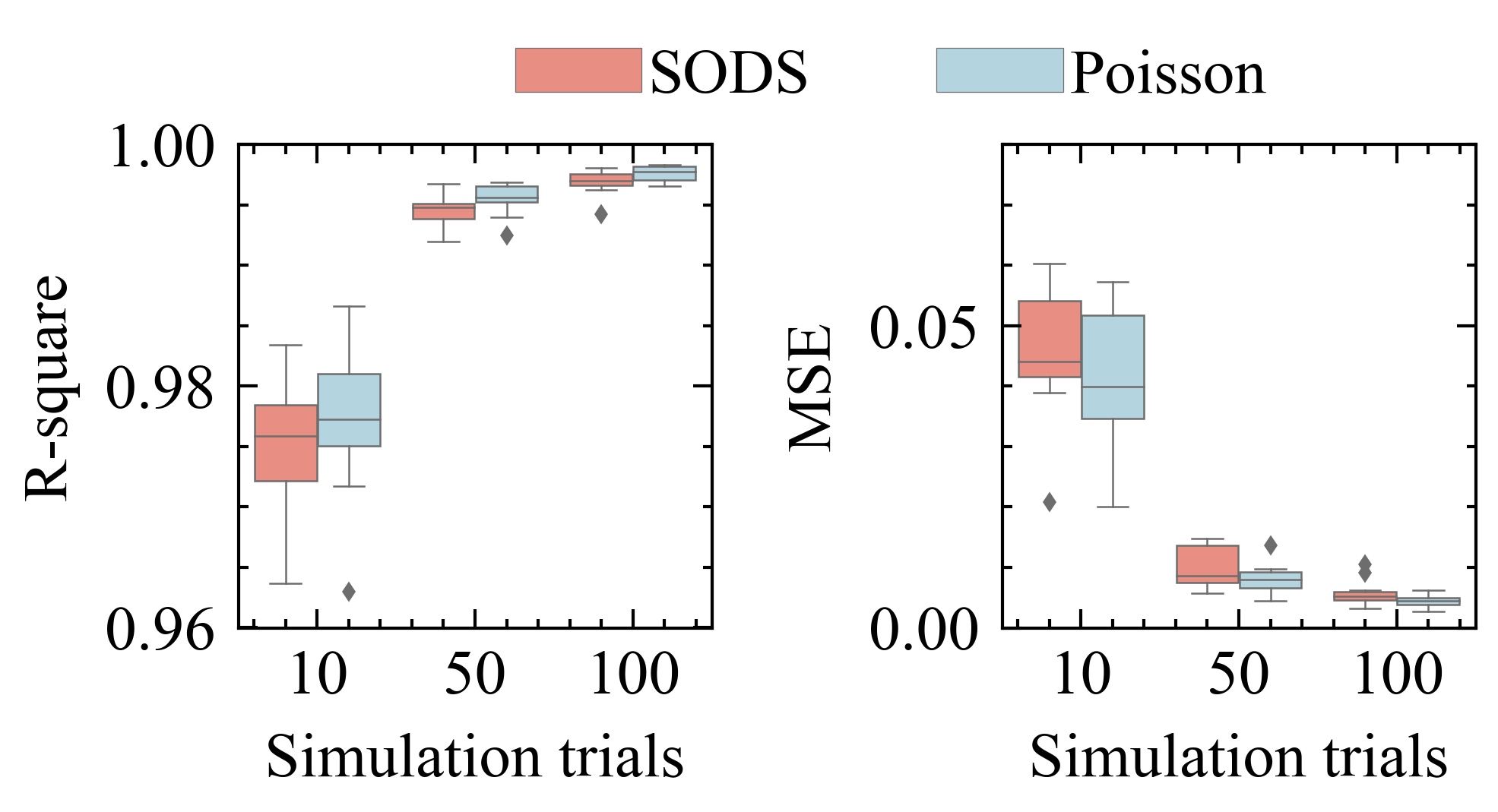}
    \vspace{-8mm}
    \caption{}
\end{subfigure}
\begin{subfigure}{0.8\columnwidth}
     \includegraphics[width=\columnwidth]{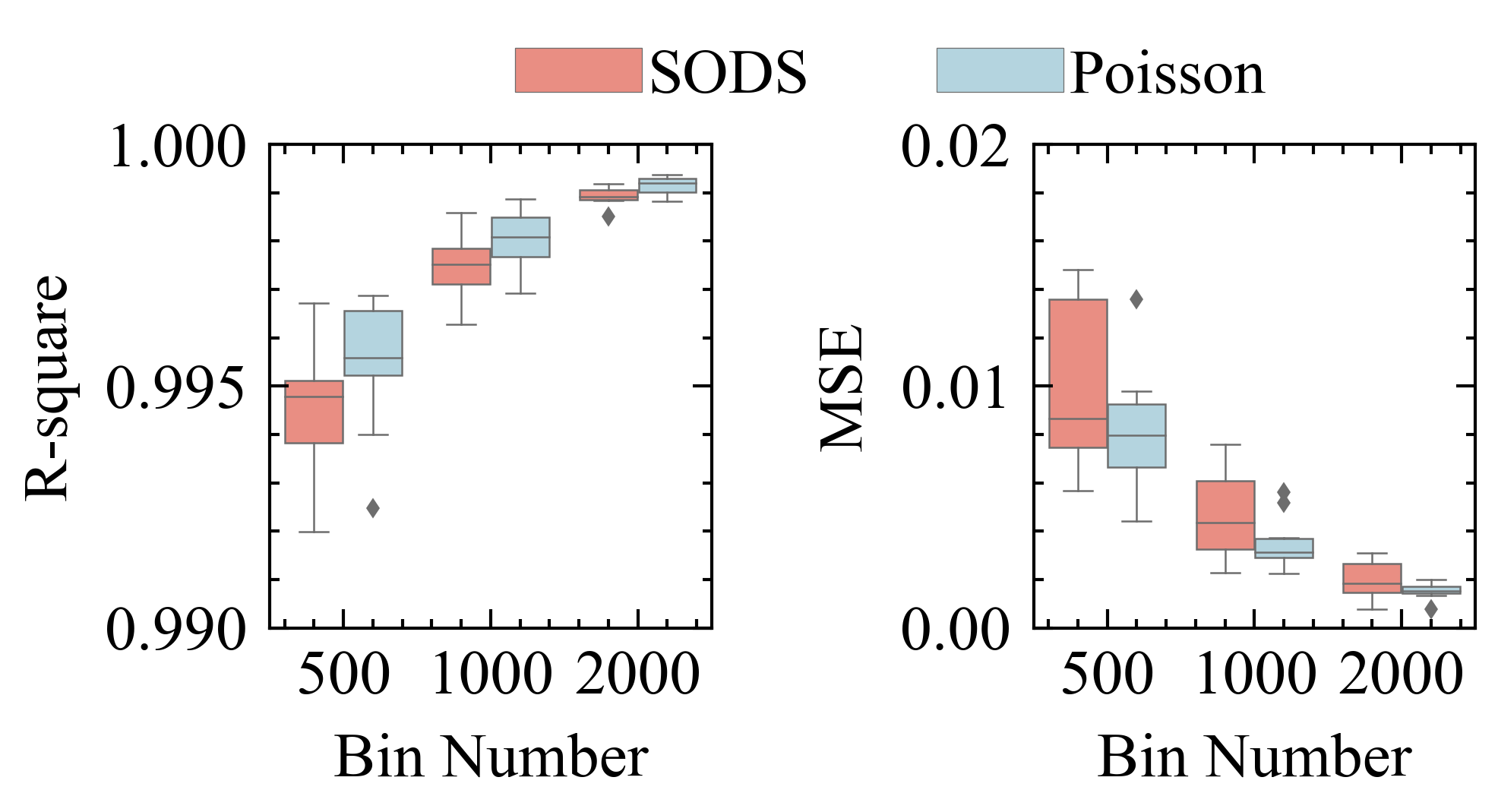}
     \vspace{-8mm}
     \caption{}
\end{subfigure}
\caption{Box-plots of the R-squared and MSE value for ``$\SODS$" estimator and Poisson-like $\GLM$, when estimating the mean spike counts from Poisson-like $\GLM$ simulated data of different simulation trials $N_{s}$ (a), and different data length $K$ (b). Red:``$\SODS$" estimator; Blue: Poisson-like GLM estimator. $K=500$ in (a), $N_{s}=50$ in (b).}
\label{fig:poisson_bin_sim_r2_mse}
\end{figure}

In addition to simulating data from ``$\SODS$" model, we further evaluated the ``$\SODS$" estimator when the simulation data were generated from NB-$\GLM$ and Poisson-like $\GLM$.
As seen in Fig.~\ref{fig:nbglm_bin_sim_r2_mse} and Fig.~\ref{fig:poisson_bin_sim_r2_mse}, the R-squared values achieved by the ``$\SODS$" estimator is higher than 0.95 for mean spike counts estimation. This indicates the ``$\SODS$" estimator can recover the true mean spike counts of the simulated data, even when they are generated from NB-$\GLM$ and Poisson-like $\GLM$. As expected, the performance of ``$\SODS$" estimator is not as good as those inferences using the matched model. The reduction in performance of ``$\SODS$" estimator was diminished when using more data for estimation.

\subsection{Estimation with Missing Neurons}
\begin{figure}[tbp]
\centering
  \includegraphics[width=0.8\columnwidth]{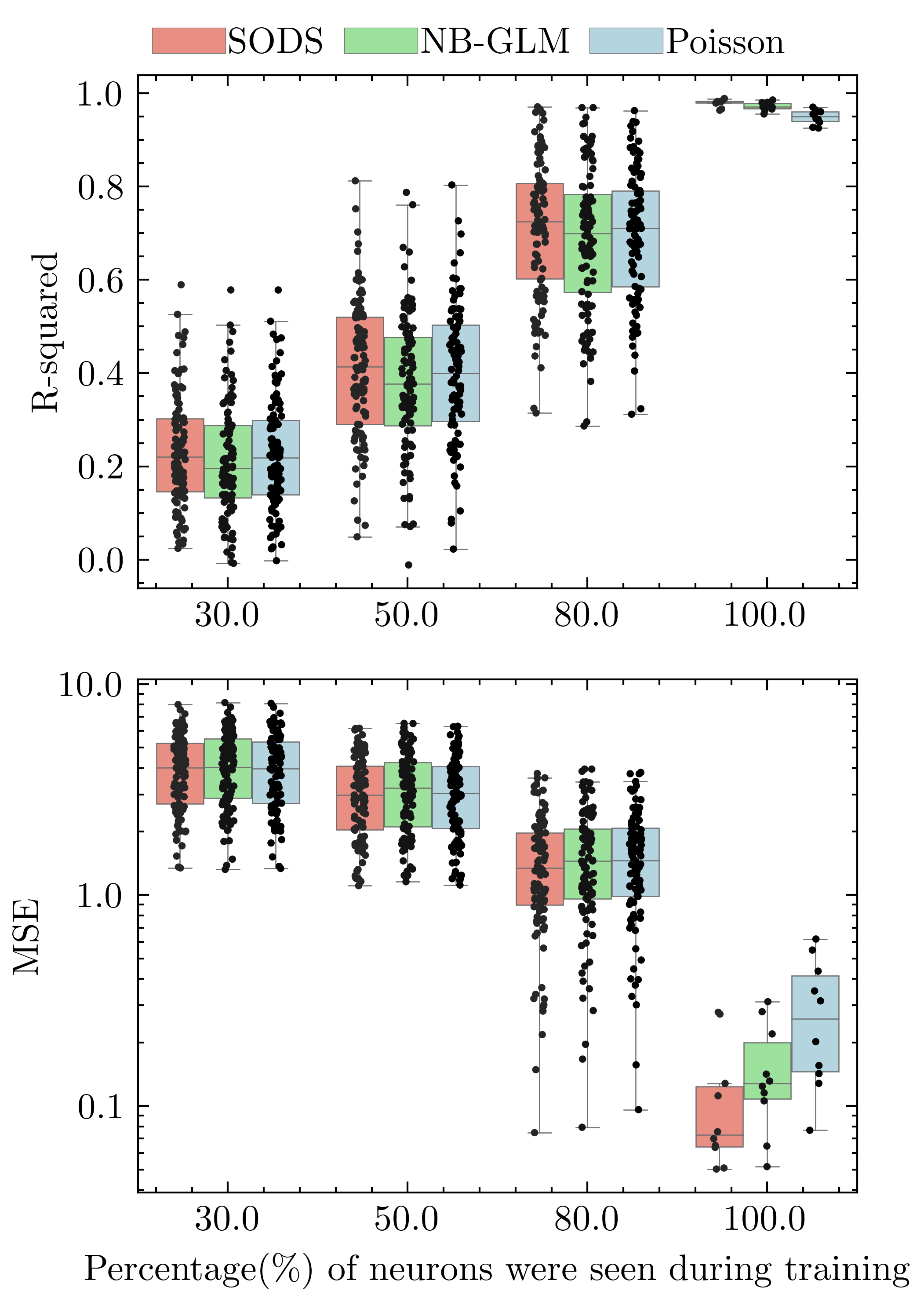}
\caption{Performance of estimators when only partial neurons were observed during training. The R-squared value (a) and MSE (b) were measured between the estimated and the true mean spike counts for testing data. The performance decreased as more neurons were missing during training. Simulation Parameters: $\bm{r}=5, \bm{\sigma}=50, \bm{\gamma}=7, N_{s}=50, K =1000$. No random sampling was done for the training data with 100\% observable neurons.}
\label{fig:missing_neuron}
\end{figure}

We cannot record spiking data from all neurons in the brain. When only partial neurons are observed, a better spike train model should flexibly capture the interaction between the observed spiking neurons. Therefore, we evaluated the performance of ``$\SODS$" estimation when using spiking data from partial neurons. Similarly, ten pairs of training and testing data were simulated from ``$\SODS$" model. The parameters and regressor $\bm{x}$ were initiated in similar ways to those simulated from ``$\SODS$" model. Then we randomly sampled a portion of the simulated neurons from the training data for estimation. The estimated parameters were used to predict the mean spike counts of the test data. These sampling, estimation, and prediction were repeated ten times for each pair of data. We compared the estimated mean spike counts with the ground truth of the test data. Results are shown in Fig.~\ref{fig:missing_neuron}. As the percentage of observed neurons decreased during estimation, all estimators have decreased R-squared value and increased MSE. When the observed neuron percentage reduced from 100\% to 30\%, the median R-squared values decreased from above 0.9 to about 0.2. Although ``$\SODS$" model had slightly higher median R-squared values and smaller median MSE when partial neurons were used for estimation, the differences among the three estimators were not significantly different.

\begin{figure}[tbp]
\centering
\vspace{-3mm}
\begin{subfigure}{0.8\columnwidth}
\centering
    \includegraphics[width=\columnwidth]{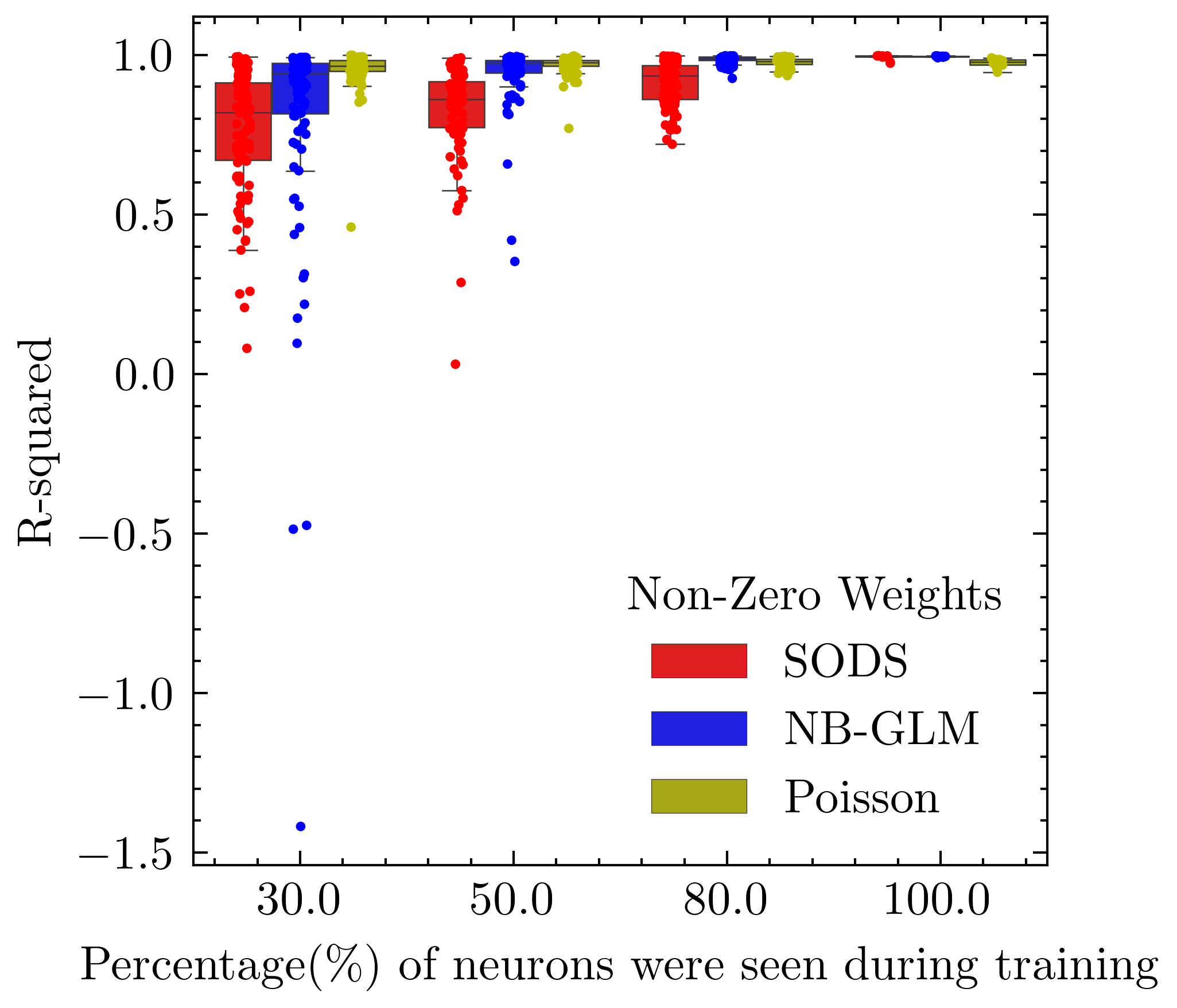}
    \caption{}
\end{subfigure}
\begin{subfigure}{0.8\columnwidth}
\centering
     \includegraphics[width=\columnwidth]{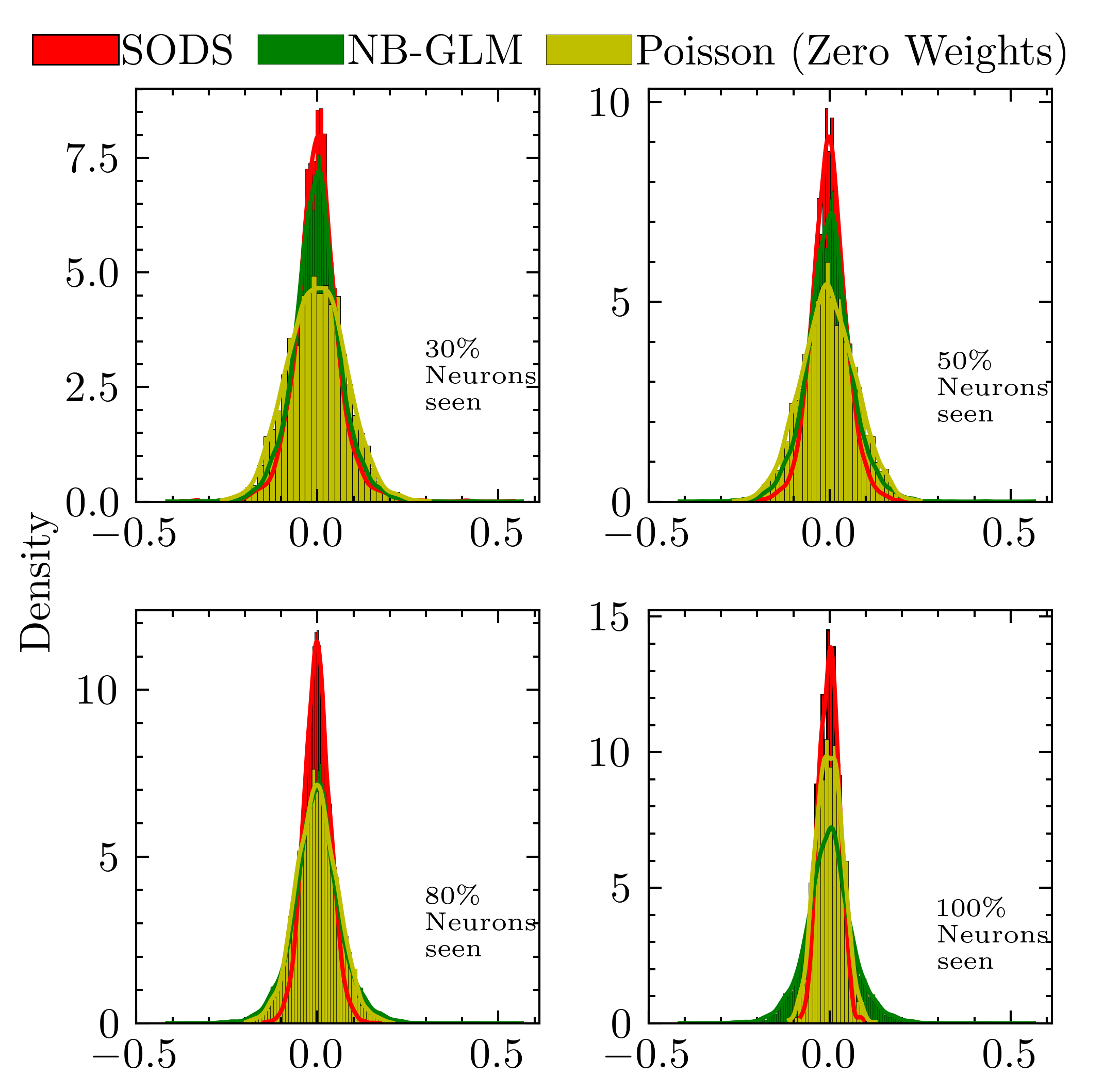}
     \caption{}
\end{subfigure}
\caption{Weights estimation when only partial neurons were observed during training. The non-zero and zero weights were evaluated separately. (a) R-squared value between non-zero weights and the corresponding estimated weights. (b) The density of estimated weights when the true weights were zero.  Simulation from ``$\SODS$" with parameters: $\bm{r}=5, \bm{\sigma}=50, \bm{\gamma}=7, N_{s}=50, K = 1000$.}
\label{fig:missing_neuron_weights}
\end{figure}

The estimated weights $\bm{\omega}$ were compared with the ground truth in missing neuron condition. We separated the estimated weights based on their corresponding ground truth values. For non-zero weights in $\bm{\omega}$, all three models decreased in the R-squared value when more neurons were missing. ``$\SODS$" showed more reduction in the R-squared value than the other two models(Fig.~\ref{fig:missing_neuron_weights}a). However, ``$\SODS$" can better estimate when the ground truth weights equal to zero. As shown in Fig.~\ref{fig:missing_neuron_weights}b, the estimated weights of ``$\SODS$" model were more likely to be closed to zero than the other two models.
\begin{figure}[tbp]
\centering
\includegraphics[width=2in]{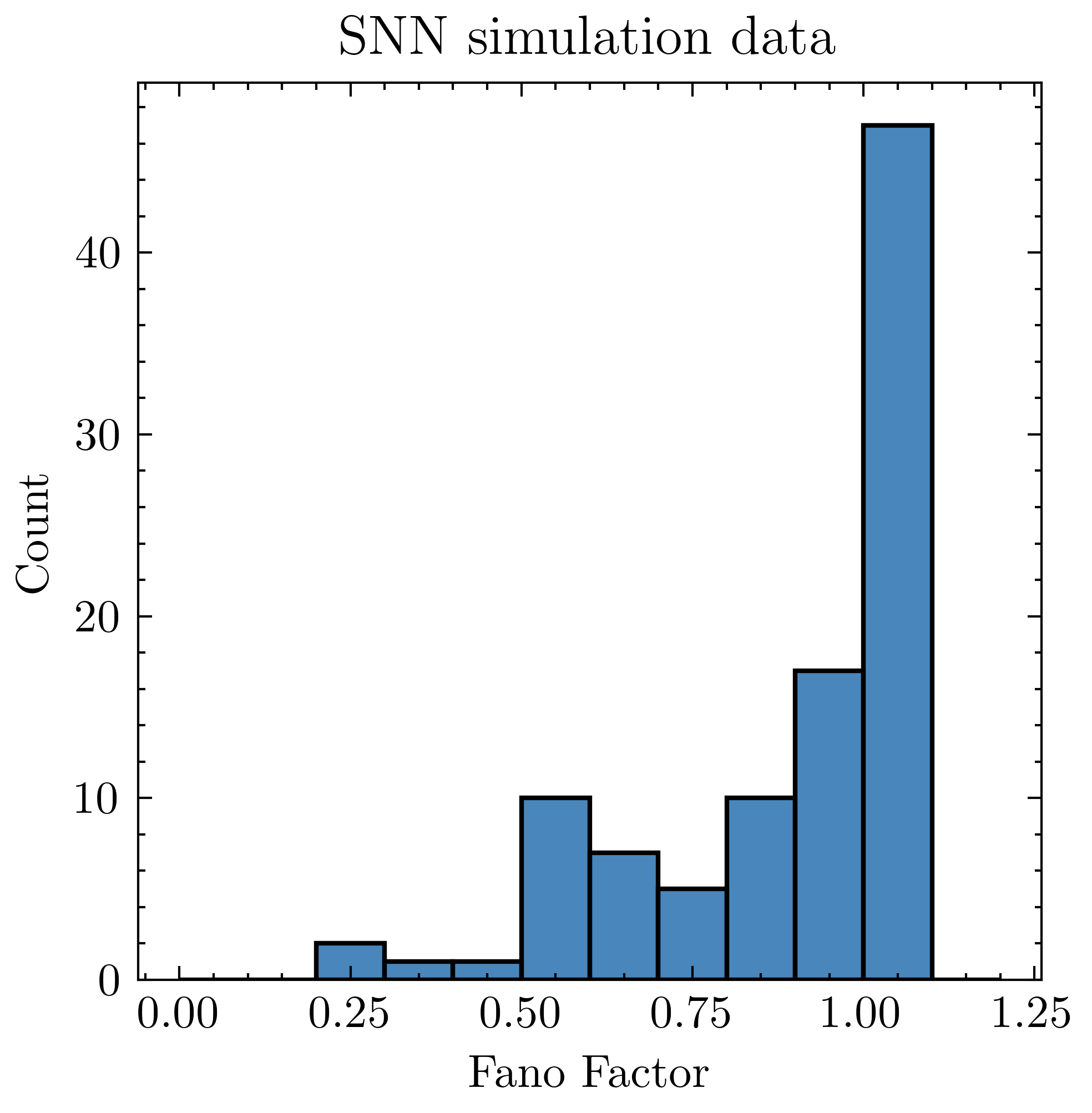}
\caption{The fano factor histogram of the 100 SNN-simulated target neurons. The fano factors varied from 0.29 to 1.07 in the target neurons.}
\label{fig:snn_overdispersion}
\end{figure} 
\begin{figure}[tbp]
\centering
\begin{subfigure}{0.8\columnwidth}
    \includegraphics[width=\columnwidth]{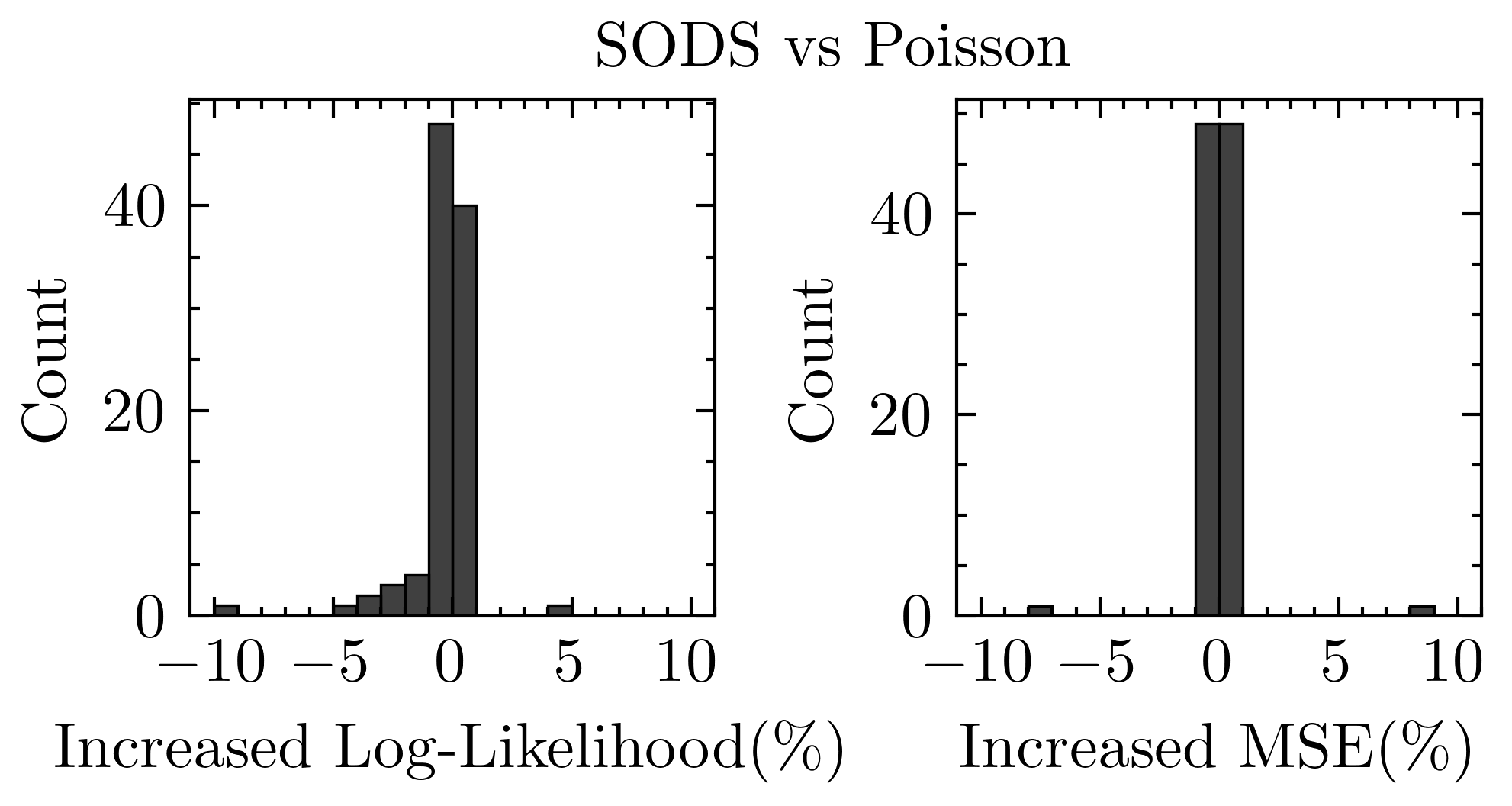}
    \caption{}
\end{subfigure}
\begin{subfigure}{0.8\columnwidth}
     \includegraphics[width=\columnwidth]{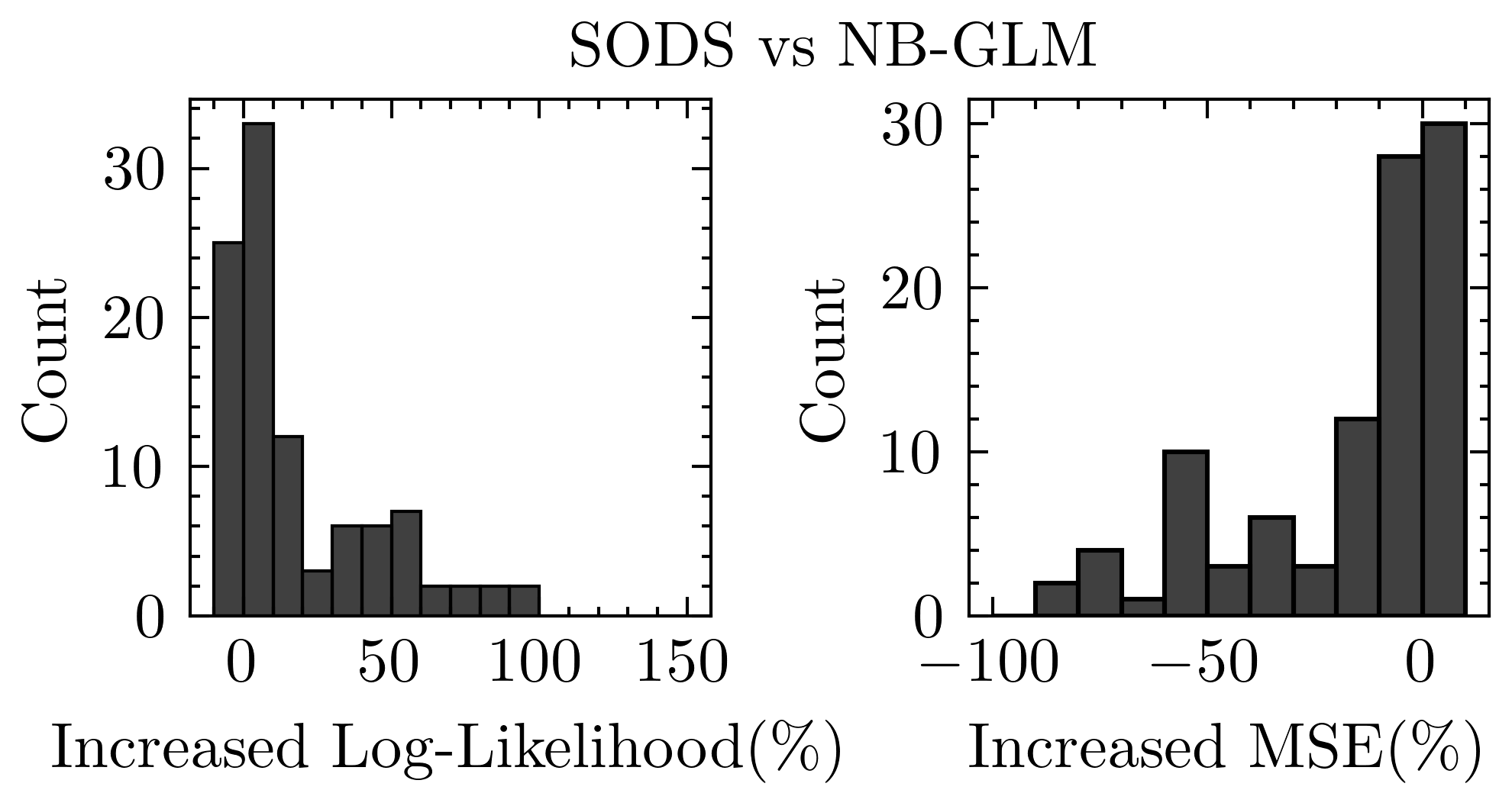}
     \caption{}
\end{subfigure}
\caption{Comparisons of ``$\SODS$" and other $\GLM$ models in SNN data estimation. Both the predictive log-likelihood and MSE were averaged across each fold of the 5-fold cross-validation. The positive value for the increased log-likelihood percentage indicates higher held-out data log-likelihood from ``$\SODS$" estimator; while negative value for the increased MSE percentage indicates lower held-out data MSE from ``$\SODS$" estimator. (a) The histogram of the increased log-likelihood percentage (left) and the increased MSE percentage (right) between ``$\SODS$" and Poisson models. (b)The histogram of the increased log-likelihood percentage (left) and the increased MSE percentage (right) between ``$\SODS$" and NB-GLM models.}
\label{fig:snn_poisson_nbglm}
\end{figure} 
\subsection{Estimation Results of Simulated Data from SNN Models}
We calculated the fano factors for all target neurons from the SNN model simulation. Results in Fig.~\ref{fig:snn_overdispersion} showed that 47\% of neurons have fano factor larger than 1, and 53\% of them have fano factor less than 1. We then compared the estimation performance for ``$\SODS$", NB-$\GLM$ and Poisson-like $\GLM$ methods. The estimation was done for each target neuron independently. The regressors $\bm{x}$ are the mean spike counts from the 100 neurons in the first layer of SNN. The target $\bm{y}$ is the mean spike counts of the target neuron from the second layer of SNN. The target $\bm{y}$ is one-bin later than the regressors $\bm{x}$. Firstly, we utilized 5-fold cross-validation to select the optimal parameters for the elastic-net regularization. Then another 5-fold cross-validation was used to estimate the predictive log-likelihood of the held-out data given different models, and MSE between the held-out data and the estimated mean spike counts. 
A higher predictive log-likelihood and a smaller MSE indicate the model better estimates the SNN data. We denote $\ell_{\mathrm{SODS}}$, $\ell_{\mathrm{NB}}$, $\ell_{\mathrm{Poisson}}$ as the predictive log-likelihoods of each model; and $MSE_{\mathrm{SODS}}$, $MSE_{\mathrm{NB}}$, $MSE_{\mathrm{Poisson}}$ as the predictive MSE. The percentage log-likelihood increase is calculated by $\frac{\ell_{\mathrm{SODS}}-\ell_{\mathrm{NB}}}{|\ell_{\mathrm{NB}}|}\times 100\%$ and $\frac{\ell_{\mathrm{SODS}}-\ell_{\mathrm{Poisson}}}{|\ell_{\mathrm{Poisson}}|}\times 100\%$.
The percentage MSE increase is calculated by $\frac{MSE_{\mathrm{SODS}}-MSE_{\mathrm{NB}}}{MSE_{\mathrm{NB}}}\times 100\%$ and $\frac{MSE_{\mathrm{SODS}}-MSE_{\mathrm{Poisson}}}{MSE_{\mathrm{Poisson}}}\times 100\%$.

Fig.~\ref{fig:snn_poisson_nbglm}a revealed that the log-likelihood and the MSE value of the held-out test data were similar between Poisson-like $\GLM$ and ``$\SODS$"estimator, as most of the differences were closed to zero.
But around 70\% of neurons showed higher log-likelihood and smaller MSE value in ``$\SODS$" estimator when compared to NB-$\GLM$, as seen in Fig.~\ref{fig:snn_poisson_nbglm}b.

Similar estimation performance between Poisson-$\GLM$ and  ``$\SODS$" estimators may be because the SNN spiking data distribution is very likely to be a Poisson distribution. Some neurons have fano factors very closed to 1. Our ``$\SODS$" estimator can achieve similar performance as Poisson-like $\GLM$ and outperform NB-$\GLM$ when the spiking data is not over-dispersed. 
\begin{figure}[tbp]
\centering
\begin{subfigure}{0.8\columnwidth}
    \includegraphics[width=\columnwidth]{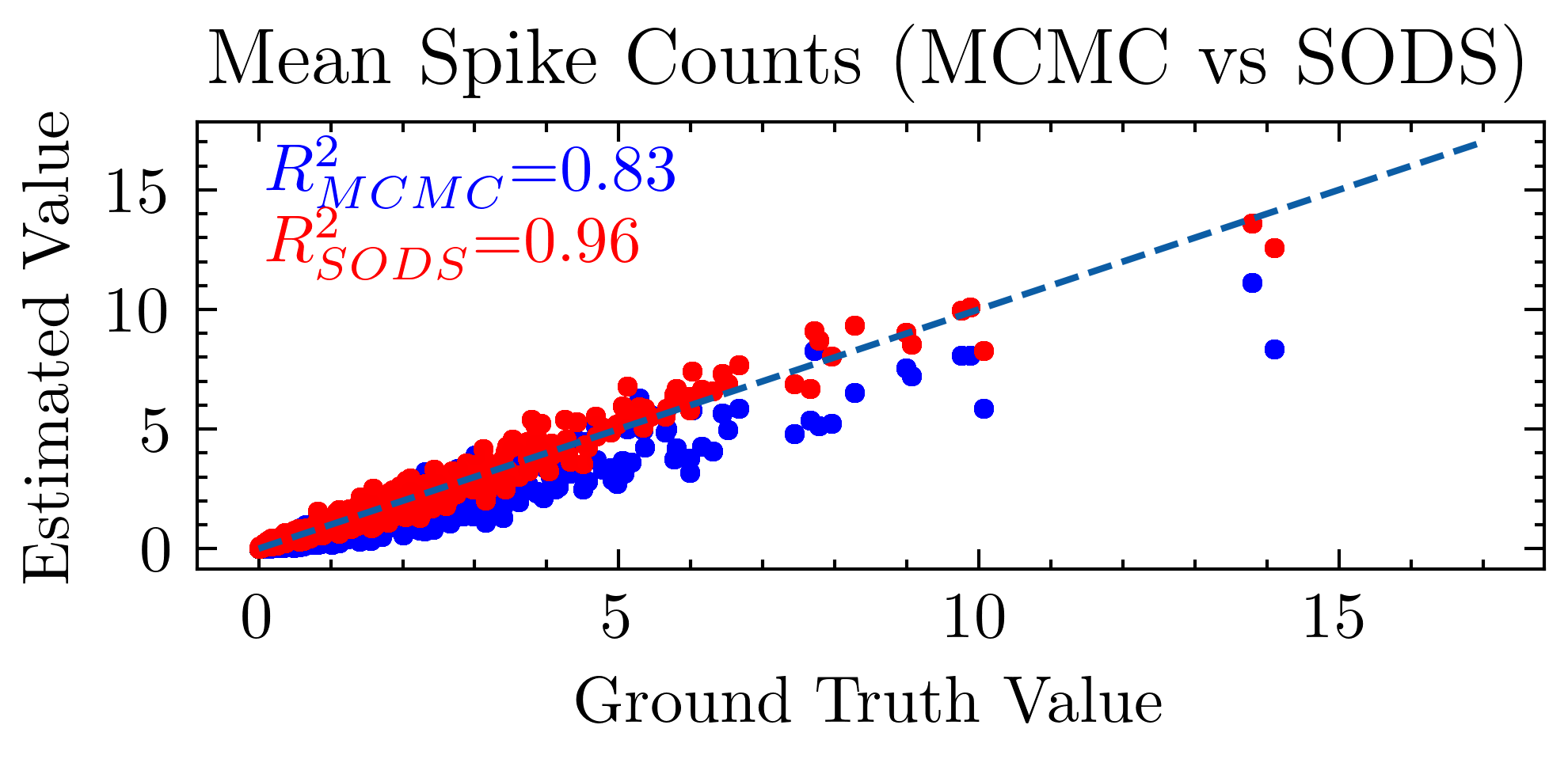}
    \caption{}
\end{subfigure}
\begin{subfigure}{0.8\columnwidth}
     \includegraphics[width=\columnwidth]{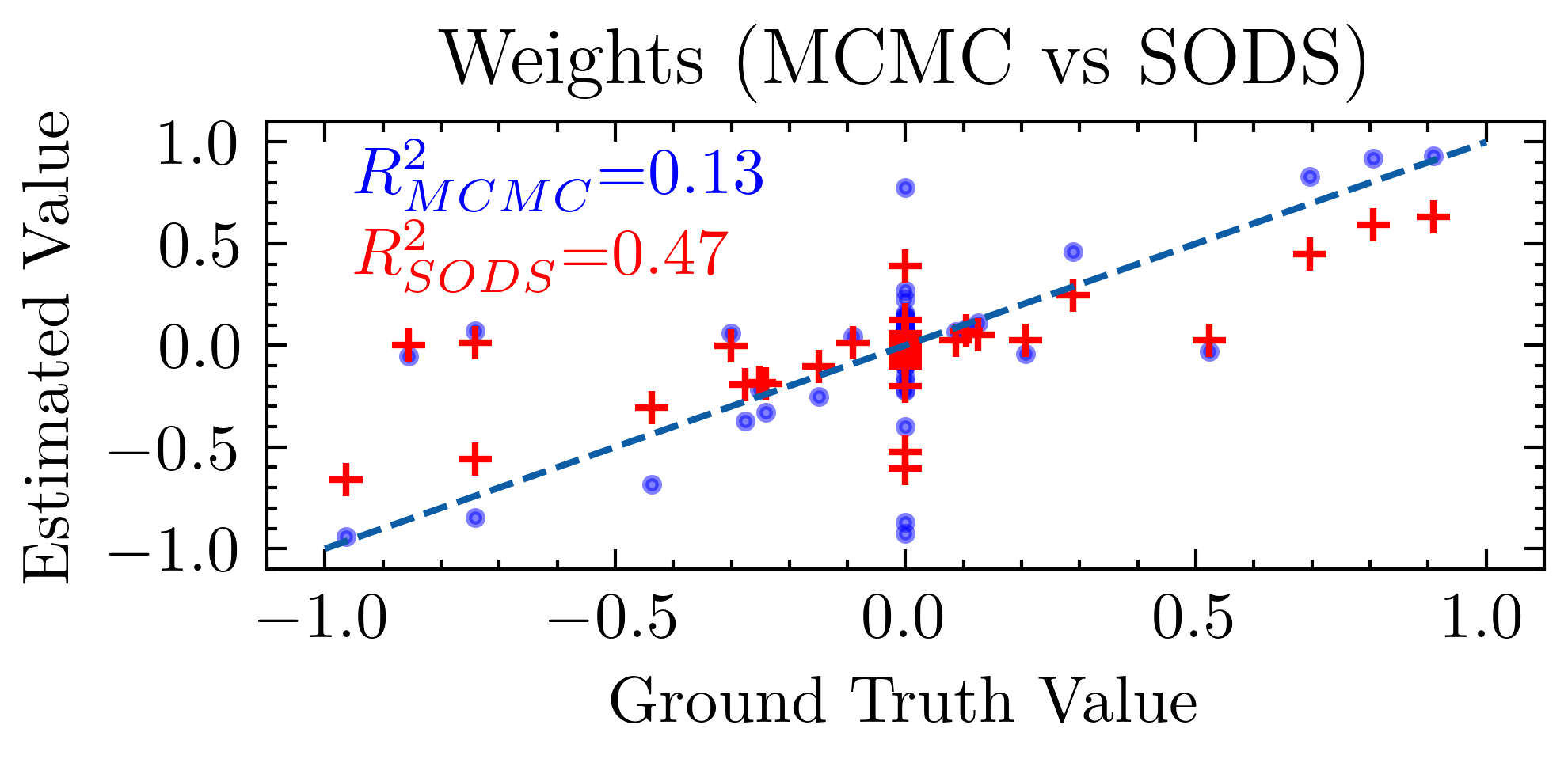}
     \caption{}
\end{subfigure}
\caption{Comparison of the MCMC inference and ``$\SODS$" estimation. (a) Comparison between the ground truth mean spike counts and the estimated mean spike counts from the MCMC inference (blue) and ``$\SODS$" estimation (red). R-squared value is higher in SODS ($R^{2}_{SODS} = 0.96$) than in the MCMC inference ($R^{2}_{MCMC} = 0.83$). (b) Comparison between the ground truth weights and the estimated weights from the MCMC inference (blue) and ``$\SODS$" estimation (red). R-squared value for weight estimation is higher in ``$\SODS$" ($R^{2}_{SODS} = 0.47$) than in the MCMC ($R^{2}_{MCMC} = 0.13$). Spiking data were generated by ``$\SODS$" simulation with parameters used for : $\bm{r} = 5, \bm{s} = 50, \bm{\gamma}=7, N_{s} = 50, K = 500$. We only used one set of training and test data for this comparison.}
\label{fig:mcmc}
\end{figure} 
\begin{figure}[tbp]
\centering
\begin{subfigure}{0.8\columnwidth}
\centering
    \includegraphics[width=\columnwidth]{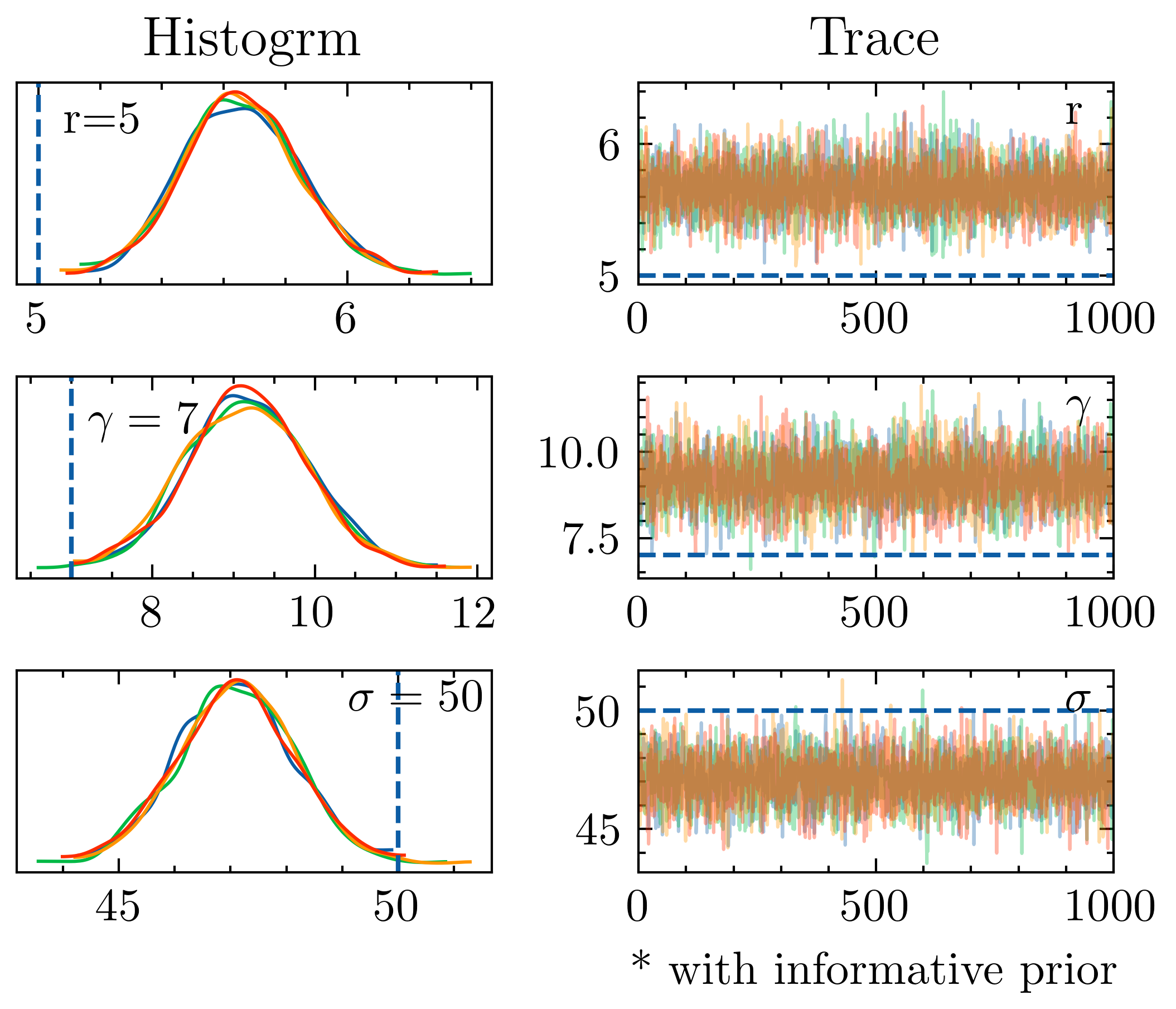}
    \caption{}
\end{subfigure}
\begin{subfigure}{0.8\columnwidth}
\centering
     \includegraphics[width=\columnwidth]{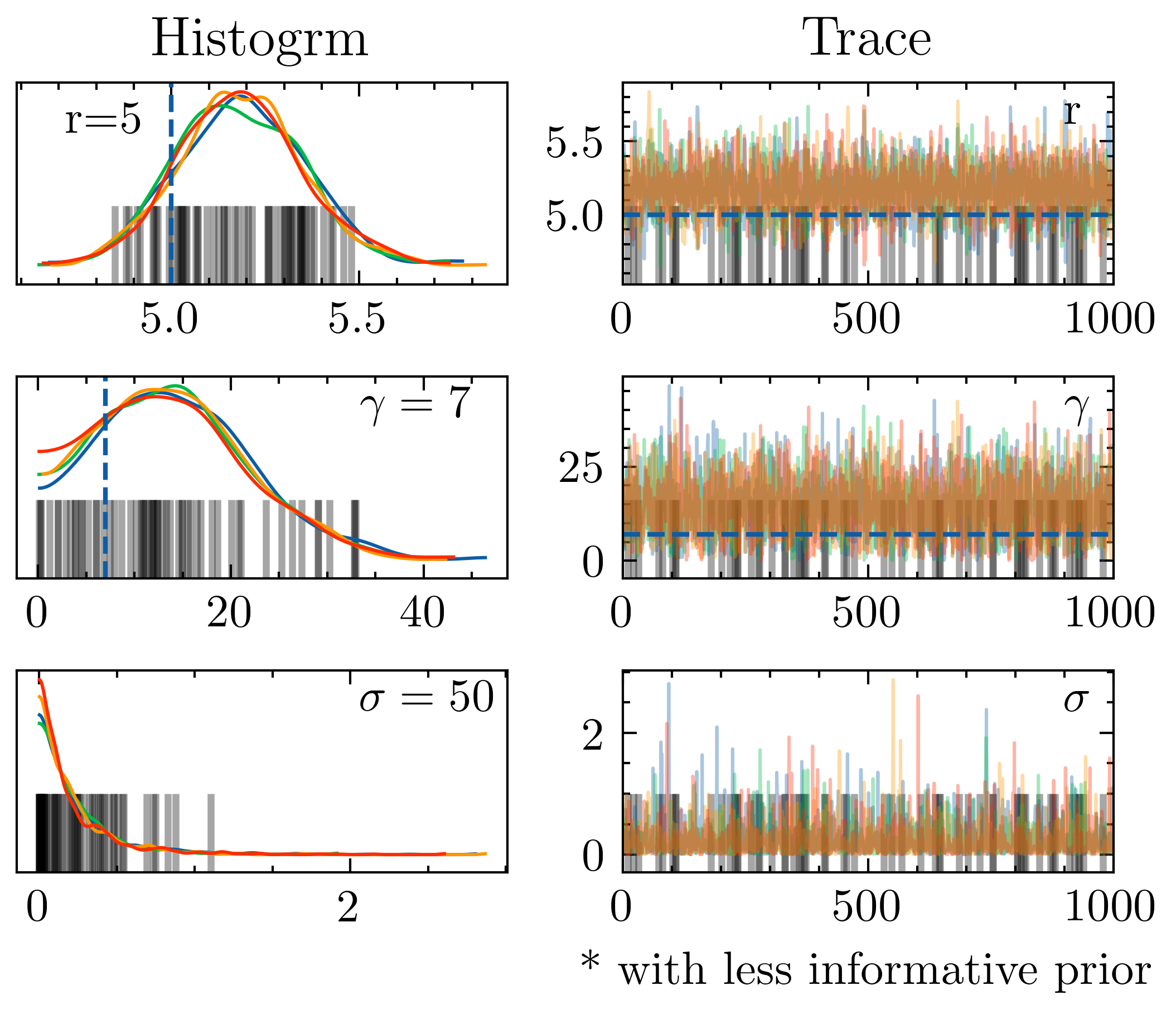}
     \caption{}
\end{subfigure}
\caption{The trace plot of MCMC inference with informative prior (a), and less informative prior (b). For informative prior, The sigma of each normal distribution in Eq~\eqref{eq:mcmc} equals to 1. For less informative prior, the sigma of each normal distribution equals to 10. The left column is the smoothed histogram (using kernel density estimation) of the marginal posteriors of three parameters $\bm{r}, \bm{\gamma}, \bm{\sigma}$. The dashed lines are the true values for the simulation parameters. The right column is the samples of the Markov chain plotted in sequential order.}
\label{fig:mcmc_convergence}
\end{figure} 
\subsection{Comparison between Empirical Bayes and fully Bayes}
The advantage of empirical Bayes inferences is its lower computational cost than fully Bayes approach. We use Basin-Hopping and L-BFGS-B in our empirical Bayes inferences. The computational complexity for L-BFGS-B is $O(mn)$. Here $m=10$ indicates $m$ recent gradients are used by L-BFGS-B for Hessian approximation\cite{nocedal2006numerical}. For fully Bayes, the process is based on the MCMC algorithm with Not U-Turn Sampler(NUTS), which has computational complexity as $O(n^\frac{5}{4})$, according to \cite{hoffman2014no}.

We evaluated the estimated mean spike counts and weights from the fully Bayes and the ``$\SODS$" model. We first set the informative hyperprior distribution for the parameters based on the simulation setting:
\begin{eqnarray}
     \bm{r} & \in & Normal(\mu=5, \sigma=1), lb=0 \nonumber \\
     \bm{\sigma} & \in & Normal(\mu=50, \sigma =1), lb=1 \nonumber \\
     \bm{\gamma} & \in & Normal(\mu=7, \sigma =1), lb=0 \nonumber\\ 
     \bm{\omega} & \in & Normal(\mu=0, \sigma =1),lb=-1, ub=1 \nonumber\\
\label{eq:mcmc}
\end{eqnarray}
The normal distribution here is truncated normal distribution with bounds. $lb$ denotes the lower bound, and $ub$ denotes the upper bound. We evaluated the estimation for one pair of training and testing data. Results are shown in Fig.~\ref{fig:mcmc}. ``$\SODS$" estimation have better performance than MCMC inference, in both mean spike counts (Fig.~\ref{fig:mcmc}a) and weights estimation (Fig.~\ref{fig:mcmc}b), even when the parameter hyperpriors were chosen arbitrarily to match the simulation setting during MCMC inference.

The posterior predictive of the parameters $\bm{r}, \bm{\sigma}, \bm{\gamma}$ showed convergence across 4 chains, but the estimated values were very much different from the true value (Fig.~\ref{fig:mcmc_convergence}a). When we used a less informative hyperprior by increasing the sigma value from 1 to 10 in the hyperprior distribution of the parameters $\bm{r}, \bm{\sigma}, \bm{\gamma}$, the posterior predictive of $\bm{\sigma}$ shifts greatly to zero (Fig.~\ref{fig:mcmc_convergence}b). Moreover, the convergence of the MCMC method become much worse when the hyperprior is less informative. It is challenging to use MCMC inference in our hierarchical model, as we can not design the hyperprior distribution for real experimental data. In contrast, we can achieve convergence with empirical Bayes inference. The typical iteration number for L-BFGS-B is less than 20 (when estimating the ``$\SODS$" simulation data with simulation trials $N_s=50$, data length $K=500$ or $K=1000$, Neurons number $N= 100$), which is much less than the maximum 5000 iteration we set for L-BFGS-B. Results in Fig.\ref{fig:weights} show that when using enough simulation trials and data length, we can achieve R-squared value higher than $0.95$ for weights estimate. The good fit of estimated and true weights indicates the convergence of the optimization in empirical Bayes inference. 
\subsection{Estimation Results of Experimental Data}
\label{ssec:experimental}
\begin{figure}[tbp]
\centering
\includegraphics[width=1.8in]{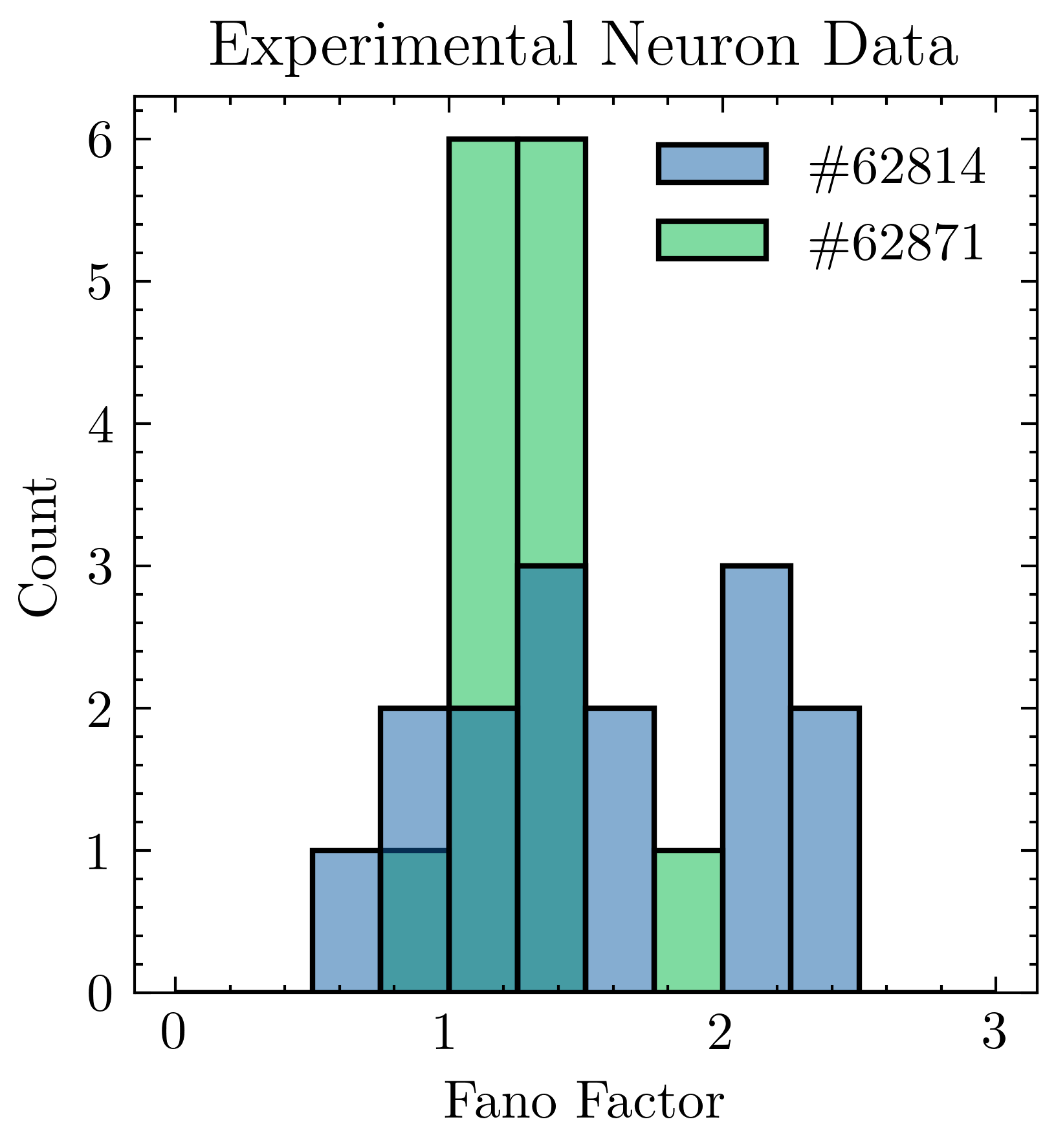}
\caption{The fano factor histogram of the experimental dataset neurons. Neurons with fano factor \textgreater 1 or fano factor \textless 1 were found in both datasets.}
\label{fig:real_data_overdispersion}
\end{figure}

The experimental data used here is taken from multi-unit recordings of retinal ganglion cells from the $\mathrm{ret}$-$1$ database~\cite{lefebvre2008gamma,zhang2014multi}, curated at $\mathrm{CRCNS.org}$. This database has single-unit neural responses recorded using a 61-electrode array from isolated retina of mice. The recordings were taken in response to various visual stimuli with the aim of aiding understanding of how different visual stimuli influence the spiking activity of retina cells. For population activity, network models using the $\GLM$ framework are quite popular~\cite{chen2011statistical,song2015identification,pillow2008spatio}. Therefore, we test our framework with state-of-the-art methods on $2$ datasets containing $15$ and $14$ neurons, respectively. The experimental data (spike counts) were binned into $16$ ms bins. This bin size is a trade-off between how finely time is discretized and the computational costs.
Fig.~\ref{fig:real_data_overdispersion} shows the experimental neurons' fano factors, which are varied from 0.45 to 2.68.
\begin{figure}[tbp]
    \centering
    \includegraphics[width=0.8\columnwidth]{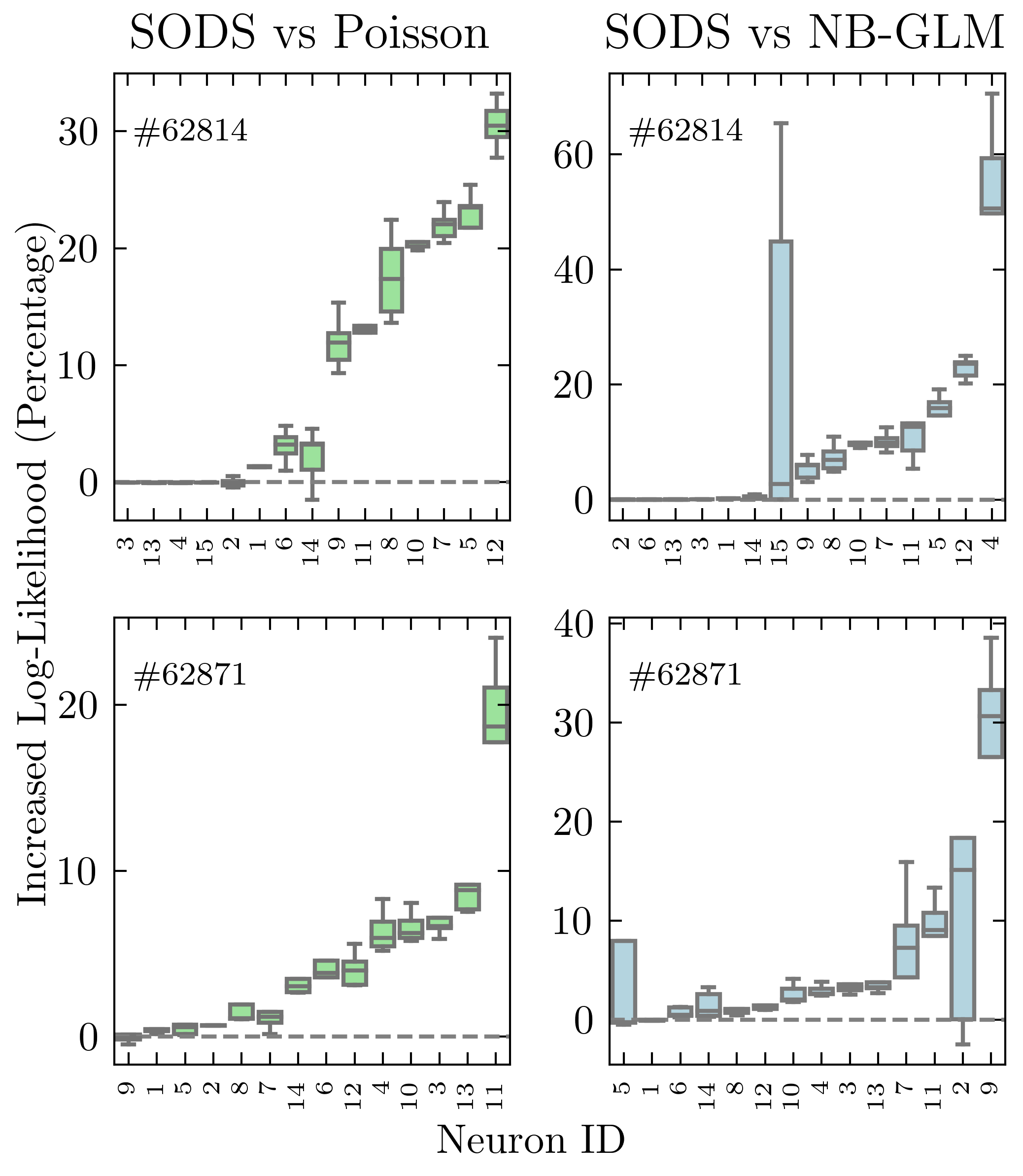}
    \caption{The box-plots of the percentage increase in held-out test data log-likelihood (sorted by median value). The log-likelihood is calculated by 5-fold cross-validation, using spiking data in dataset \#62814 (15 neurons) and \#62871 (14 neurons). Left column: ``$\SODS$" compared with Poisson-like $\GLM$; Right column:``$\SODS$" compared with NB-$\GLM$}
    \label{fig:exp_ll}
\end{figure}
\begin{figure}[tbp]
    \centering
    \includegraphics[width=0.8\columnwidth]{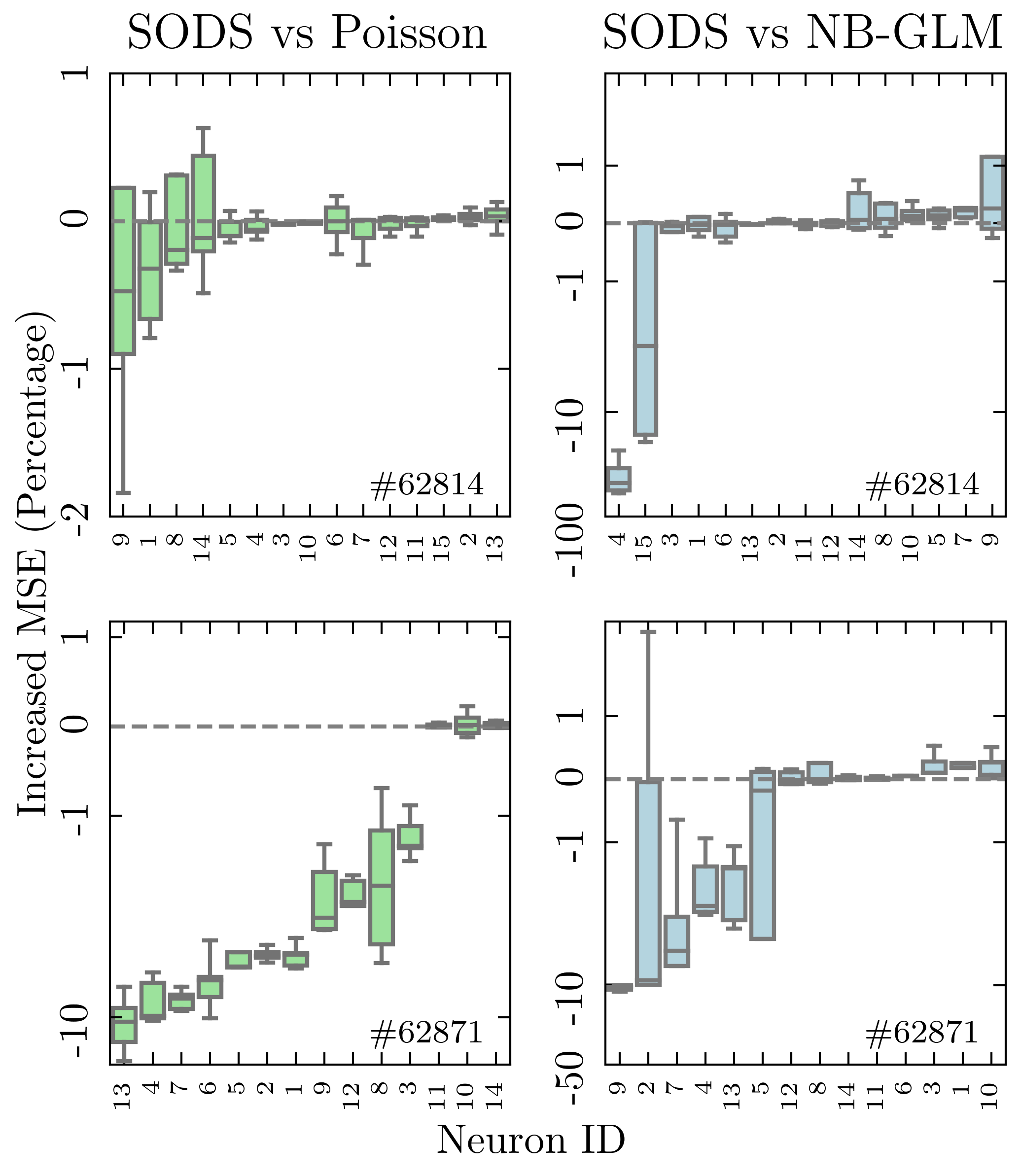}
    \caption{The box-plots of the percentage increase in held-out test data MSE (sorted by median value). The MSE is calculated between the estimated and actual mean spike counts, via 5-fold cross-validation, using spiking data in dataset \#62814 (15 neurons) and \#62871 (14 neurons). Left column: ``$\SODS$" compared with Poisson-like $\GLM$; Right column: ``$\SODS$" compared with NB-$\GLM$}
    \label{fig:exp_mse}\vspace{-3mm}
\end{figure}
\begin{figure}[tbp]
    \centering
    \includegraphics[width=0.8\columnwidth]{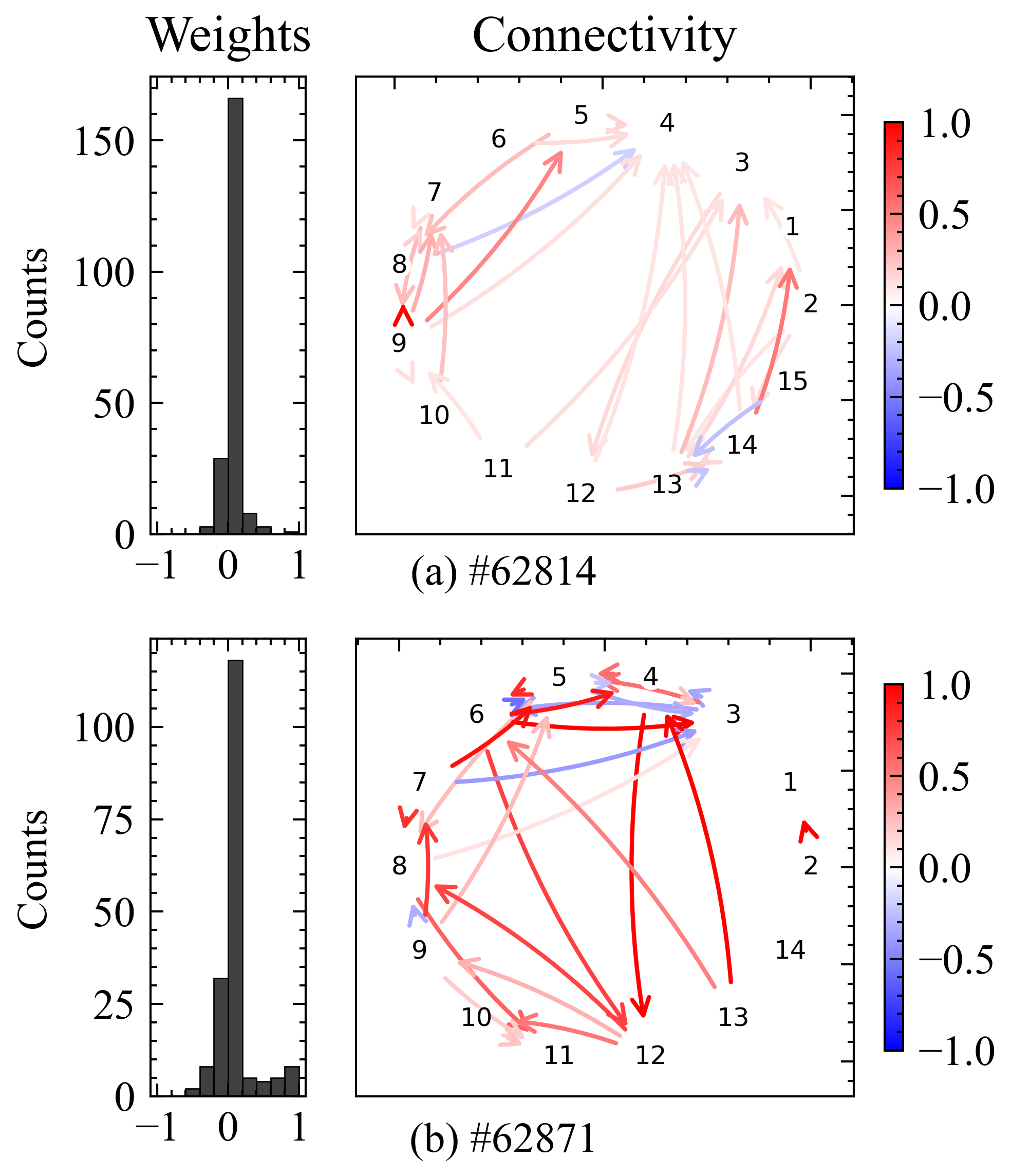}
    \caption{Estimated network weights of two experimental datasets \#62814(a) and \#62871 (b) using ``$\SODS$" model. The neural interactions recovered from two dataset showed sparsity. Left: histogram of the estimated weights; right: the network visualization of the estimated weights. Each number indicating one neuron. Red/blue lines highlight the positive/negative weights, and the color intensity indicates the weight strength. The edges with weights lower than 0.05 were not presented.}
    \label{fig:real_connection}
\end{figure}

The estimation using experimental data is similar to the one in SNN data. The mean spike counts of an individual neuron are estimated from other neurons within the same dataset. 5-fold cross-validation was used to find the optimal parameters for elastic-net regularization. Another 5-fold cross-validation was used to estimate the predictive log-likelihoods and MSE. 

``$\SODS$" estimator outperforms Poisson-like $\GLM$ and NB-$\GLM$ in the predictive log-likelihood comparisons. Fig.~\ref{fig:exp_ll} shows 66.67\% of neurons in \#62814 and 92.86\% of neurons in \#62871 present higher predictive log-likelihoods in ``$\SODS$" estimator compared to Poisson-like $\GLM$. 93.33\% of neurons in \#62814 and 85.71\% of neurons in \#62871 have higher averaged predictive log-likelihoods when using the ``$\SODS$" model versus using NB-$\GLM$. The comparison of MSE values for three estimators shows that 53.33\% of neurons in \#62814 and 78.57\% of neurons in \#62871 have lower MSE when using ``$\SODS$" model versus using the Poisson-like GLM (Fig.~\ref{fig:exp_mse}). But only 40.00\% of neurons in \#62814 and 78.57\%  of neurons in \#62871 showed lower MSE with the ``$\SODS$" estimator than with NB-$\GLM$ estimator.

We estimated the network weights for two experimental datasets with ``$\SODS$". Fig.~\ref{fig:real_connection} shows the network weights estimated using ``$\SODS$" for two experimental datasets. Around $87.63\%$ of total weights strength in \#62814 dataset and $79.56\%$ in \#62871 dataset is positive $\left(\frac{|\bm{\omega_{+}}|}{|\bm{\omega_{+}}|+|\bm{\omega_{-}}|}\right)$. The weights histogram in Fig.~\ref{fig:real_connection} showed most of the weights were closed to zeros, which indicating that it may be a sparse network. More accurate inference in the coupling weights under neural circuits provides more insights about how the neuronal population process the information, as the neural circuits utilize the balanced or imbalanced excitation and inhibition to facilitate the information processing \cite{malina2013imbalance,higley2006balanced}.
\section{Discussion and Conclusion}
\label{sec:Discus}
The bio-signal processing community has shown great interest in multivariate regression methods~\cite{schmidt2011estimating,giurcuaneanu2014new,dai2015generalized,rangayyan2015biomedical,sheikhattar2016recursive}. These methods can provide a clear view of the nature of neuronal interactions. Linderman \emph{et al.}~\cite{linderman2016bayesian} developed a fully Bayesian inference method for Negative Binomial responses that yields regularized estimations for all of the hyperparameters. Although it can have uncertainties (probability distributions) on all the parameters, applying fully Bayesian approaches to hierarchical models is computationally intensive. As an alternative, empirical Bayes can provide a bias-variance trade-off which can achieve a small mean square error at a lower computational cost. To estimate the unknown parameters of the model, Paninski \emph{et al.}~\cite{paninski2004maximum} used maximum likelihood estimation, but when the dataset is small, the estimation becomes biased. \textit{The ``$\SODS$" estimator developed here, to model over-dispersed spiking behaviour and extract latent interactions among neural populations, combines both of the above methods}. It has the benefit of providing a bias-variance trade-off estimator for Negative Binomial responses, while not needing the intensive computation of fully Bayesian inference. 

We took advantage of the beneficial properties of both $\GLMs$ and empirical Bayes inference to propose the ``$\SODS$" estimator. We used the Negative Binomial distribution to model the spike counts of each neuron. The Negative Binomial distribution was selected as it allows for over-dispersed spike counts using a dispersion parameter superior to the standard Poisson model. The beta distribution is employed as the prior information for the probability parameter in the Negative Binomial distribution, which allows for a closed-form posterior distribution. We propose a flexible link function family in order to model the prior mean using regressors. By using the recorded data from other neurons as the covariates, we can then infer the functional weights among the neural population. Unlike fully Bayesian inference, which utilizes informative hyperpriors in our cases, we instead estimate the hyperparameters by maximizing the marginal likelihood. The proposed ``$\SODS$" estimator is a shrinkage estimator and the weights we estimate can be viewed as the hidden functional dependences. By taking the neurons as nodes in our functional neural network, and their spike-train data as the observations, our empirical Bayes inference method can be used to identify the neural interactions, including excitatory and inhibitory behaviours. 

We have validated our method using both simulated data and experimental retinal neuron data. Compared with two of the most widely used regression methods: Poisson and Negative Binomial regressions, ``$\SODS$" outperforms by accurately estimating simulated spiking data from different simulation systems. The performance of ``$\SODS$" remains excellent when handling spiking data simulated from NB-$\GLM$ and Poisson-$\GLM$ system. The performance of ``$\SODS$" is no less than Poisson-$\GLM$ and much better than NB-$\GLM$ in estimating spiking data simulated from SNN model. Moreover, we found that ``$\SODS$" can accurately recover the neuronal dependency by estimating the weights parameter from the simulation data. With elastic-net regularization, ``$\SODS$" can estimate the sparse properties of neural network. Although performance decreased when partial neurons were missing during estimation, ``$\SODS$" model is still good at identifying the zeros weights in the connectivity. For the experimental neurons, there was a substantial improvement in the predictive log-likelihood of the held-out data when compared with NB-$\GLM$ and Poisson-$\GLM$ methods. 

While the results presented here are promising, going forward, we are interested in extending our model. For instance, the incorporation of Hebbian learning rules could account for time-varying weights. Applying prior knowledge regarding network structure, such as random, small world or scale-free networks, could also be a promising avenue for future research. External covariates can also be incorporated into the model to capture neural patterns of stimulus dependence. Finally, the ability of our model to operate in data-limited cases would open possibilities for future applications to real-time settings, such as for closed-loop experiments or improved brain-machine interface (BMI) devices.
\section*{Acknowledgements}
The authors thank Prof. J. T. Kwok, who provided valuable comments and insightful discussions on the derivations, experiments in the manuscript. The work described in this paper was fully supported by grants from the Research Grants Council of the Hong Kong Special Administrative Region, China [Project No. CityU 11214020 and CityU C1020-19E]. 


%

%
%
%
%
%

\ifCLASSOPTIONcaptionsoff
  \newpage
\fi



%
\bibliographystyle{IEEEtran}
{\small{\bibliography{Ref}}}

\begin{thebibliography}{10}
\providecommand{\url}[1]{#1}
\csname url@samestyle\endcsname
\providecommand{\newblock}{\relax}
\providecommand{\bibinfo}[2]{#2}
\providecommand{\BIBentrySTDinterwordspacing}{\spaceskip=0pt\relax}
\providecommand{\BIBentryALTinterwordstretchfactor}{4}
\providecommand{\BIBentryALTinterwordspacing}{\spaceskip=\fontdimen2\font plus
\BIBentryALTinterwordstretchfactor\fontdimen3\font minus
  \fontdimen4\font\relax}
\providecommand{\BIBforeignlanguage}[2]{{%
\expandafter\ifx\csname l@#1\endcsname\relax
\typeout{** WARNING: IEEEtran.bst: No hyphenation pattern has been}%
\typeout{** loaded for the language `#1'. Using the pattern for}%
\typeout{** the default language instead.}%
\else
\language=\csname l@#1\endcsname
\fi
#2}}
\providecommand{\BIBdecl}{\relax}
\BIBdecl

\bibitem{park2013kernel}
I.~M. Park, S.~Seth, A.~R. Paiva, L.~Li, and J.~C. Principe, ``Kernel methods
  on spike train space for neuroscience: a tutorial,'' \emph{IEEE Signal
  Processing Magazine}, vol.~30, no.~4, pp. 149--160, 2013.

\bibitem{chen2011statistical}
Z.~Chen, D.~F. Putrino, S.~Ghosh, R.~Barbieri, and E.~N. Brown, ``Statistical
  inference for assessing functional connectivity of neuronal ensembles with
  sparse spiking data,'' \emph{IEEE Transactions on Neural Systems and
  Rehabilitation Engineering}, vol.~19, no.~2, pp. 121--135, 2011.

\bibitem{okatan2005analyzing}
M.~Okatan, M.~A. Wilson, and E.~N. Brown, ``Analyzing functional connectivity
  using a network likelihood model of ensemble neural spiking activity,''
  \emph{Neural Computation}, vol.~17, no.~9, pp. 1927--1961, 2005.

\bibitem{song2015identification}
D.~Song, R.~H. Chan, B.~S. Robinson, V.~Z. Marmarelis, I.~Opris, R.~E. Hampson,
  S.~A. Deadwyler, and T.~W. Berger, ``Identification of functional synaptic
  plasticity from spiking activities using nonlinear dynamical modeling,''
  \emph{Journal of Neuroscience Methods}, vol. 244, pp. 123--135, 2015.

\bibitem{pillow2008spatio}
J.~W. Pillow, J.~Shlens, L.~Paninski, A.~Sher, A.~M. Litke, E.~Chichilnisky,
  and E.~P. Simoncelli, ``Spatio-temporal correlations and visual signalling in
  a complete neuronal population,'' \emph{Nature}, vol. 454, no. 7207, pp.
  995--999, 2008.

\bibitem{seghouane2014sparse}
A.-K. Seghouane and A.~Shah, ``Sparse estimation of the hemodynamic response
  functionin functional near infrared spectroscopy,'' in \emph{Acoustics,
  Speech and Signal Processing (ICASSP), 2014 IEEE International Conference
  on}.\hskip 1em plus 0.5em minus 0.4em\relax IEEE, 2014, pp. 2074--2078.

\bibitem{she2015reconstruction}
Q.~She, W.~K. So, and R.~H. Chan, ``Reconstruction of neural network topology
  using spike train data: Small-world features of hippocampal network,'' in
  \emph{2015 37th Annual International Conference of the IEEE Engineering in
  Medicine and Biology Society (EMBC)}.\hskip 1em plus 0.5em minus 0.4em\relax
  IEEE, 2015, pp. 2506--2509.

\bibitem{she2016evaluating}
Q.~She, G.~Chen, and R.~H. Chan, ``Evaluating the small-world-ness of a sampled
  network: Functional connectivity of entorhinal-hippocampal circuitry,''
  \emph{Scientific Reports}, vol.~6, 2016.

\bibitem{she2016effective}
Q.~She, W.~K. So, and R.~H. Chan, ``Effective connectivity matrix for neural
  ensembles,'' in \emph{2016 38th Annual International Conference of the IEEE
  Engineering in Medicine and Biology Society (EMBC)}.\hskip 1em plus 0.5em
  minus 0.4em\relax IEEE, 2016, pp. 1612--1615.

\bibitem{lakshmanan2015extracting}
K.~C. Lakshmanan, P.~T. Sadtler, E.~C. Tyler-Kabara, A.~P. Batista, and M.~Y.
  Byron, ``Extracting low-dimensional latent structure from time series in the
  presence of delays,'' \emph{Neural Computation}, vol.~27, no.~9, pp.
  1825--1856, 2015.

\bibitem{cichocki2015tensor}
A.~Cichocki, D.~Mandic, L.~De~Lathauwer, G.~Zhou, Q.~Zhao, C.~Caiafa, and H.~A.
  Phan, ``Tensor decompositions for signal processing applications: From
  two-way to multiway component analysis,'' \emph{IEEE Signal Processing
  Magazine}, vol.~32, no.~2, pp. 145--163, 2015.

\bibitem{palmerston2018weighted}
J.~B. Palmerston, Q.~She, and R.~H. Chan, ``Weighted network density predicts
  range of latent variable model accuracy,'' in \emph{2018 40th Annual
  International Conference of the IEEE Engineering in Medicine and Biology
  Society (EMBC)}.\hskip 1em plus 0.5em minus 0.4em\relax IEEE, 2018, pp.
  2414--2417.

\bibitem{she2018reduced}
Q.~She, Y.~Gao, K.~Xu, and R.~Chan, ``Reduced-rank linear dynamical systems,''
  in \emph{Proceedings of the AAAI Conference on Artificial Intelligence},
  vol.~32, no.~1, 2018.

\bibitem{she2020neural}
Q.~She and A.~Wu, ``Neural dynamics discovery via gaussian process recurrent
  neural networks,'' in \emph{Uncertainty in Artificial Intelligence}.\hskip
  1em plus 0.5em minus 0.4em\relax PMLR, 2020, pp. 454--464.

\bibitem{pillow2003maximum}
J.~W. Pillow, L.~Paninski, and E.~P. Simoncelli, ``Maximum likelihood
  estimation of a stochastic integrate-and-fire neural model.'' in
  \emph{{Advances in Neural Information Processing Systems}}, 2003, pp.
  1311--1318.

\bibitem{hawkes1971spectra}
A.~G. Hawkes, ``Spectra of some self-exciting and mutually exciting point
  processes,'' \emph{Biometrika}, vol.~58, no.~1, pp. 83--90, 1971.

\bibitem{blundell2012modelling}
C.~Blundell, J.~Beck, and K.~A. Heller, ``Modelling reciprocating relationships
  with hawkes processes,'' in \emph{Advances in Neural Information Processing
  Systems}, 2012, pp. 2600--2608.

\bibitem{moore2016hawkes}
M.~G. Moore and M.~A. Davenport, ``A {H}awkes' eye view of network information
  flow,'' in \emph{Statistical Signal Processing Workshop (SSP), 2016
  IEEE}.\hskip 1em plus 0.5em minus 0.4em\relax IEEE, 2016, pp. 1--5.

\bibitem{chen2017multivariate}
S.~Chen, A.~Shojaie, E.~Shea-Brown, and D.~Witten, ``The multivariate hawkes
  process in high dimensions: Beyond mutual excitation,'' \emph{arXiv preprint
  arXiv:1707.04928}, 2017.

\bibitem{mark2018network}
B.~Mark, G.~Raskutti, and R.~Willett, ``Network estimation from point process
  data,'' \emph{IEEE Transactions on Information Theory}, vol.~65, no.~5, pp.
  2953--2975, 2018.

\bibitem{linderman2014discovering}
S.~W. Linderman and R.~P. Adams, ``Discovering latent network structure in
  point process data,'' in \emph{International Conference on Machine Learning},
  2014, pp. 1413--1421.

\bibitem{macke2011empirical}
J.~H. Macke, L.~Buesing, J.~P. Cunningham, M.~Y. Byron, K.~V. Shenoy, and
  M.~Sahani, ``Empirical models of spiking in neural populations,'' in
  \emph{{Advances in Neural Information Processing Systems}}, 2011, pp.
  1350--1358.

\bibitem{churchland2010stimulus}
M.~M. Churchland, M.~Y. Byron, J.~P. Cunningham, L.~P. Sugrue, M.~R. Cohen,
  Corrado \emph{et~al.}, ``Stimulus onset quenches neural variability: a
  widespread cortical phenomenon,'' \emph{Nature Neuroscience}, vol.~13, no.~3,
  pp. 369--378, 2010.

\bibitem{goris2014partitioning}
R.~L. Goris, J.~A. Movshon, and E.~P. Simoncelli, ``Partitioning neuronal
  variability,'' \emph{Nature Neuroscience}, vol.~17, no.~6, pp. 858--865,
  2014.

\bibitem{gao2015high}
Y.~Gao, L.~Busing, K.~V. Shenoy, and J.~P. Cunningham, ``High-dimensional
  neural spike train analysis with generalized count linear dynamical
  systems,'' in \emph{{Advances in Neural Information Processing Systems}},
  2015, pp. 2044--2052.

\bibitem{stevenson2016flexible}
I.~H. Stevenson, ``Flexible models for spike count data with both over-and
  under-dispersion,'' \emph{Journal of Computational Neuroscience}, pp. 1--15,
  2016.

\bibitem{lawless1987negative}
J.~F. Lawless, ``Negative binomial and mixed poisson regression,''
  \emph{Canadian Journal of Statistics}, vol.~15, no.~3, pp. 209--225, 1987.

\bibitem{chen1998mean}
B.~Chen, ``Mean-value deviance detection of transient signals modeled as
  overdispersed {DFT} data,'' \emph{IEEE Transactions on Signal Processing},
  1998.

\bibitem{gao2016linear}
Y.~Gao, E.~W. Archer, L.~Paninski, and J.~P. Cunningham, ``Linear dynamical
  neural population models through nonlinear embeddings,'' in \emph{Advances in
  Neural Information Processing Systems}, 2016, pp. 163--171.

\bibitem{brown2004multiple}
E.~N. Brown, R.~E. Kass, and P.~P. Mitra, ``Multiple neural spike train data
  analysis: state-of-the-art and future challenges,'' \emph{Nature
  Neuroscience}, vol.~7, no.~5, p. 456, 2004.

\bibitem{spira2013multi}
M.~E. Spira and A.~Hai, ``Multi-electrode array technologies for neuroscience
  and cardiology,'' \emph{Nature Nanotechnology}, vol.~8, no.~2, pp. 83--94,
  2013.

\bibitem{mauk2004neural}
M.~D. Mauk and D.~V. Buonomano, ``The neural basis of temporal processing,''
  \emph{Annual Review of Neuroscience}, vol.~27, pp. 307--340, 2004.

\bibitem{bertschinger2004real}
N.~Bertschinger and T.~Natschl{\"a}ger, ``Real-time computation at the edge of
  chaos in recurrent neural networks,'' \emph{Neural Computation}, vol.~16,
  no.~7, pp. 1413--1436, 2004.

\bibitem{Charles165670}
A.~S. Charles, M.~Park, J.~P. Weller, G.~D. Horwitz, and J.~W. Pillow,
  ``Dethroning the fano factor: a flexible, model-based approach to
  partitioning neural variability,'' \emph{Neural computation}, vol.~30, no.~4,
  pp. 1012--1045, 2018.

\bibitem{scott2012fully}
J.~Scott and J.~W. Pillow, ``Fully {B}ayesian inference for neural models with
  negative-binomial spiking,'' in \emph{{Advances in Neural Information
  Processing Systems}}, 2012, pp. 1898--1906.

\bibitem{pessoa2014understanding}
L.~Pessoa, ``Understanding brain networks and brain organization,''
  \emph{Physics of life reviews}, vol.~11, no.~3, pp. 400--435, 2014.

\bibitem{nelder1972generalized}
J.~A. Nelder and R.~J. Baker, \emph{Generalized Linear Models}.\hskip 1em plus
  0.5em minus 0.4em\relax Wiley Online Library, 1972.

\bibitem{friedman2010regularization}
J.~Friedman, T.~Hastie, and R.~Tibshirani, ``Regularization paths for
  generalized linear models via coordinate descent,'' \emph{Journal of
  Statistical Software}, vol.~33, no.~1, p.~1, 2010.

\bibitem{shanno1985broyden}
D.~F. Shanno, ``On {Broyden-Fletcher-Goldfarb-Shanno} method,'' \emph{Journal
  of Optimization Theory and Applications}, vol.~46, no.~1, pp. 87--94, 1985.

\bibitem{liu1989limited}
D.~C. Liu and J.~Nocedal, ``On the limited memory {BFGS} method for large scale
  optimization,'' \emph{Mathematical Programming}, vol.~45, no. 1-3, pp.
  503--528, 1989.

\bibitem{jensen1906fonctions}
J.~L. W.~V. Jensen \emph{et~al.}, ``Sur les fonctions convexes et les
  in{\'e}galit{\'e}s entre les valeurs moyennes,'' \emph{Acta mathematica},
  vol.~30, pp. 175--193, 1906.

\bibitem{olson2012basin}
B.~Olson, I.~Hashmi, K.~Molloy, and A.~Shehu, ``Basin hopping as a general and
  versatile optimization framework for the characterization of biological
  macromolecules.'' \emph{Advances in Artificial Intelligence (16877470)},
  2012.

\bibitem{wales1997global}
D.~J. Wales and J.~P. Doye, ``Global optimization by basin-hopping and the
  lowest energy structures of lennard-jones clusters containing up to 110
  atoms,'' \emph{The Journal of Physical Chemistry A}, vol. 101, no.~28, pp.
  5111--5116, 1997.

\bibitem{scipytool}
\BIBentryALTinterwordspacing
E.~Jones, T.~Oliphant, P.~Peterson \emph{et~al.}, ``{SciPy}: Open source
  scientific tools for {Python},'' 2001--. [Online]. Available:
  \url{http://www.scipy.org/}
\BIBentrySTDinterwordspacing

\bibitem{park2014encoding}
I.~M. Park, M.~L. Meister, A.~C. Huk, and J.~W. Pillow, ``Encoding and decoding
  in parietal cortex during sensorimotor decision-making,'' \emph{Nature
  Neuroscience}, vol.~17, no.~10, pp. 1395--1403, 2014.

\bibitem{brette2007simulation}
R.~Brette, M.~Rudolph, T.~Carnevale, M.~Hines, D.~Beeman, J.~M. Bower,
  M.~Diesmann, A.~Morrison, P.~H. Goodman, F.~C. Harris \emph{et~al.},
  ``Simulation of networks of spiking neurons: a review of tools and
  strategies,'' \emph{Journal of Computational Neuroscience}, vol.~23, no.~3,
  pp. 349--398, 2007.

\bibitem{stimberg2019brian}
M.~Stimberg, R.~Brette, and D.~F. Goodman, ``Brian 2, an intuitive and
  efficient neural simulator,'' \emph{Elife}, vol.~8, p. e47314, 2019.

\bibitem{nocedal2006numerical}
J.~Nocedal and S.~Wright, \emph{Numerical optimization}.\hskip 1em plus 0.5em
  minus 0.4em\relax Springer Science \& Business Media, 2006.

\bibitem{hoffman2014no}
M.~D. Hoffman and A.~Gelman, ``The no-u-turn sampler: adaptively setting path
  lengths in hamiltonian monte carlo.'' \emph{J. Mach. Learn. Res.}, vol.~15,
  no.~1, pp. 1593--1623, 2014.

\bibitem{lefebvre2008gamma}
J.~L. Lefebvre, Y.~Zhang, M.~Meister, X.~Wang, and J.~R. Sanes,
  ``$\gamma$-protocadherins regulate neuronal survival but are dispensable for
  circuit formation in retina,'' \emph{Development}, vol. 135, no.~24, pp.
  4141--4151, 2008.

\bibitem{zhang2014multi}
Y.~Zhang, H.~Asari, and M.~Meister, ``Multi-electrode recordings from retinal
  ganglion cells,'' \emph{CRCNS. org., 2014b. URL http://dx. doi.
  org/10.6080/K0RF5RZT}, 2014.

\bibitem{malina2013imbalance}
K.~C.-K. Malina, M.~Jubran, Y.~Katz, and I.~Lampl, ``Imbalance between
  excitation and inhibition in the somatosensory cortex produces postadaptation
  facilitation,'' \emph{Journal of Neuroscience}, vol.~33, no.~19, pp.
  8463--8471, 2013.

\bibitem{higley2006balanced}
M.~J. Higley and D.~Contreras, ``Balanced excitation and inhibition determine
  spike timing during frequency adaptation,'' \emph{Journal of Neuroscience},
  vol.~26, no.~2, pp. 448--457, 2006.

\bibitem{schmidt2011estimating}
D.~F. Schmidt and E.~Makalic, ``Estimating the order of an autoregressive model
  using normalized maximum likelihood,'' \emph{IEEE Transactions on Signal
  Processing}, vol.~59, no.~2, pp. 479--487, 2011.

\bibitem{giurcuaneanu2014new}
C.~D. Giurc{\u{a}}neanu and F.~A.~A. Saip, ``New insights on {AR} order
  selection with information theoretic criteria based on localized
  estimators,'' \emph{Digital Signal Processing}, vol.~32, pp. 37--47, 2014.

\bibitem{dai2015generalized}
W.~Dai, H.~Xiong, J.~Wang, S.~Cheng, and Y.~F. Zheng, ``Generalized context
  modeling with multi-directional structuring and {MDL}-based model selection
  for heterogeneous data compression,'' \emph{IEEE Transactions on Signal
  Processing}, vol.~63, no.~21, pp. 5650--5664, 2015.

\bibitem{rangayyan2015biomedical}
R.~M. Rangayyan, \emph{Biomedical Signal Analysis}.\hskip 1em plus 0.5em minus
  0.4em\relax John Wiley \& Sons, 2015, vol.~33.

\bibitem{sheikhattar2016recursive}
A.~Sheikhattar, J.~B. Fritz, S.~A. Shamma, and B.~Babadi, ``Recursive sparse
  point process regression with application to spectrotemporal receptive field
  plasticity analysis,'' \emph{IEEE Transactions on Signal Processing},
  vol.~64, no.~8, pp. 2026--2039, 2016.

\bibitem{linderman2016bayesian}
S.~Linderman, R.~P. Adams, and J.~W. Pillow, ``Bayesian latent structure
  discovery from multi-neuron recordings,'' in \emph{Advances in Neural
  Information Processing Systems}, 2016, pp. 2002--2010.

\bibitem{paninski2004maximum}
L.~Paninski, ``Maximum likelihood estimation of cascade point-process neural
  encoding models,'' \emph{Network: Computation in Neural Systems}, vol.~15,
  no.~4, pp. 243--262, 2004.

\end{thebibliography}

\end{document}